\documentclass{statsoc}
\usepackage{natbib}
\usepackage{amsmath,amssymb,subfigure,graphicx,setspace,mathptmx}
\usepackage[text={6in,9in},right=1in]{geometry}

\newcommand{\bftheta}{\hbox{\boldmath$\theta$}}
\newcommand{\bfL}{\hbox{\boldmath$L$}}
\newcommand{\bfX}{\hbox{\boldmath$X$}}
\newcommand{\bfv}{\hbox{\boldmath$v$}}
\newcommand{\bfu}{\hbox{\boldmath$u$}}
\newcommand{\dnorm}[2]{\ensuremath{\mathcal{N}\left(#1, #2\right)}}
\newcommand{\dgamma}[2]{\ensuremath{\text{Gamma}\left(#1, #2\right)}}
\newcommand{\bfalpha}{\hbox{\boldmath$\alpha$}}
\newcommand{\bfmu}{\hbox{\boldmath$\mu$}}

\title{Bayesian Analysis of Loss Ratios Using the Reversible Jump Algorithm}

\author{G. O. Brown\footnote{Corresponding author, Email:\texttt{gob20@statslab.cam.ac.uk}}} 
\address{Statistical Laboratory, Centre for Mathematical Sciences, 
  Cambridge CB3 0WB, UK}

\author[]{S. P. Brooks}
\address{Statistical Laboratory, Centre for Mathematical Sciences, 
  Cambridge CB3 0WB, UK}

\begin{document}
\maketitle

\begin{abstract}
  In this paper we consider the problem of model choice for a set of
  insurance loss ratios. We use a reversible jump algorithm for our model
  discrimination and show how the vanilla reversible jump algorithm can be
  improved on using recent methodological advances in reversible jump
  computation.
\end{abstract}

\clearpage
\section{Introduction}
In traditional insurance settings model selection and uncertainty are usually
not treated, even though model selection problems have been actively
researched in statistics. Recently this shortcoming has been addressed by
several authors: \citet{cairns2000} in general insurance and risk theory;
\citet{keatinge1999} in estimating the number of components in a mixture of
exponentials for estimating claims amounts; also \citet{harris1999} considers
the problem of model selection for vector autoregression for financial time
series. In the field of credibility theory, \citet{buhlmann1999} considers
the selection of variables in certain regression credibility models.

In this paper we consider the problem of parameter estimation and model
selection in the analysis of workers' compensation loss ratios.  This paper
is motivated by the analysis of data consisting of workers' compensation loss
ratios arising over a seven year period.  The data are part of a set
containing frequency counts on workers' compensation insurance. The number of
claims against the workers' compensation insurance scheme is recorded,
together with the corresponding exposure values.  The exposures are scaled
payroll totals and provide a measure of the size of the exposed group.

The model we fit to the data is described in Section~\ref{sec:ds3fullmodel},
%%and is a variation of \citeN{gerber1975},
we then introduce two additional models, both of which are sub-models of the
first. Using the reversible jump method described in
Section~\ref{sec:reversible} we discriminate between the three models. We
also use the efficient proposals method of \citet{brooks2003} to derive
proposals for our reversible jump updates.  The results are compared with the
pilot-tuned vanilla reversible jump of \citet{green1995}.

\section{The Data and Model}\label{sec:ds3fullmodel}
\begin{figure}
  \centering
  \makebox{
    \includegraphics[width=0.8\textwidth]{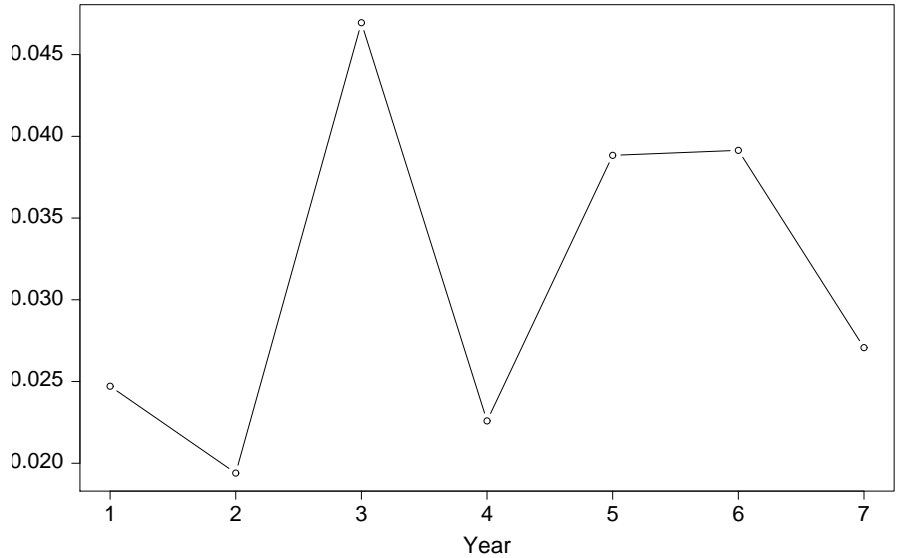}}
  \caption{\label{fig:ds3-data}A plot of the loss ratios against year.}
\end{figure}

We denote the number of claims for year $j$ by $L_j$ and the corresponding
exposure values by $E_j$ for $j=1,\ldots,n$. For this particular dataset we
have $n=7$. The loss ratios which we propose to model are then defined as the
number of losses per unit exposure and will be denoted by $R_j$, where
$R_j=L_j/E_j$.  Let $R^n$ denote the collective loss ratios and $E^n$ denote
the collective exposure values.  Here, we use a hierarchical normal model to
describe the loss ratios, so that
\begin{equation}\label{eq:model}
  R_j\sim\dnorm{\alpha_j}{(\sigma E_j)^{-1}}\, j = 1,\ldots, n
\end{equation}
where $\alpha_j$ denotes some underlying time-varying process which describes
the progression of ratio level over time. Here, we follow \citet{klugman1992}
and adopt the following model for the $\alpha_j$ process
%%which is similar to the Markov model of \citeN{bonsdorff1990}
%% introduced in Section~\ref{sec:recursiveprocesses}.
\begin{equation}\label{eq:hyper}
  \alpha_j\sim\dnorm{\rho\alpha_{j-1}+(1-\rho)\eta}{\tau^{-1}}\,j=1,\ldots,n.
\end{equation}
For $\alpha_0$, $\rho$, and $\eta$ we use standard normal \dnorm{0}{1} priors
and for the precision (inverse variance) parameters $\sigma$ and $\tau$ we
use \dgamma{a}{b} priors. The literature provides empirical evidence to
support the introduction of this model for describing loss ratios
\citep{ledolter1991} and a simple plot of the data in
Figure~\ref{fig:ds3-data} confirms that the observed behaviour can be
described by a model of this sort.  However, one disadvantage of this model
is that whereas the loss ratios are always non-negative, the normal model has
support extending across the entire real line so that negative values could,
in theory, occur.  One way around this would be to restrict the normal model
in \eqref{eq:model} to loss ratios within the region $[0,\infty)$ and/or to
impose similar restrictions on the $\alpha_j$ process.  These restrictions
are very easily implemented as a trivial extension of the scheme we describe
here but, since by adopting the more general model, serious failures in the
ability of the model to describe the observed data can be detected when
negative estimates are obtained, the more general model provides a useful
check for the adequacy of our modelling scheme.

\section{A Gibbs Sampling Algorithm}\label{sec:postcond}
For the model presented in Equations~\eqref{eq:model} and \eqref{eq:hyper} we
use Gibbs updates to obtain samples from the posterior distribution of the
model parameters. The full joint posterior distribution of all the model
parameters is given by
\begin{equation*}
  \pi(\bfalpha, \tau, \sigma, \rho, \eta | R^n,  E^n) \propto
  \bfL(R^n | \bfalpha, \sigma, E^n) p(\bfalpha | \rho,\eta,\tau,\alpha_0)
  p(\sigma)p(\tau)p(\rho)p(\eta)p(\alpha_0),
\end{equation*}
where $\bfalpha$ denotes the collection $(\alpha_1, \cdots, \alpha_n)$, 
$p(\cdot)$ denotes the prior distribution for the corresponding
parameter and the likelihood term
\begin{equation*}
  \bfL(R^n | \bfalpha, \sigma,  E^n) =
  \prod_{j=1}^n f (R_j\vert\bfalpha,\sigma,  E^n) .
\end{equation*}
We now derive the full posterior conditional for each of the model parameters
in turn. These conditional distributions will then be used to implement a
Gibbs update algorithm. The full posterior conditional distribution for
$\sigma$ is
\begin{align*}%%\label{eq:postsigma}
  \pi(\sigma\vert \bfalpha) &\propto
  p (\sigma)\bfL(R^n\vert\bfalpha,\sigma, E^n) \nonumber\\
  &\propto \sigma^{a+\frac{n}{2} - 1}
  \exp\left\lbrace -\sigma\left(b+\tfrac1{2}\sum\nolimits_{j=1}^n
    E_j(R_j-\alpha_j)^2
  \right)\right\rbrace,
\end{align*}
which is a gamma distribution with shape parameter $a+\tfrac{n}{2}$ and scale
parameter $b+\tfrac1{2}\sum\nolimits_{j=1}^nE_j(R_j-\alpha_j)^2$. The full
posterior conditional for $\tau$ is
\begin{align*}%%\label{eq:posttau}
  \pi(\tau\vert\bfalpha,\rho,\eta ) &\propto
  p(\tau) p (\bfalpha\vert\alpha_0,\rho,\eta,\tau) \nonumber\\
  &\propto \tau^{a+\frac{n}{2}-1}
  \exp\left\lbrace-\tau\left( b + \tfrac1{2}\sum\nolimits_{j=1}^n
      (\alpha_j-\rho\alpha_{j-1}-(1-\rho)\eta)^2
    \right)\right\rbrace,
\end{align*}
which is a gamma distribution with shape parameter $a+\tfrac{n}{2}$ and scale
parameter $b+\tfrac1{2}\sum\nolimits_{j=1}^n
(\alpha_j-\rho\alpha_{j-1}-(1-\rho)\eta)^2$. The full posterior conditional
distribution for $\rho$ is
\begin{align*}%%\label{eq:postrho}
  \pi(\rho\vert\bfalpha,\eta,\tau)& \propto
  p(\rho)p(\bfalpha\vert\alpha_0,\rho,\eta)\nonumber\\
  &\propto \exp\left\lbrace
    -\frac{(1+\tau\sum\nolimits_{j=1}^n(\eta-\alpha_{j-1})^2)}{2}
    \left(
      \rho -
      \frac{\tau\sum_{j=1}^n(\eta-\alpha_j)(\eta-\alpha_{j-1})}
      {1+\tau\sum_{j=1}^n(\eta-\alpha_{j-1})^2}
    \right)^2
  \right\rbrace,
\end{align*}
which is a normal distribution with mean
\begin{equation*}
(1+\tau\sum\nolimits_{j=1}^n(\eta-\alpha_{j-1})^2)^{-1}
(\tau\sum\nolimits_{j=1}^n(\eta-\alpha_j)(\eta-\alpha_{j-1})),
\end{equation*}
and variance
$(1+\tau\sum\nolimits_{j=1}^n(\eta-\alpha_{j-1})^2)^{-1}$. The full posterior
conditional distribution for $\eta$ is
\begin{align*}%%\label{eq:posteta}
  \pi(\eta\vert\bfalpha,\rho,\tau) &\propto
  p(\eta) p(\bfalpha\vert\alpha_0,\rho,\eta)\nonumber\\
  &\propto \exp\left\lbrace
    -\frac{1+n\tau(1-\rho)^2}{2}
    \left(
      \eta -
      \frac{\tau(1-\rho)\sum\nolimits_{j=1}^n(\alpha_j-\rho\alpha_{j-1})}
      {1+n\tau(1-\rho)^2}
    \right)^2
  \right\rbrace,
\end{align*}
which is a normal distribution with mean
\begin{equation*}
  (1+n\tau(1-\rho)^2)^{-1}
  (\tau(1-\rho)\sum\nolimits_{j=1}^n(\alpha_j-\rho\alpha_{j-1})),
\end{equation*}
and variance $(1+n\tau(1-\rho)^2)^{-1}$.  The full posterior conditional
distribution for $\alpha_j$ is
\begin{multline*}
  \pi(\alpha_j\vert\bfalpha_{(j)},\rho,\eta,\sigma,\tau) \propto \\
  \begin{cases}
    p(\alpha_j)p(\alpha_{j+1}\vert\alpha_{j},\rho,\eta,\tau) & j=0 \\
    p(\alpha_j\vert\alpha_{j-1},\rho,\eta,\tau)
    p(\alpha_{j+1}\vert\alpha_{j},\rho,\eta,\tau)
    p(R_j\vert\alpha_j,\sigma)& j=1,\ldots, n-1 \\
    p(\alpha_j\vert\alpha_{j-1},\rho,\eta,\tau)
    p(R_j\vert\alpha_j,\sigma) & j=n
  \end{cases},%% \nonumber \\
\end{multline*}
which is a normal distribution with mean $m_j$ and variance $v_j$ where
\begin{equation*}%%\label{eq:postalphamean}
  m_j=\begin{cases}
    \frac{
      \displaystyle (\rho\tau(\alpha_{j+1}-(1-\rho)\eta))}
    {\displaystyle(1+\rho^2\tau)} &j=0\\
    \frac{
      \displaystyle (\rho\tau(\alpha_{j+1}-(1-\rho)\eta )+
      \tau(\rho\alpha_{j-1}+ (1-\rho)\eta)+\sigma E_j R_j)}
    {\displaystyle (\rho^2\tau+\tau+\sigma E_j)} &j=1,\ldots, n-1,  \\
    \frac{
      \displaystyle(\tau(\rho\alpha_{j-1}+(1-\rho)\eta)+\sigma E_j R_j)}
    {\displaystyle (\tau+\sigma E_j)} &j=n,
  \end{cases}
\end{equation*}
and
\begin{equation*}%%\label{eq:postalphaprec}
  v_j=\begin{cases}
    \frac{\displaystyle 1}
    {\displaystyle(1+\rho^2\tau)}&j=0\\
    \frac{\displaystyle 1}
    {\displaystyle(\rho^2\tau+\tau+\sigma E_j)}&j=1,\ldots, n-1,  \\
    \frac{\displaystyle 1}
    {\displaystyle(\tau+\sigma E_j )}&j=n .
  \end{cases}
\end{equation*}
These conditional distributions are then used to simulate a dependent sample
from the posterior distribution of the model parameters given the data $R^n$
by sampling each in turn within each iteration. All the posterior
conditionals are standard distributions, hence there are no difficulties in
simulating from them. We could construct a more general Metropolis type
algorithm which updates the parameters $\bfalpha$, $\rho$ and $\eta$ at the
same time.  This would necessitate introducing an acceptance/rejection stage
to ensure stationarity.

\subsection{Simulation Results}
\begin{figure}
   \centering
   \makebox{
   \includegraphics[width=0.8\textwidth]{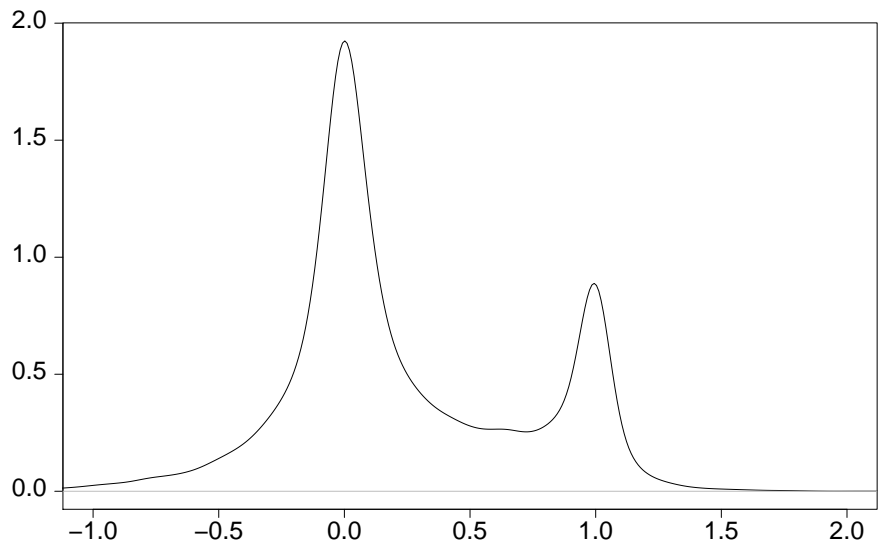}}
   \caption{\label{fig:ds3-rho-dens_M1}Posterior density of $\rho$.}
   \label{fig:posteriordensrho}
 \end{figure}

\begin{figure}
  \subfigure[Trace plot of $\rho$.]{
    \label{fig:ds3-rho}
    \begin{minipage}[b]{0.5\textwidth}
      \centering 
      \includegraphics[width=0.9\textwidth,height=0.15\textheight]{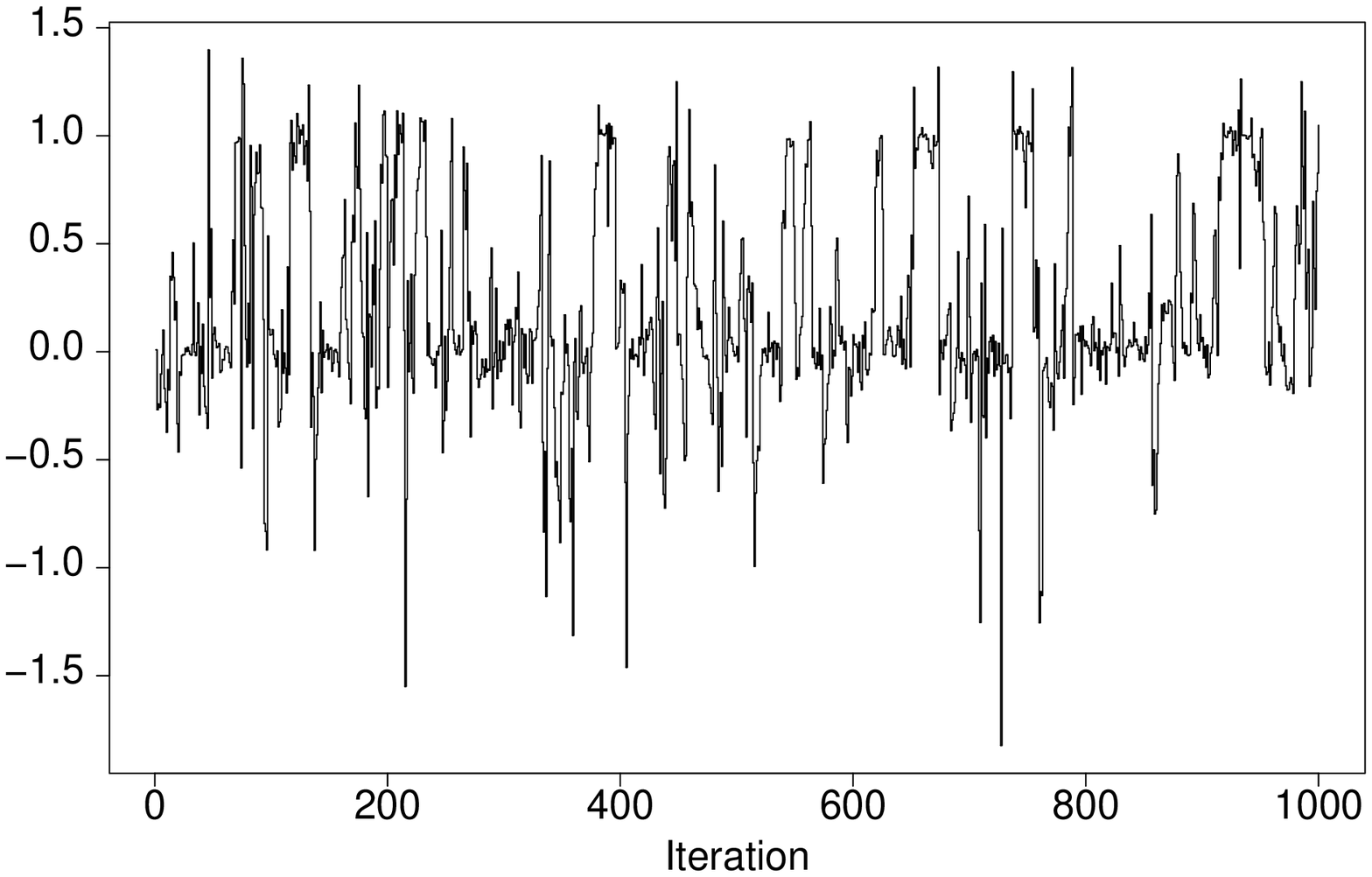}
    \end{minipage}}%
  \subfigure[Trace plot of $\eta$.]{
    \label{fig:ds3-eta}
    \begin{minipage}[b]{0.5\textwidth}
      \centering 
      \includegraphics[width=0.9\textwidth,height=0.15\textheight]{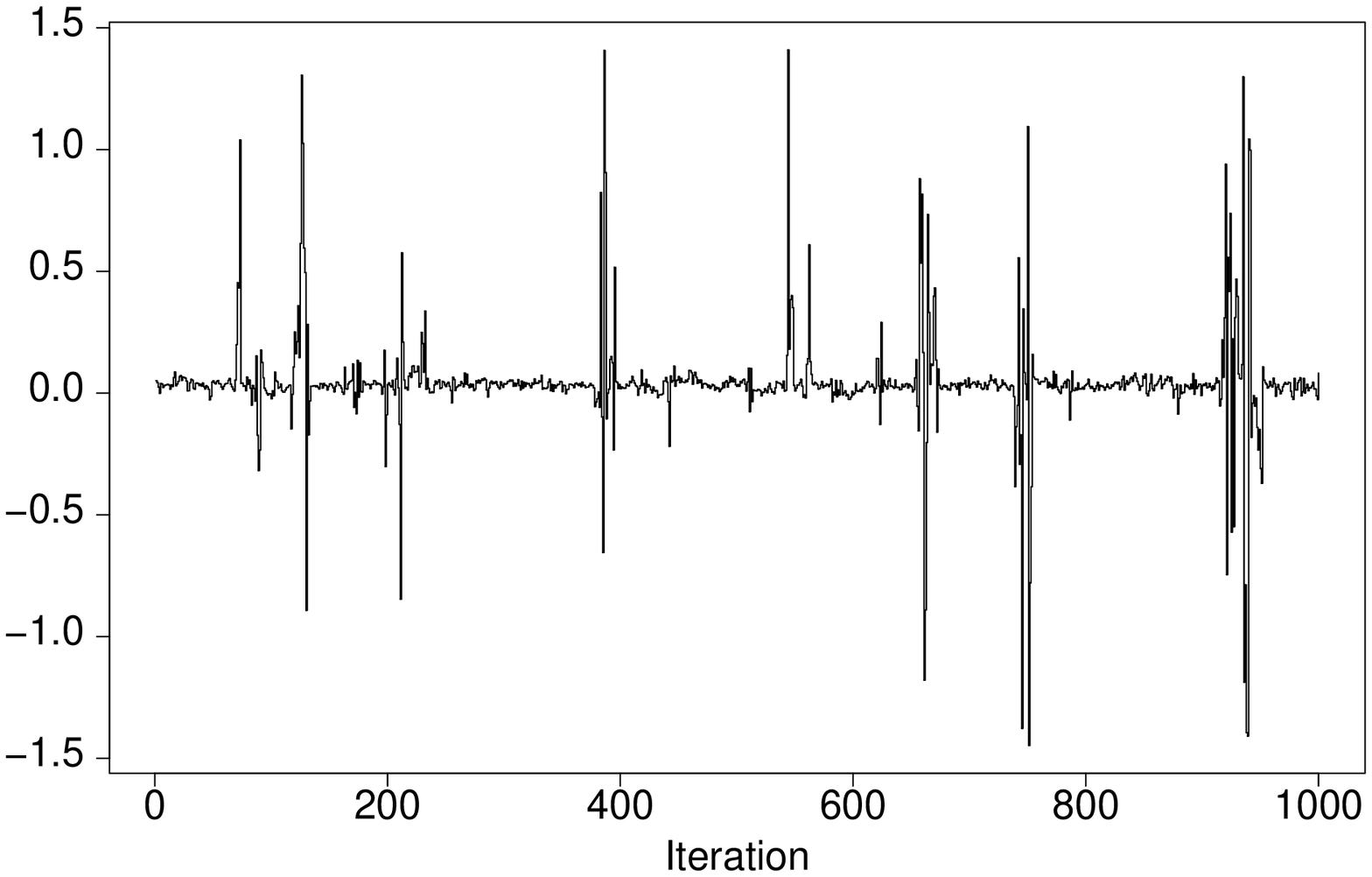}
    \end{minipage}} \\
  \subfigure[Trace plot of $\alpha_{0}$.]{
    \label{fig:ds3-alpha:0}
    \begin{minipage}[b]{0.5\textwidth}
      \centering 
      \includegraphics[width=0.9\textwidth,height=0.15\textheight]{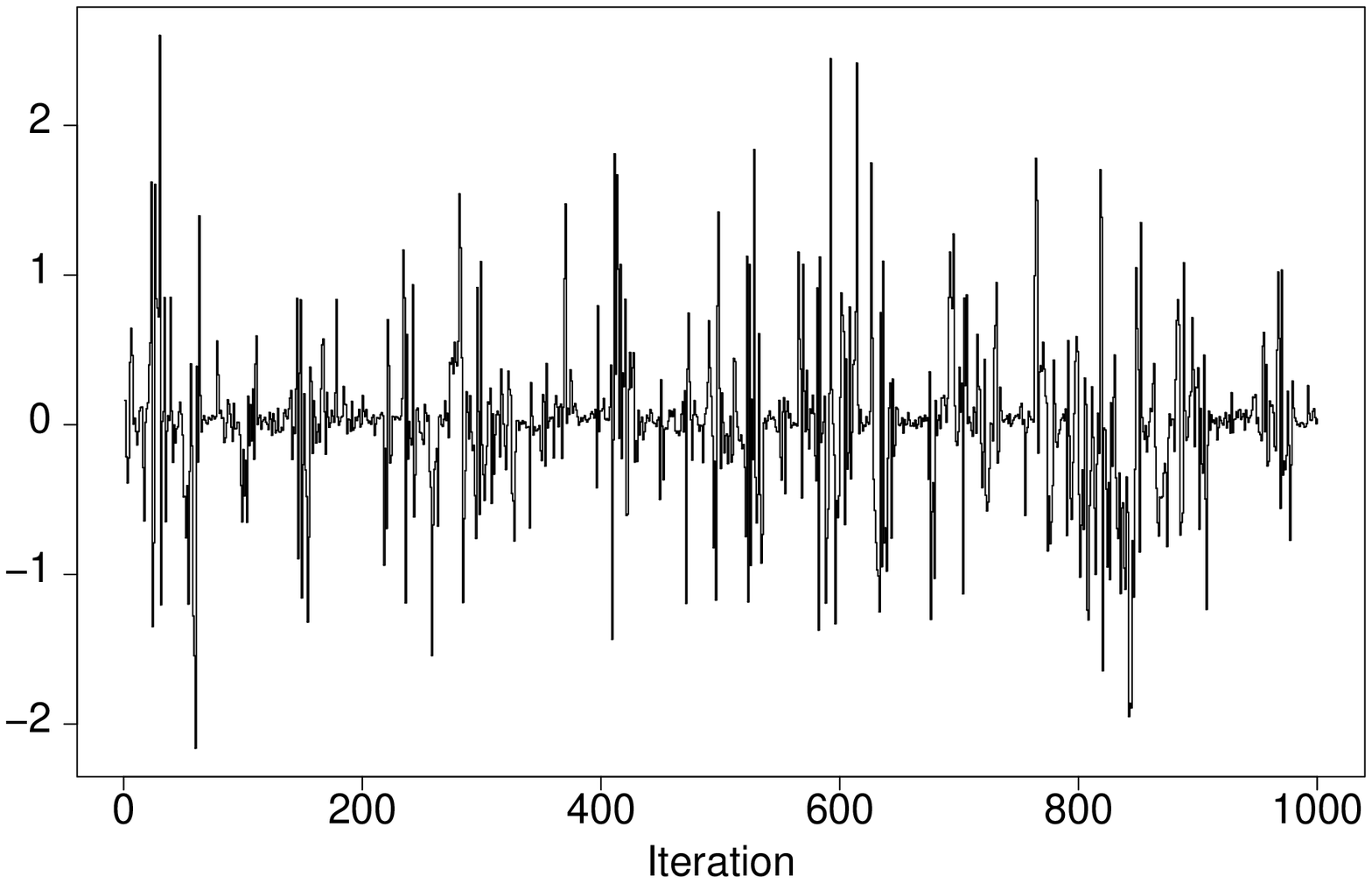}
    \end{minipage}} 
  \subfigure[Trace plot of $\alpha_{1}$.]{
    \label{fig:ds3-alpha:1}
    \begin{minipage}[b]{0.5\textwidth}
      \centering 
      \includegraphics[width=0.9\textwidth,height=0.15\textheight]{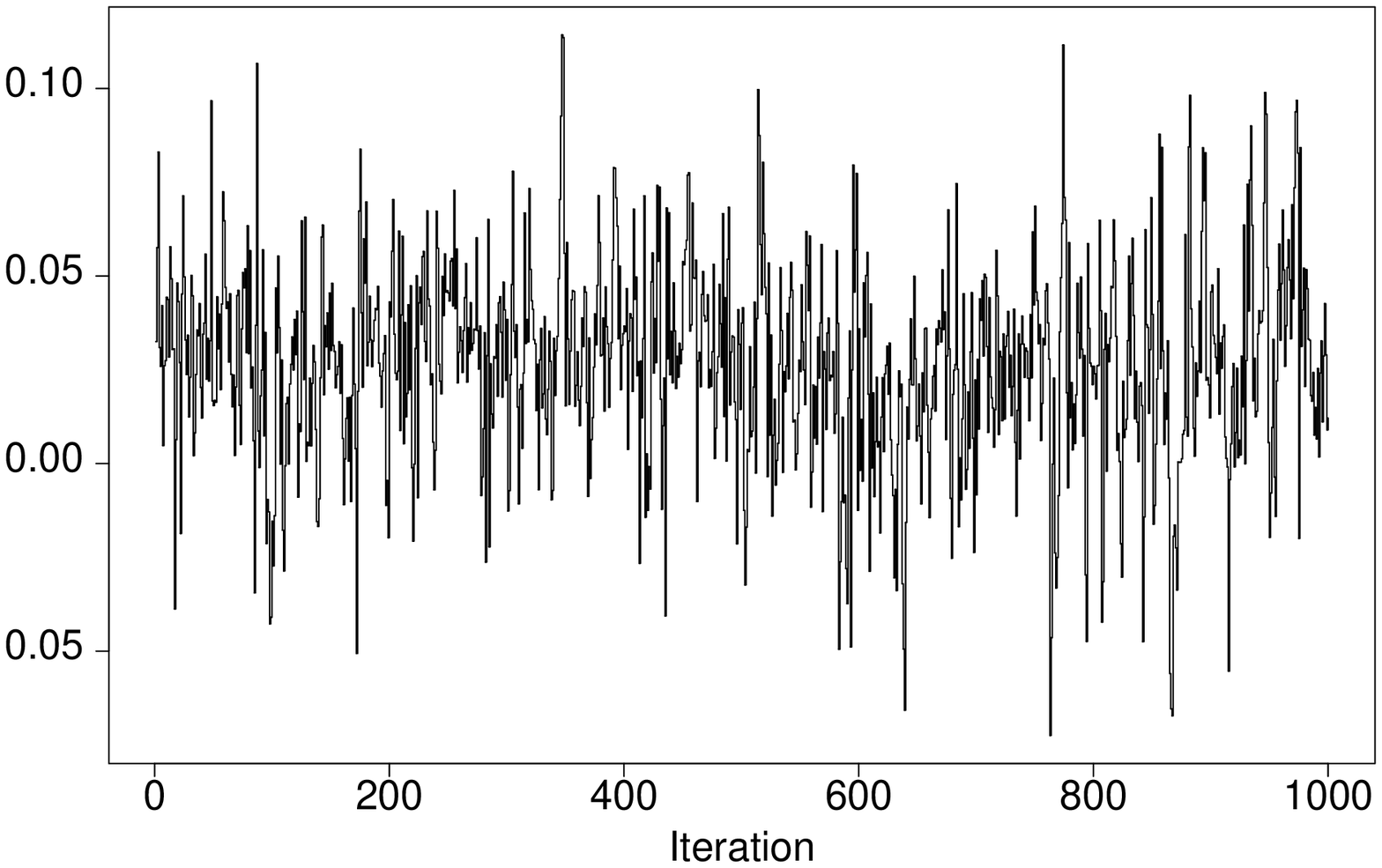}
    \end{minipage}} \\
  \subfigure[Trace plot of $\alpha_{2}$.]{
    \label{fig:ds3-alpha:2}
    \begin{minipage}[b]{0.5\textwidth}
      \centering 
      \includegraphics[width=0.9\textwidth,height=0.15\textheight]{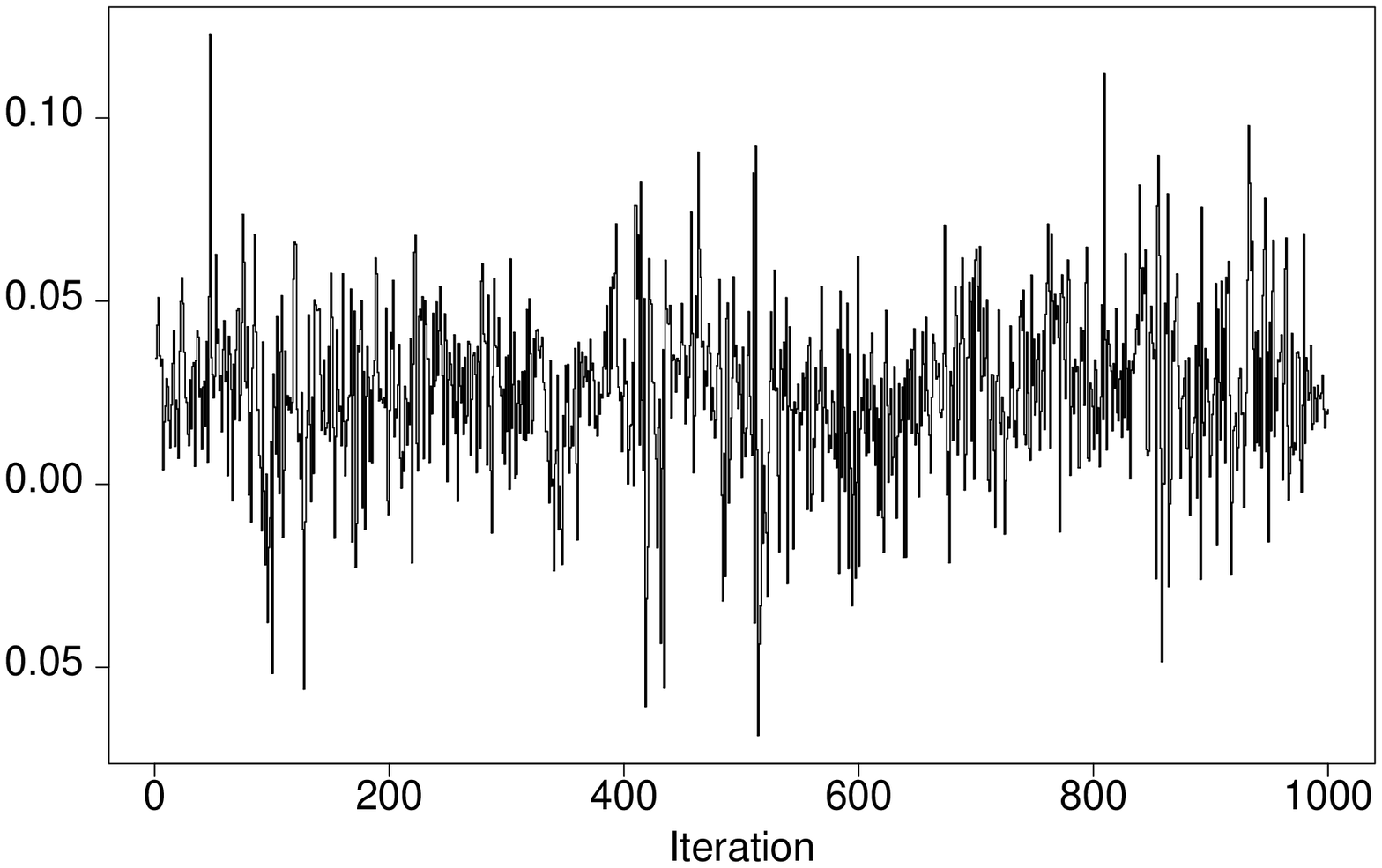}
    \end{minipage}}%
  \subfigure[Trace plot of $\alpha_{3}$.]{
    \label{fig:ds3-alpha:3}
    \begin{minipage}[b]{0.5\textwidth}
      \centering 
      \includegraphics[width=0.9\textwidth,height=0.15\textheight]{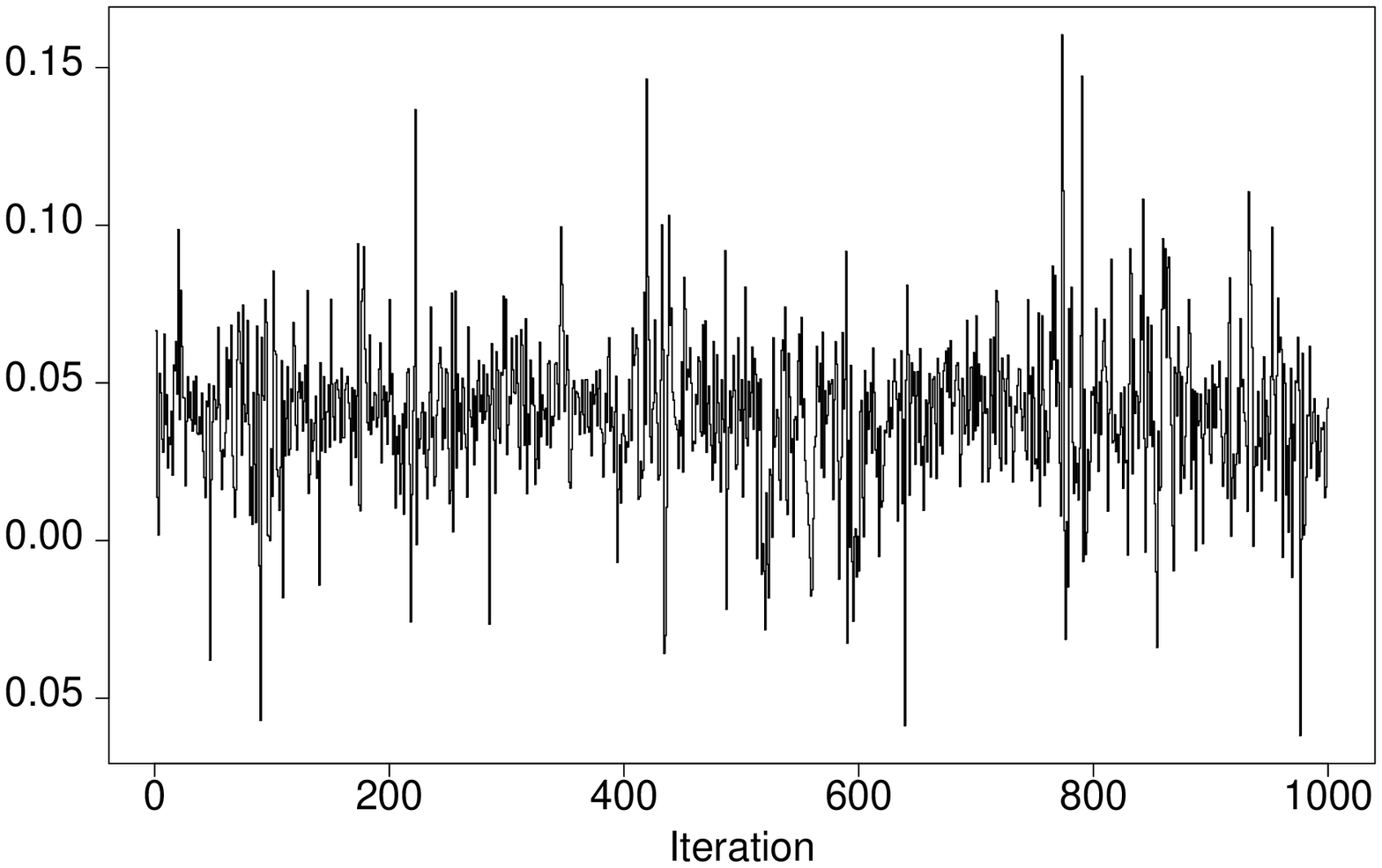}
    \end{minipage}} \\
  \subfigure[Trace plot of $\alpha_{4}$.]{
    \label{fig:ds3-alpha:4}
    \begin{minipage}[b]{0.5\textwidth}
      \centering 
      \includegraphics[width=0.9\textwidth,height=0.15\textheight]{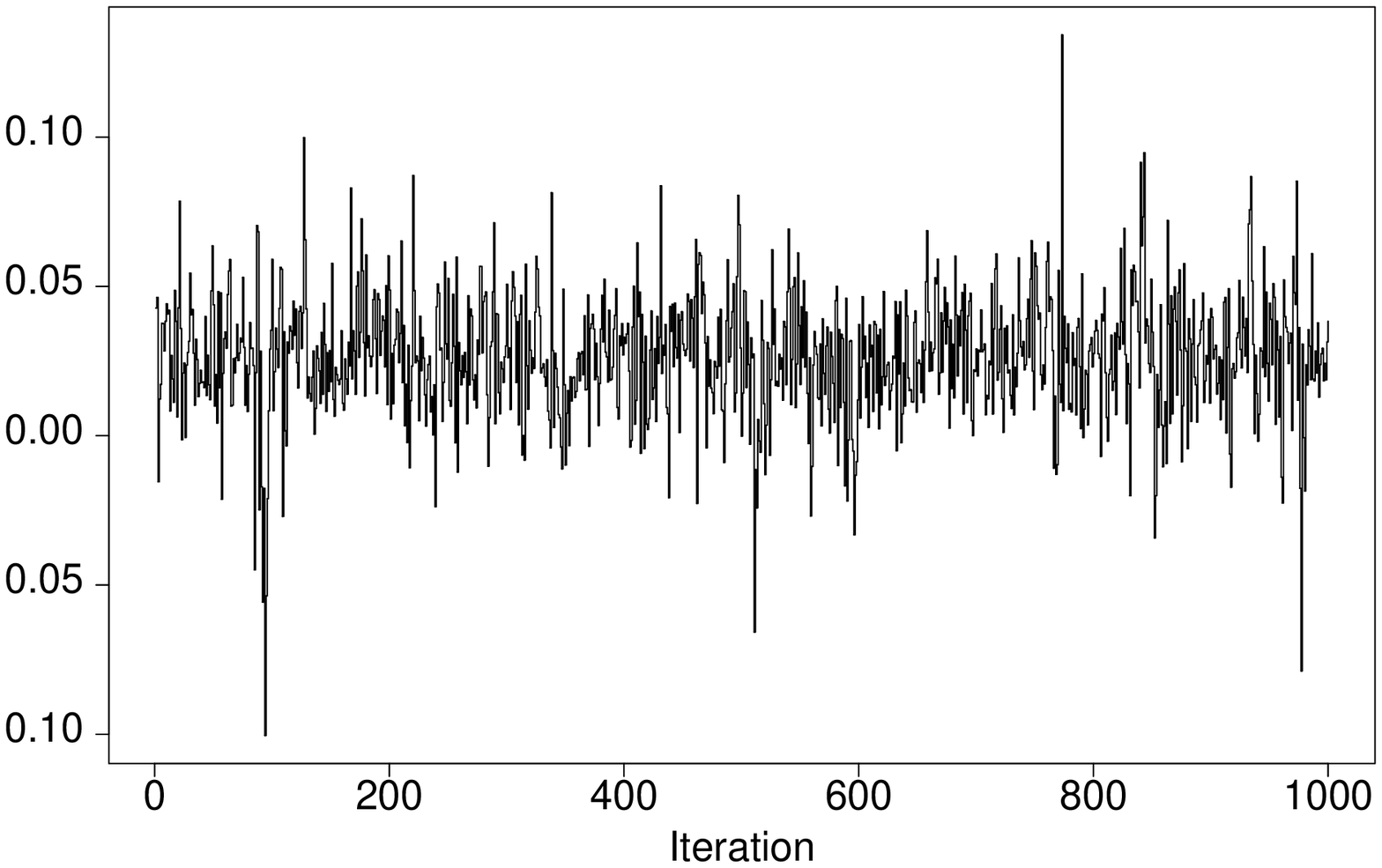}
    \end{minipage}}%
  \subfigure[Trace plot of $\alpha_{5}$.]{
    \label{fig:ds3-alpha:5}
    \begin{minipage}[b]{0.5\textwidth}
      \centering 
      \includegraphics[width=0.9\textwidth,height=0.15\textheight]{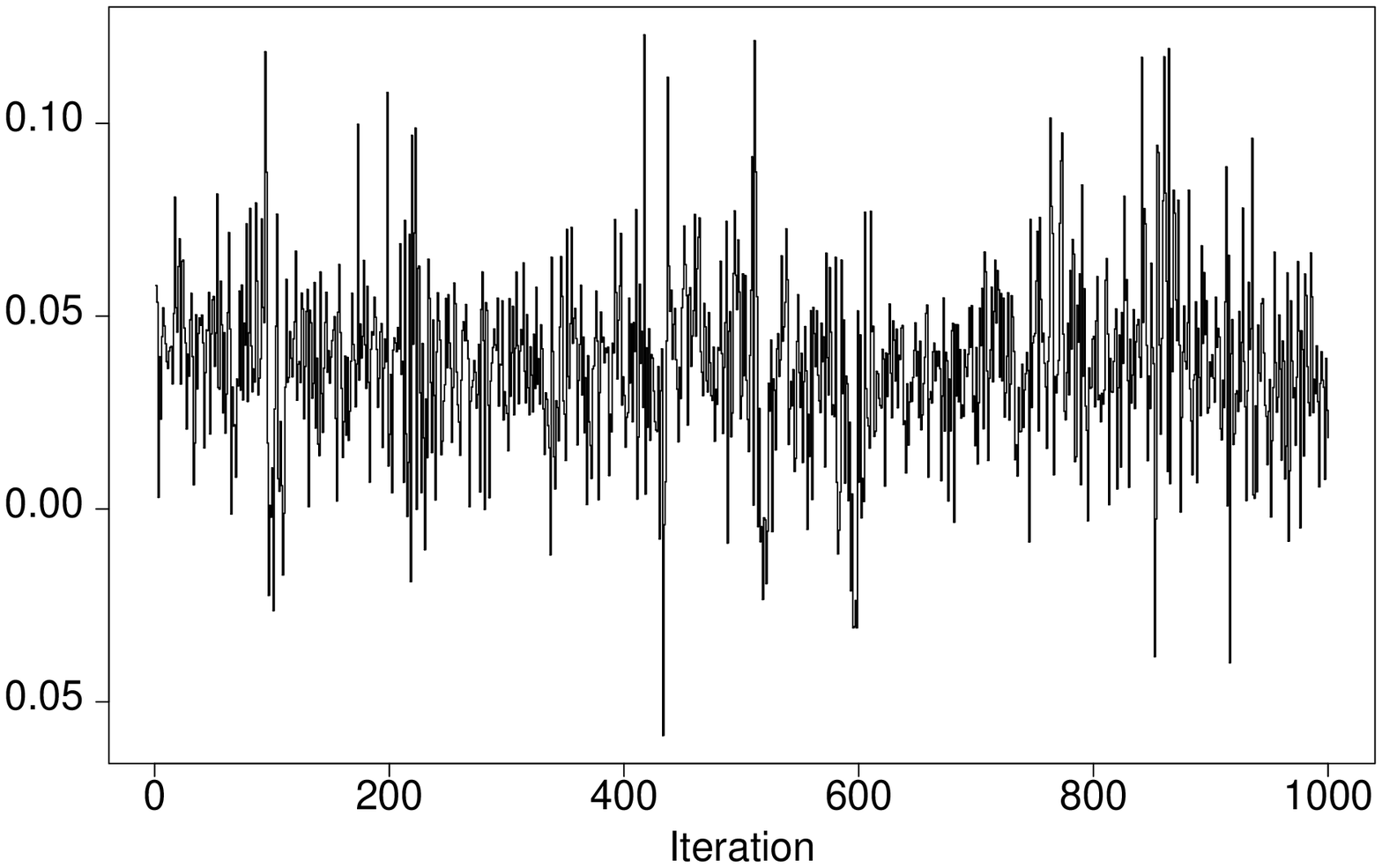}
    \end{minipage}} \\
  \subfigure[Trace plot of $\alpha_{6}$.]{
    \label{fig:ds3-alpha:6}
    \begin{minipage}[b]{0.5\textwidth}
      \centering 
      \includegraphics[width=0.9\textwidth,height=0.15\textheight]{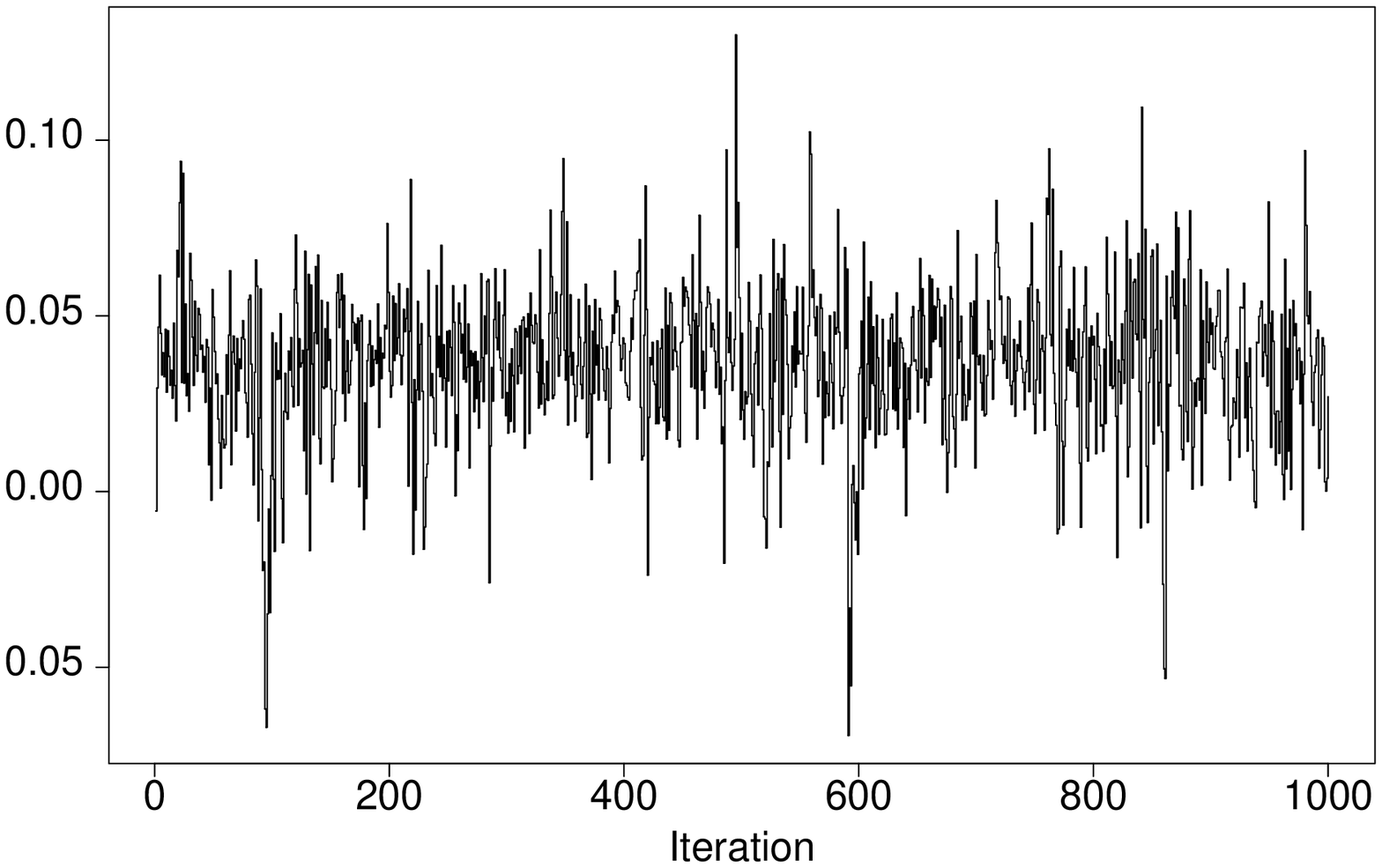}
    \end{minipage}}%
  \subfigure[Trace plot of $\alpha_{7}$.]{
    \label{fig:ds3-alpha:7}
    \begin{minipage}[b]{0.5\textwidth}
      \centering 
      \includegraphics[width=0.9\textwidth,height=0.15\textheight]{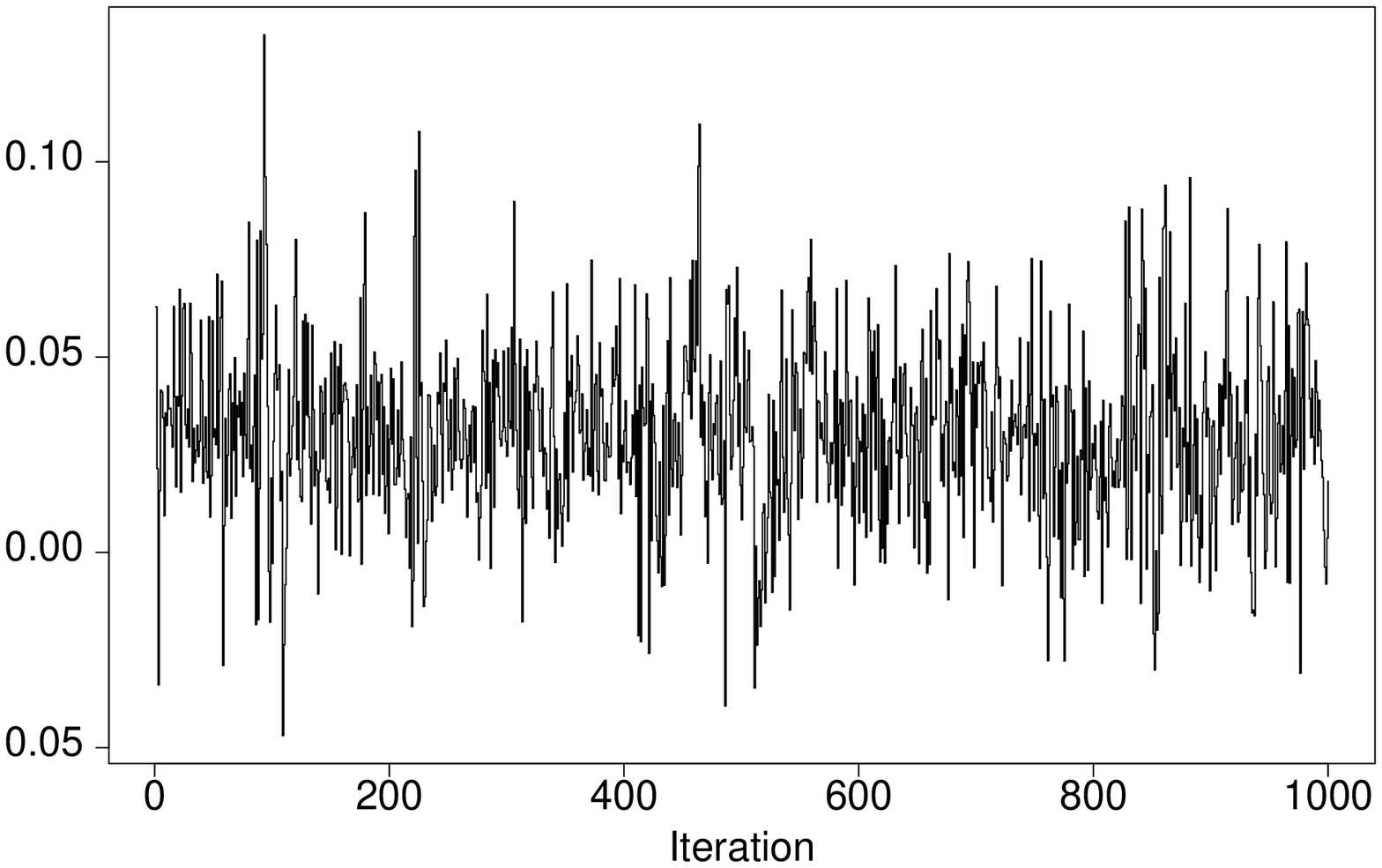}
    \end{minipage}}%
  \caption{Trace plots of model parameters.}
  \label{fig:ds3-alpha} %% label for entire figure
\end{figure}

The Gibbs model above was implemented with $a=0.001$ and $b=0.001$ so that
the precision parameters $\sigma$ and $\tau$ have vague, flat priors. The
posterior means and 95\% highest posterior density (HPD) intervals are shown
in Table~\ref{tab:ds3Fullresults} and trace plots of the model parameters are
shown in Figure~\ref{fig:ds3-alpha}. The 95\% HPD interval is the smallest
region of the parameter space which contains 95\% of the posterior
probability mass of the parameter. A plot of the marginal posterior of $\rho$
reveals that its density is bimodal with one mode near 0 and another at
$\rho=1$. The posterior density of rho is shown in
Figure~\ref{fig:posteriordensrho}. Even though the 95\% HPD interval consists
of only one interval, a corresponding 90\% HPD interval is actually a union
of two disjoint intervals, each containing one of the two modes. The effect
of the bimodality of $\rho$ can be seen by the wide intervals for the
parameters $\alpha_0$ and $\eta$.

A possible explanation is that the posterior conditional of $\alpha_0$ is the
same as its prior density, since there is no need for $\alpha_0$ if $\rho$ is
identically 0.  Similarly when $\rho$ is close to the mode at 1, the
conditional posterior of $\eta$ is almost identical to its prior density.
Consequently, these two parameters are being sampled from two distinct
posterior densities corresponding to whether $\rho$ is close to 0 or 1.  The
reason for the bimodality of $\rho$ is not entirely clear, the model may be
overparameterised since we are fitting 12 parameters to 7 data points. 
To observe the effect of the number of parameters we can reduce the effective
number of parameters being fitted by integrating out the nuisance parameters
$\sigma$ and $\tau$, then re-fitting the model and observing any differences.
The results of this new implementations are identical to the first
implementation with both $\sigma$ and $\tau$ included, as we show in the
next section.
 \begin{table}
   \caption{\label{tab:ds3Fullresults}Posterior means and 95\% HPD Intervals 
     for the model parameters with $a=b=c=0.001$.}
   \centering
   \fbox{%
     \begin{tabular}{rrr}
       \hline\hline
     Parameter & estimate & 95\% HPD Interval \\
     \hline
     $\alpha_0$ &0.0167 & (-1.1188, 1.0619) \\
     $\alpha_1$ &0.0256 & (-0.0311, 0.0817) \\
     $\alpha_2$ &0.0246 & (-0.0211, 0.0708) \\
     $\alpha_3$ &0.0398 & (-0.0067, 0.0846) \\
     $\alpha_4$ &0.0271 & (-0.0165, 0.0712) \\
     $\alpha_5$ &0.0362 & (-0.0064, 0.0792) \\
     $\alpha_6$ &0.0364 & (-0.0072, 0.0785) \\
     $\alpha_7$ &0.0296 & (-0.0170, 0.0772) \\
     $\rho$     &0.220  & (-0.483, 1.154) \\
     $\eta$     &0.0315 & (-0.3831, 0.4477) \\
     $\sigma$   &1014.9 & (0.0, 2634.6) \\
     $\tau$     &1371.2 & (0.0, 3330.0) \\
   \end{tabular}}
 \end{table}

\subsection{Integrating out the Variance Parameters}
\citet{papaspilioppoulos2003} shows that for Gaussian models similar to
that described in Equations~\ref{eq:model} and \eqref{eq:hyper} the
convergence properties are largely determined by the values of the variance
components.  In this Section we redo the analysis, however this time we
integrate out the variance parameters $\sigma$ and $\tau$. This results in
fewer parameters to be estimated, but as the results show, it also increases
the autocorrelation of the other parameters.  Also the complexity of the
model has been reduced.  The form of the conditional posteriors, however, has
been made more complex.  We use a random walk Metropolis algorithm to
simulate from the posterior distribution of the remaining unknown parameters
$\alpha_0, \alpha_1, \ldots, \alpha_n, \rho, \eta$. The results show that the
posterior estimate of $\rho$ is still bimodal.

If $\sigma\sim\dgamma{a_1}{b_1}$ and $\tau\sim\dgamma{a_2}{b_2}$, recall that
in Section~\ref{sec:postcond} we showed that the posterior conditionals of
$\sigma$ and $\tau$ are of the form of Gamma densities, and both are
independent of each other, since the posterior conditional of $\sigma$ does
not depend on $\tau$ and likewise that of $\tau$ does not depend on $\sigma$.

Since also the posterior conditionals are of standard form, we can integrate
out these two parameters leaving a density involving only the other
parameters.  Using the fact that the posterior conditionals for $\sigma$ and
$\tau$ are standard Gamma densities, we can show that
%%%
\begin{equation*}
  \pi(\bfalpha,\alpha_0,\rho,\eta\vert R^n)=
  \int\pi(\bfalpha,\alpha_0,\rho,\eta,\sigma,\tau\vert R^n)
  \, d\sigma d\tau, \nonumber
\end{equation*}
so that
\begin{multline}\label{eq:marg}
  \pi(\bfalpha,\alpha_0,\rho,\eta\vert R^n)  \propto
  p(\rho)p(\eta)p(\alpha_0) \times
  \left( b_1 + \tfrac1{2}\sum E_j(\alpha_j - R_j )^2
  \right)^{-(a_1+n/2)} \times \\
  \left( b_2+ \tfrac1{2}\sum(\alpha_j - \rho\alpha_{j-1} -(1-\rho)\eta)^2
  \right )^{-(a_2+n/2)}
\end{multline}
%%%
where $\pi(\bfalpha,\alpha_0,\rho,\eta\vert R^n)$ is the
posterior density of $\bfalpha$, $\alpha_0$, $\rho$ and $\eta$
given the data.

Now given \eqref{eq:marg} the following posterior conditionals are readily
observed
\begin{equation*}
  \pi(\rho\vert \bfalpha, \alpha_0,\eta) \propto p(\rho)
  \left( b_2+ \tfrac1{2}\sum(\alpha_j - \rho\alpha_{j-1} -(1-\rho)\eta)^2
    \right)^{-(a_2+n/2)},
\end{equation*}
and
\begin{equation*}
  \pi(\eta\vert\bfalpha,\alpha_0,\rho ) \propto p(\eta)
  \left(b_2+ \tfrac1{2}\sum(\alpha_j - \rho\alpha_{j-1} -(1-\rho)\eta)^2
  \right)^{-(a_2+n/2)},
\end{equation*}
also
\begin{multline*}
  \pi(\bfalpha,\alpha_0\vert \rho, \eta) \propto  p(\alpha_0)
  \left( b_1 + \tfrac1{2}\sum E_j(\alpha_j - R_j )^2
  \right)^{-(a_1 +n/2)} \times \\
  \left( b_2+ \tfrac1{2}\sum(\alpha_j - \rho\alpha_{j-1} -(1-\rho)\eta)^2
    \right)^{-(a_2+n/2)}.
\end{multline*}
We use this scheme because our attempts to update $\rho$, $\eta$ and
$\bfalpha$ as one block using a 10-variate normal distribution centred at
the current values did not work very well.

These are all non-standard densities and to implement this model we used a
Gibbs updating scheme with random walk Metropolis algorithms for $\eta$,
$\rho$ and $\alpha_0$ with uniform distributions centred at the current
values.  For $\rho$, $\eta$ and $\alpha_0$, the width of the proposal
interval was determined by fine tuning an initial run until the acceptance
rates were $0.27$, $0.15$ and $0.29$, respectively.  For $\alpha_1$ , \ldots,
$\alpha_7$ we used a 7-variable normal density as the proposal for a random
walk Metropolis algorithm centred at the current values of these parameters.
The covariance matrix for this proposal distribution was determined from an
initial run from which we computed the covariance of the parameters
$\alpha_1$, $\ldots$, $\alpha_7$.  With this covariance matrix the acceptance
rate of the Metropolis algorithm is $0.15$, this is smaller than would be ideal
 \citep{roberts1998,roberts2001}.

The results for this model are shown in Table~\ref{tab:ds3margresults}, they
are similar to those in Table~\ref{tab:ds3Fullresults}. The main difference
here is that the 95\% HPD intervals for $\alpha_0$ and $\eta$ are now smaller
and more concentrated around the posterior means, which represents an
improvement on those given in Table~\ref{tab:ds3Fullresults}.

\begin{table}
  \caption{\label{tab:ds3margresults}Posterior means and 95\% HPD Intervals 
    for the model parameters, after integrating out the variance parameters 
    with $a_1=a_2=b_1=b_2=0.001$.}
  \centering
  \fbox{%
  \begin{tabular}{rrr}
    \hline\hline
    Parameter & Estimate & 95\% HPD Interval \\
    \hline
    $\alpha_0$   &0.01597       &(-1.1332, 1.0953)\\
    $\alpha_1$   &0.02551       &(-0.0312, 0.0830)\\
    $\alpha_2$   &0.02439       &(-0.0204, 0.0706)\\
    $\alpha_3$   &0.03987       &(-0.0063, 0.0842)\\
    $\alpha_4$   &0.02703       &(-0.0161, 0.0705)\\
    $\alpha_5$   &0.03630       &(-0.0086, 0.0781)\\
    $\alpha_6$   &0.03654       &(-0.0072, 0.0776)\\
    $\alpha_7$   &0.02942       &(-0.0169, 0.0777)\\
       $\rho$    &0.21410       &(-0.5585, 1.1177)\\
       $\eta$    &0.02775       &(-0.4214, 0.4264)\\
  \end{tabular}}
\end{table}

\begin{figure}
  \subfigure[$\rho$.]{
    \label{fig:ds3-acfrho}
    \begin{minipage}[b]{0.5\textwidth}
      \centering 
      \includegraphics[width=0.9\textwidth,height=0.15\textheight]{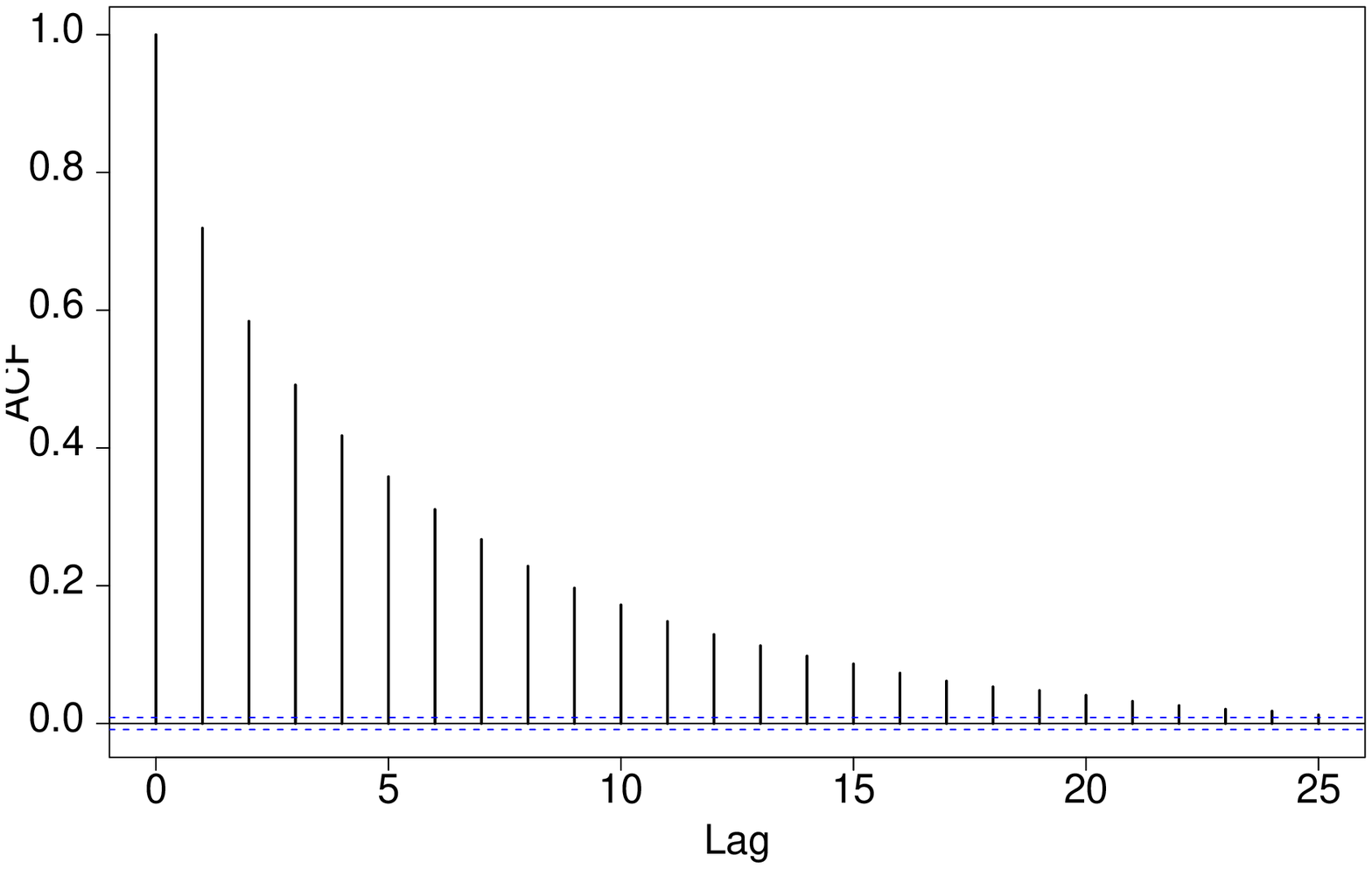}
    \end{minipage}}%
  \subfigure[$\eta$.]{
    \label{fig:ds3-acfeta}
    \begin{minipage}[b]{0.5\textwidth}
      \centering 
      \includegraphics[width=0.9\textwidth,height=0.15\textheight]{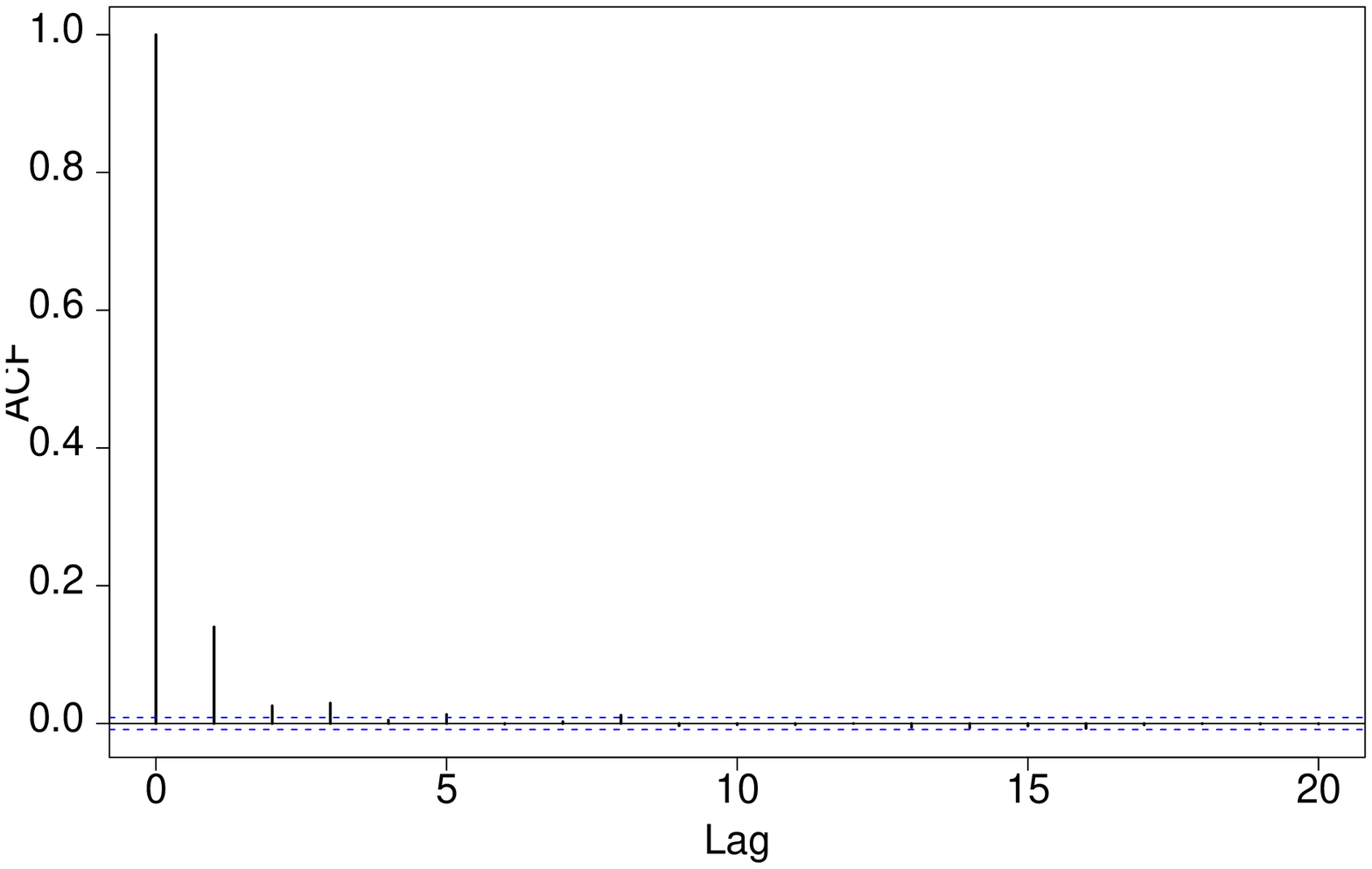}
    \end{minipage}} \\
  \subfigure[$\alpha_{0}$.]{
    \label{fig:ds3-acfalpha:0}
    \begin{minipage}[b]{0.5\textwidth}
      \centering 
      \includegraphics[width=0.9\textwidth,height=0.15\textheight]{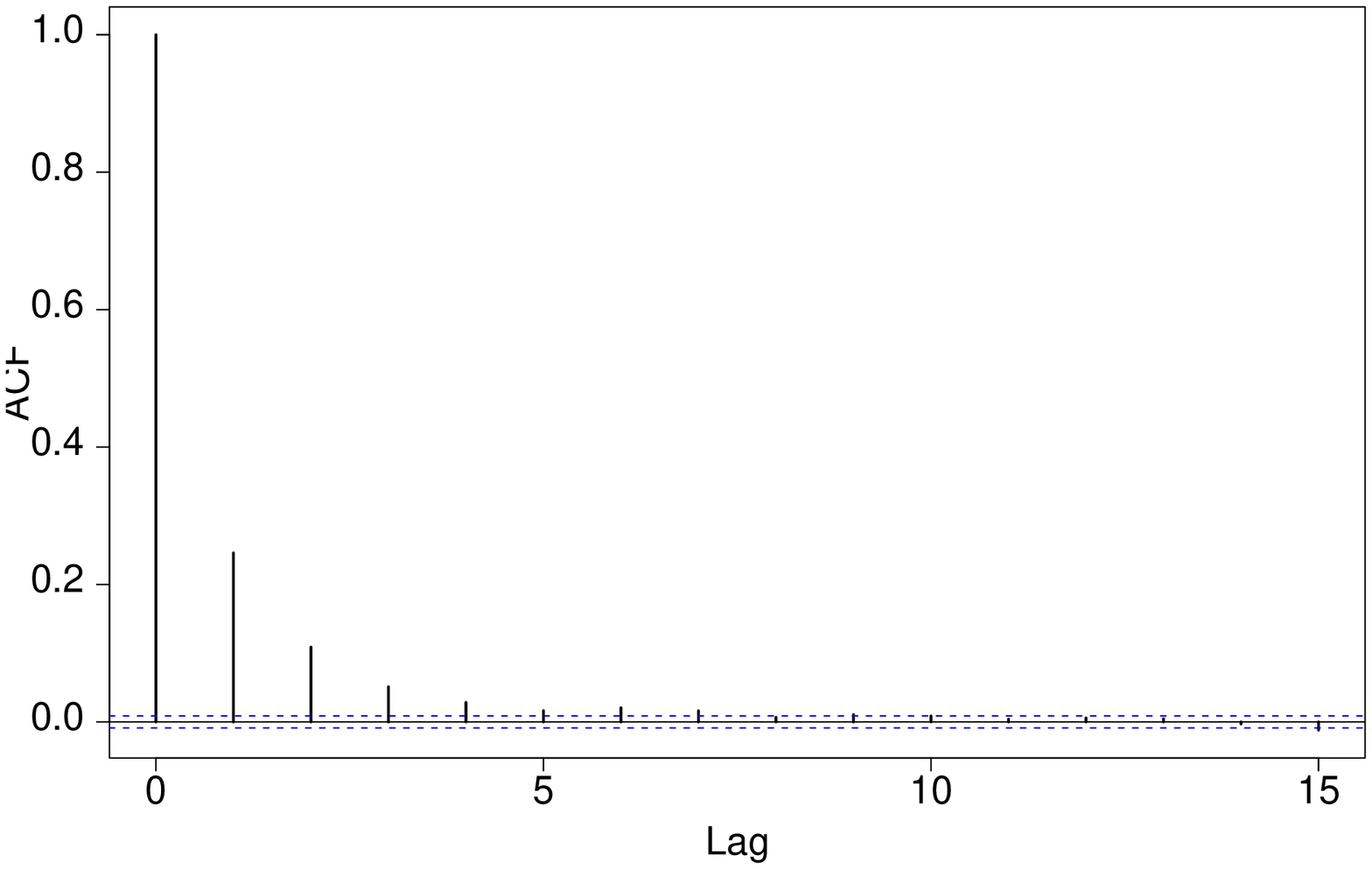}
    \end{minipage}} 
  \subfigure[$\alpha_{1}$.]{
    \label{fig:ds3-acfalpha:1}
    \begin{minipage}[b]{0.5\textwidth}
      \centering 
      \includegraphics[width=0.9\textwidth,height=0.15\textheight]{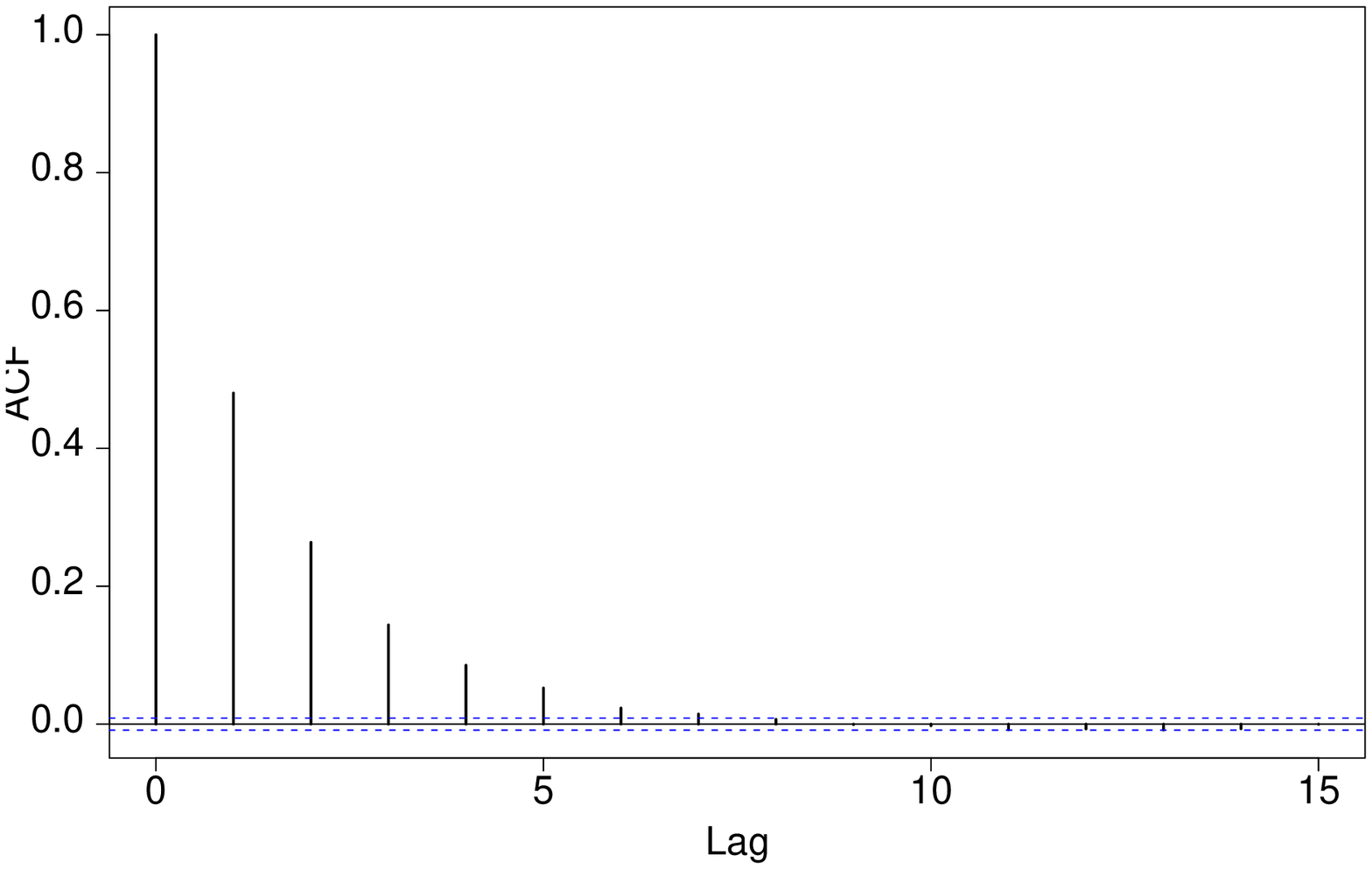}
    \end{minipage}} \\
  \subfigure[$\alpha_{2}$.]{
    \label{fig:ds3-acfalpha:2}
    \begin{minipage}[b]{0.5\textwidth}
      \centering 
      \includegraphics[width=0.9\textwidth,height=0.15\textheight]{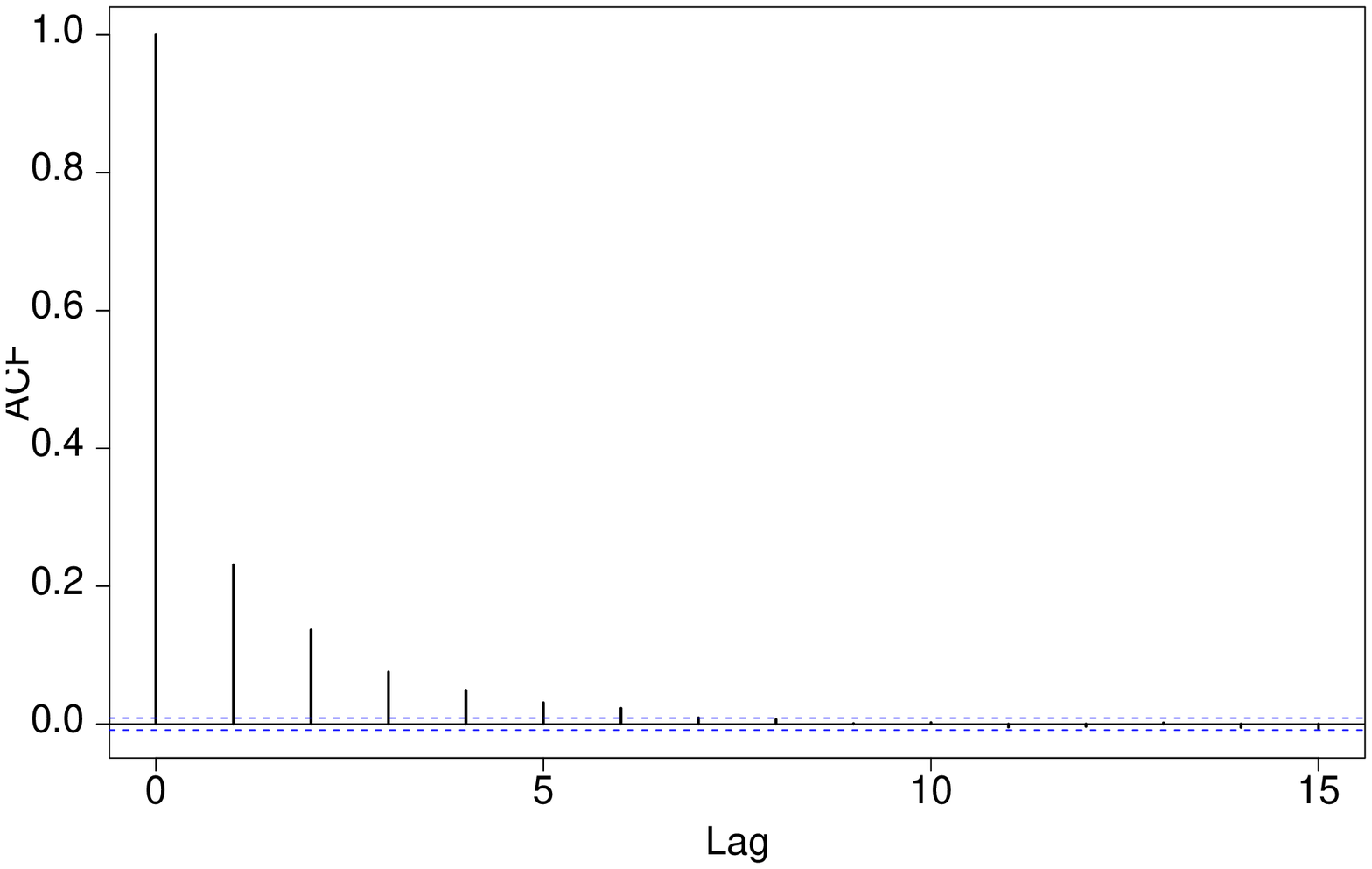}
    \end{minipage}}%
  \subfigure[$\alpha_{3}$.]{
    \label{fig:ds3-acfalpha:3}
    \begin{minipage}[b]{0.5\textwidth}
      \centering 
      \includegraphics[width=0.9\textwidth,height=0.15\textheight]{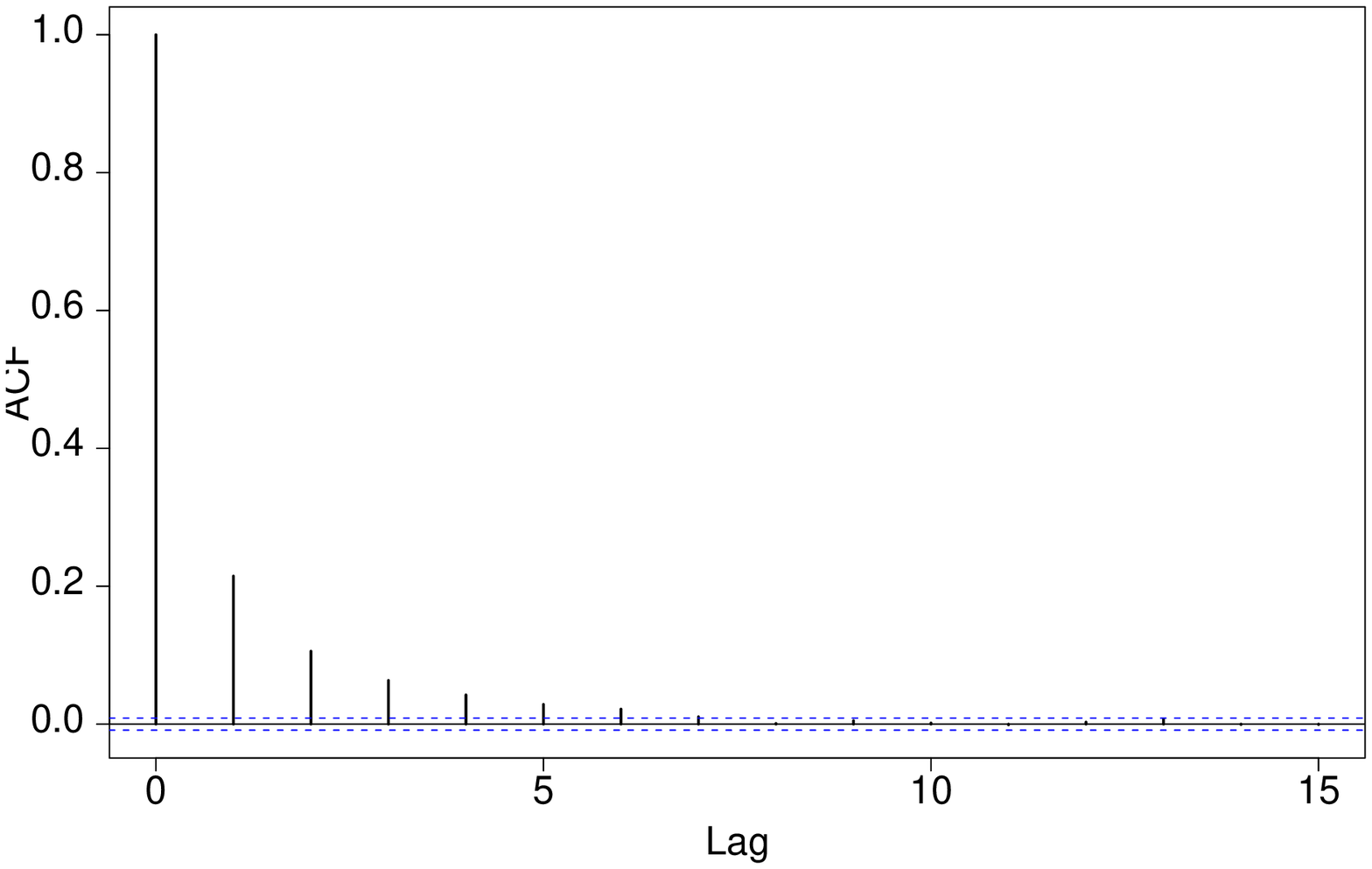}
    \end{minipage}}  \\
  \subfigure[$\alpha_{4}$.]{
    \label{fig:ds3-acfalpha:4}
    \begin{minipage}[b]{0.5\textwidth}
      \centering 
      \includegraphics[width=0.9\textwidth,height=0.15\textheight]{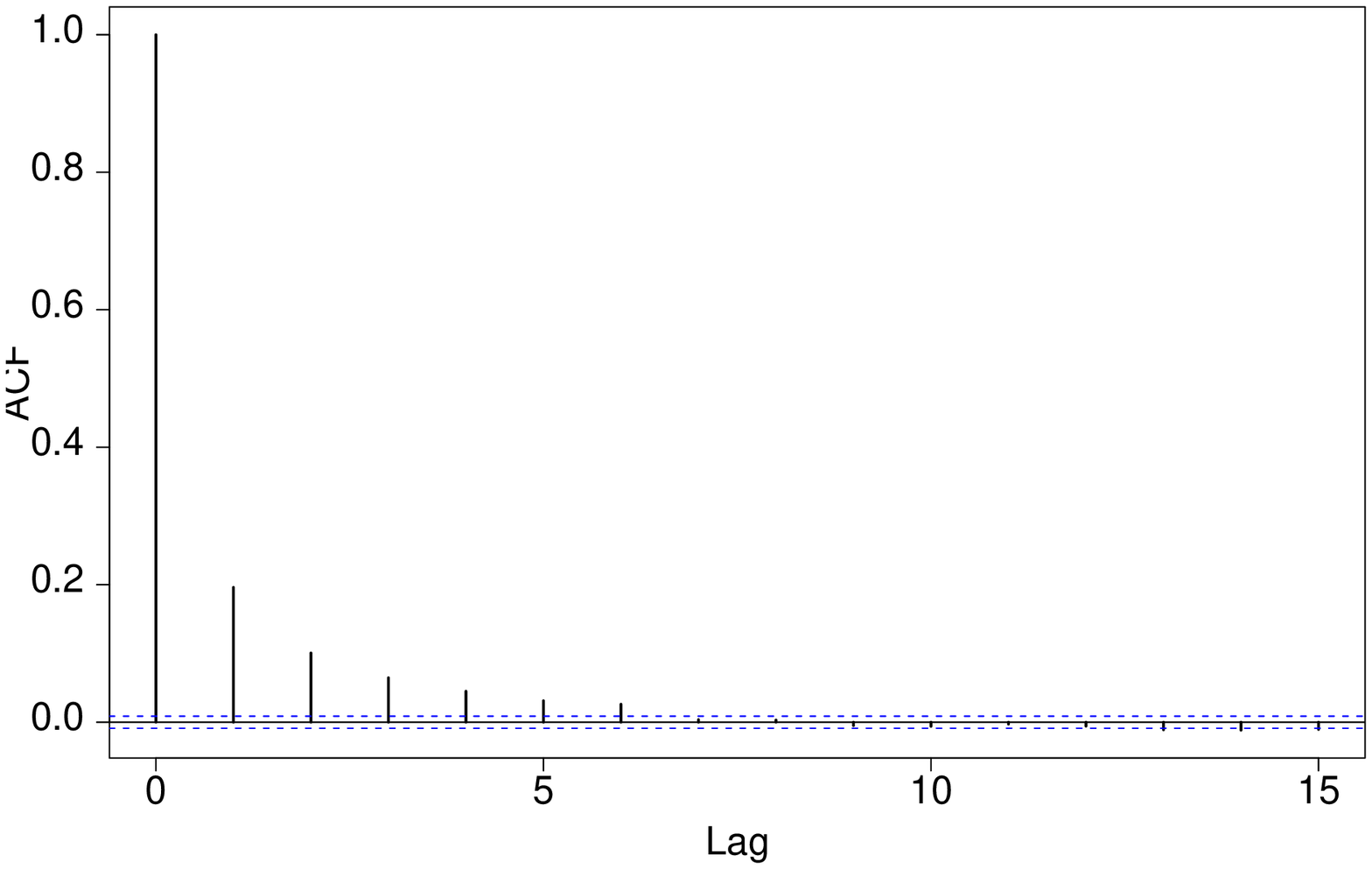}
    \end{minipage}}%
  \subfigure[$\alpha_{5}$.]{
    \label{fig:ds3-acfalpha:5}
    \begin{minipage}[b]{0.5\textwidth}
      \centering 
      \includegraphics[width=0.9\textwidth,height=0.15\textheight]{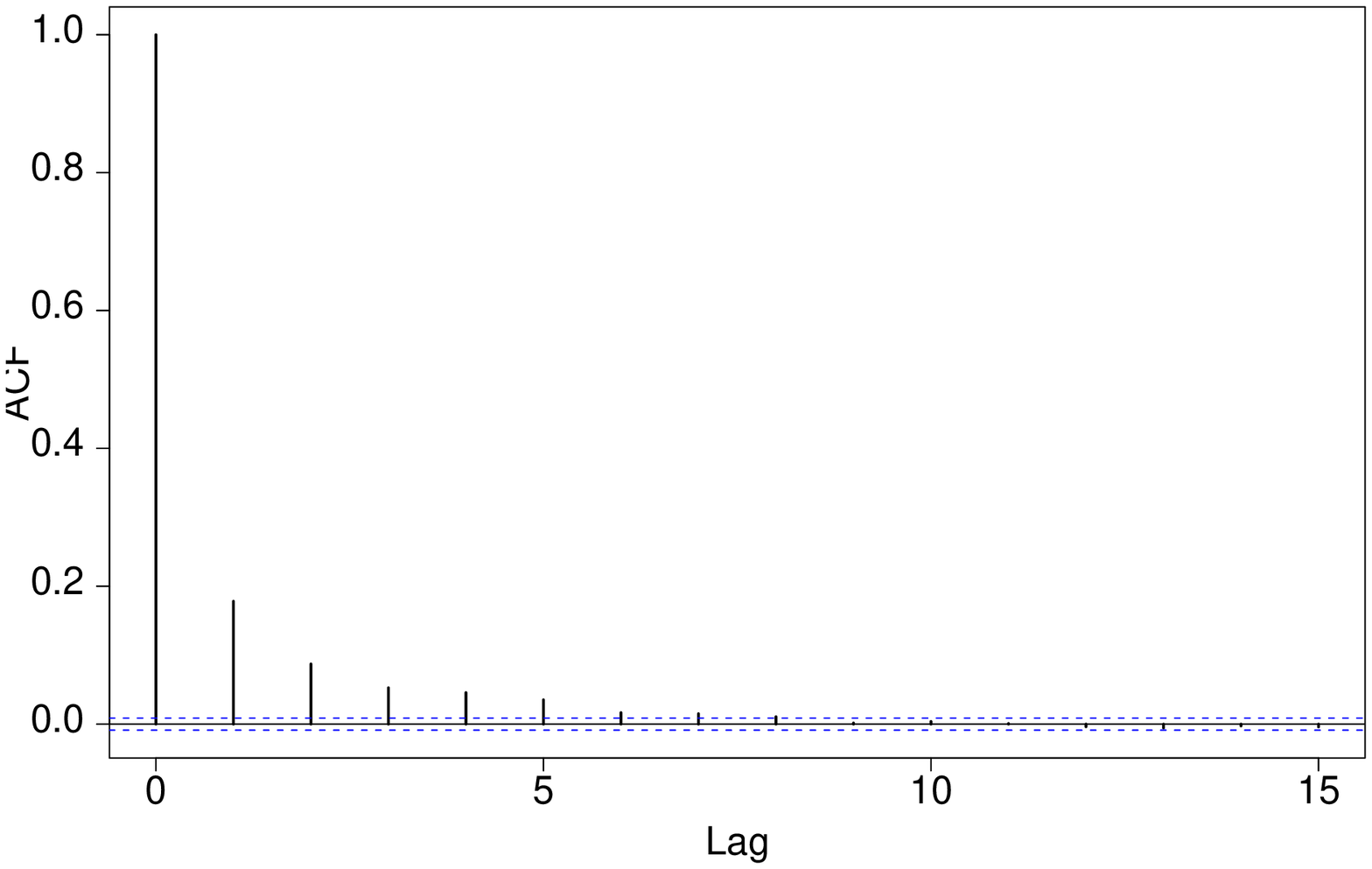}
    \end{minipage}} \\
  \subfigure[$\alpha_{6}$.]{
    \label{fig:ds3-acfalpha:6}
    \begin{minipage}[b]{0.5\textwidth}
      \centering 
      \includegraphics[width=0.9\textwidth,height=0.15\textheight]{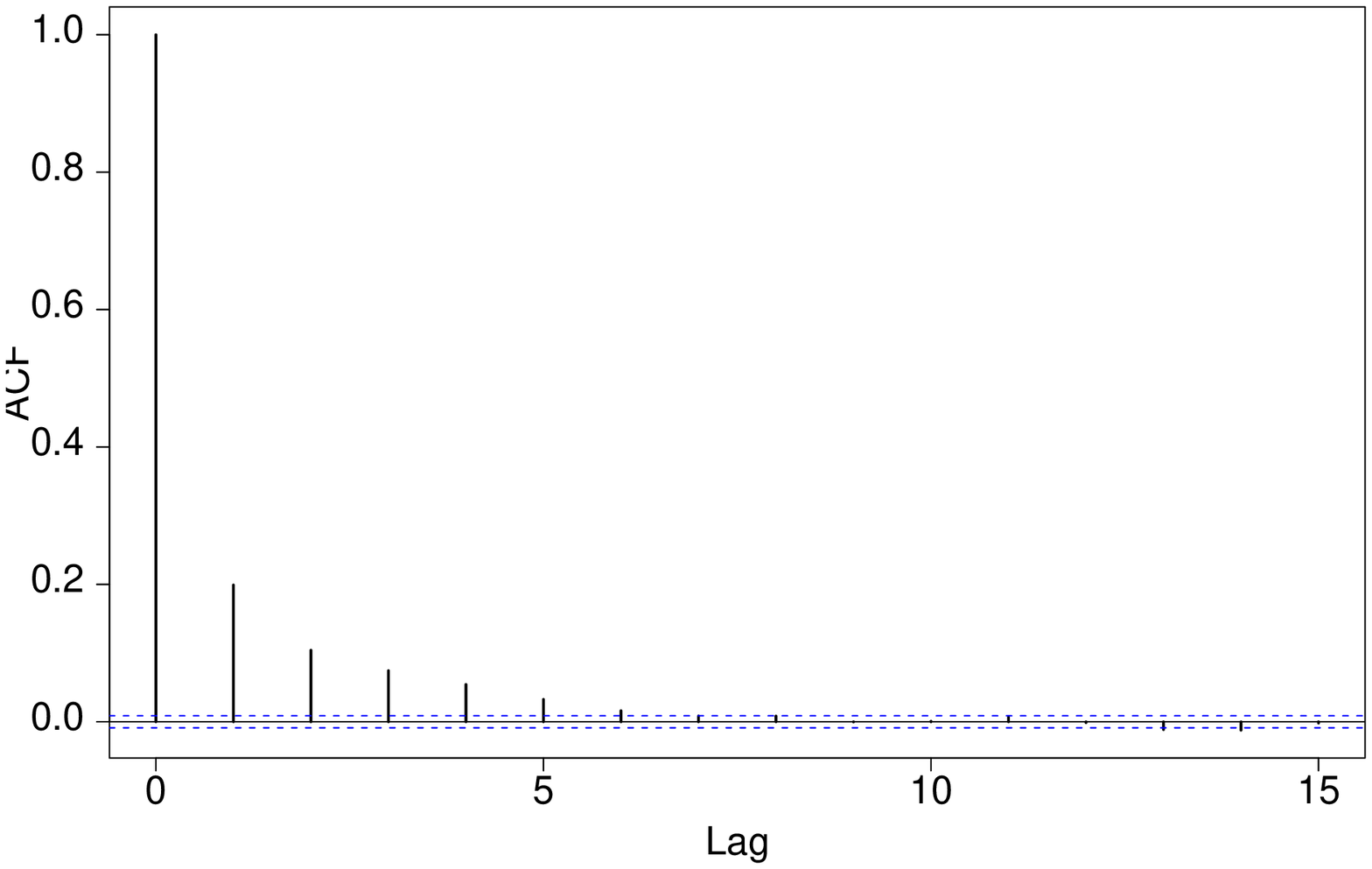}
    \end{minipage}} 
  \subfigure[$\alpha_{7}$.]{
    \label{fig:ds3-acfalpha:7}
    \begin{minipage}[b]{0.5\textwidth}
      \centering
      \includegraphics[width=0.9\textwidth,height=0.15\textheight]{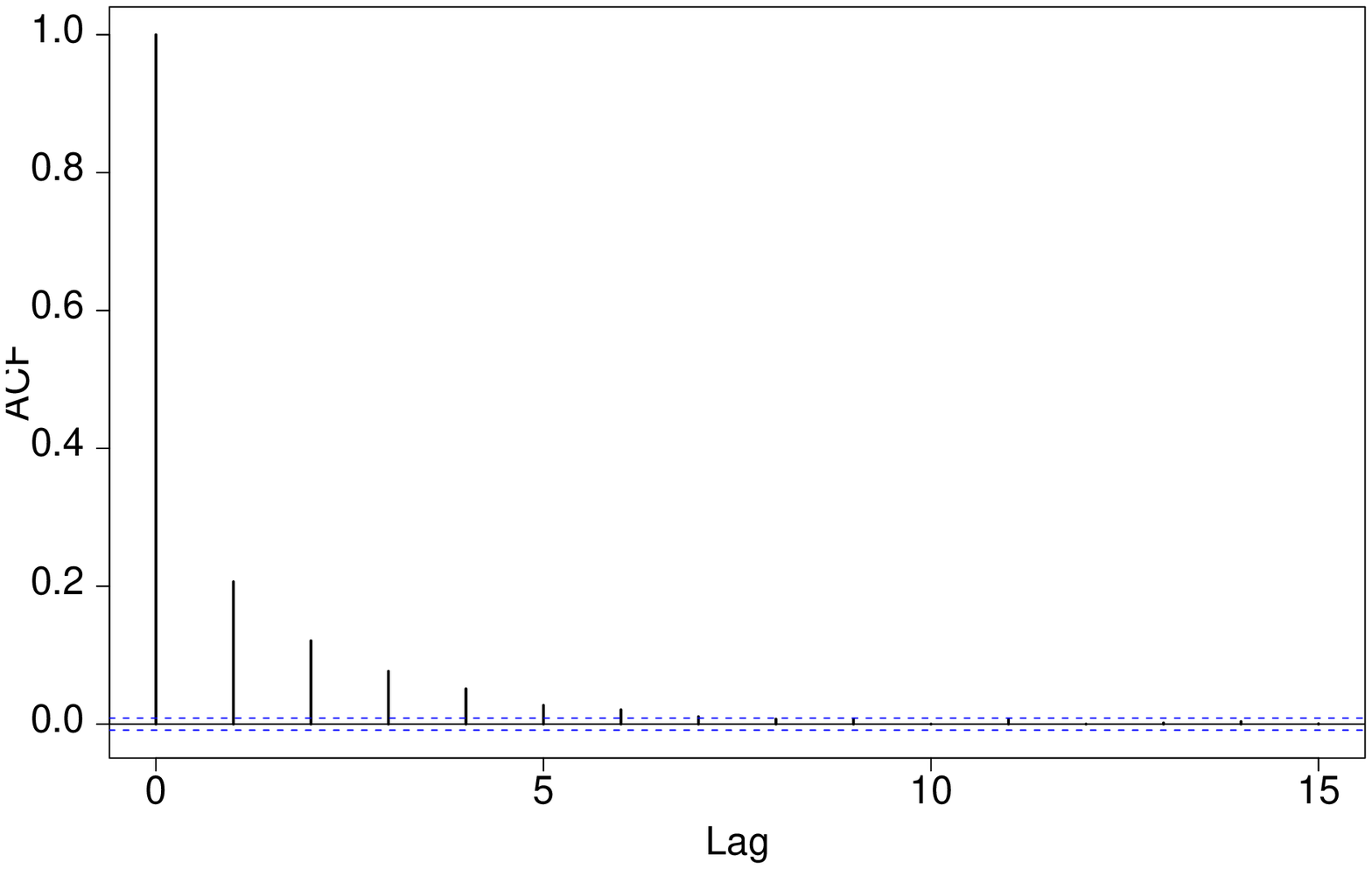}
    \end{minipage}}% 
  \caption{Autocorrelation plots for $\rho$, $\eta$ and $\bfalpha$.}
  \label{fig:ds3-acf} %% label for entire figure 
\end{figure}

\begin{figure}
  \subfigure[$\rho$.]{
    \label{fig:ds3-nacfrho}
    \begin{minipage}[b]{0.5\textwidth}
      \centering \includegraphics[width=0.9\textwidth,height=0.15\textheight]{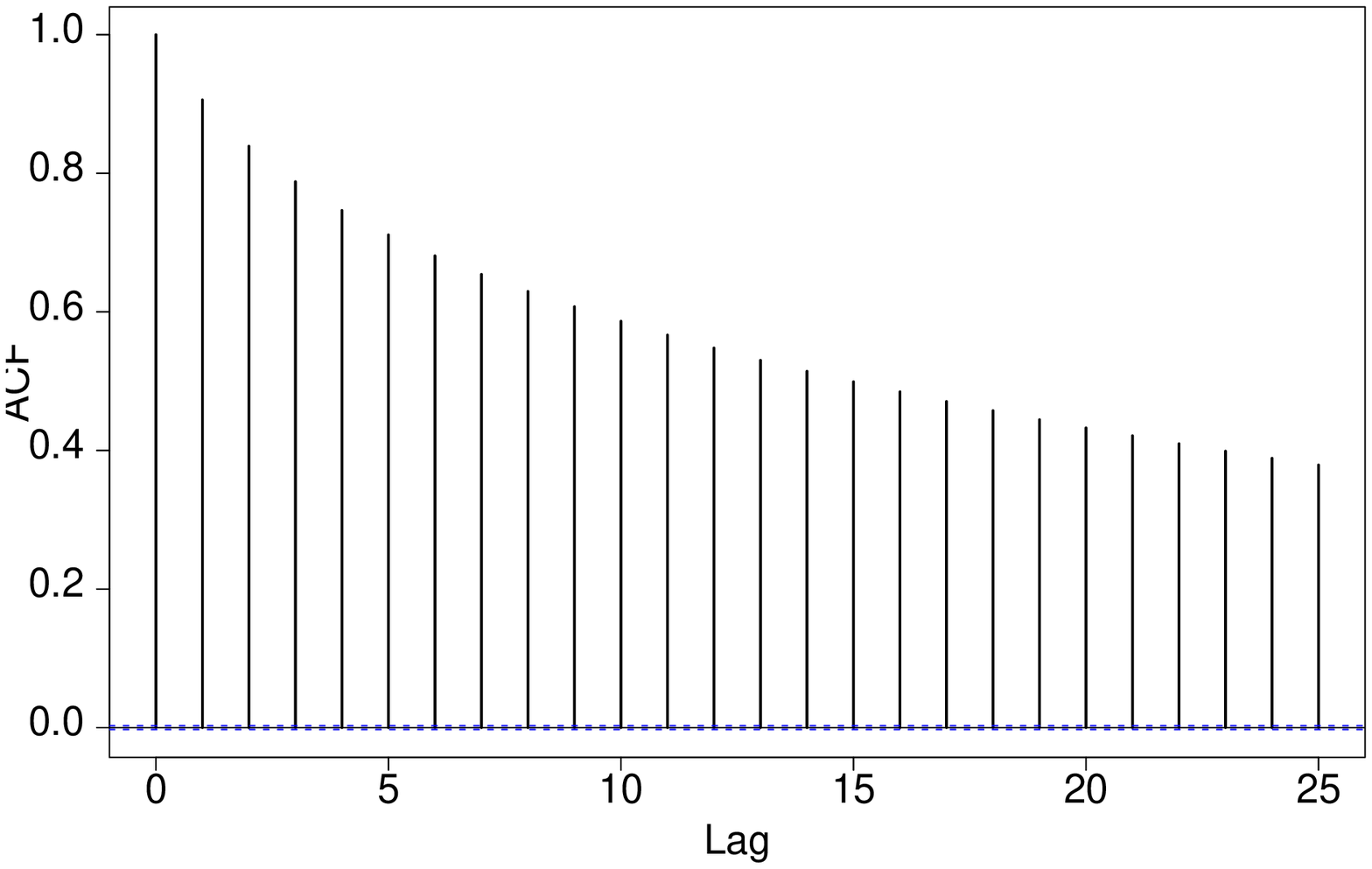}
    \end{minipage}}%
  \subfigure[$\eta$.]{
    \label{fig:ds3-nacfeta}
    \begin{minipage}[b]{0.5\textwidth}
      \centering \includegraphics[width=0.9\textwidth,height=0.15\textheight]{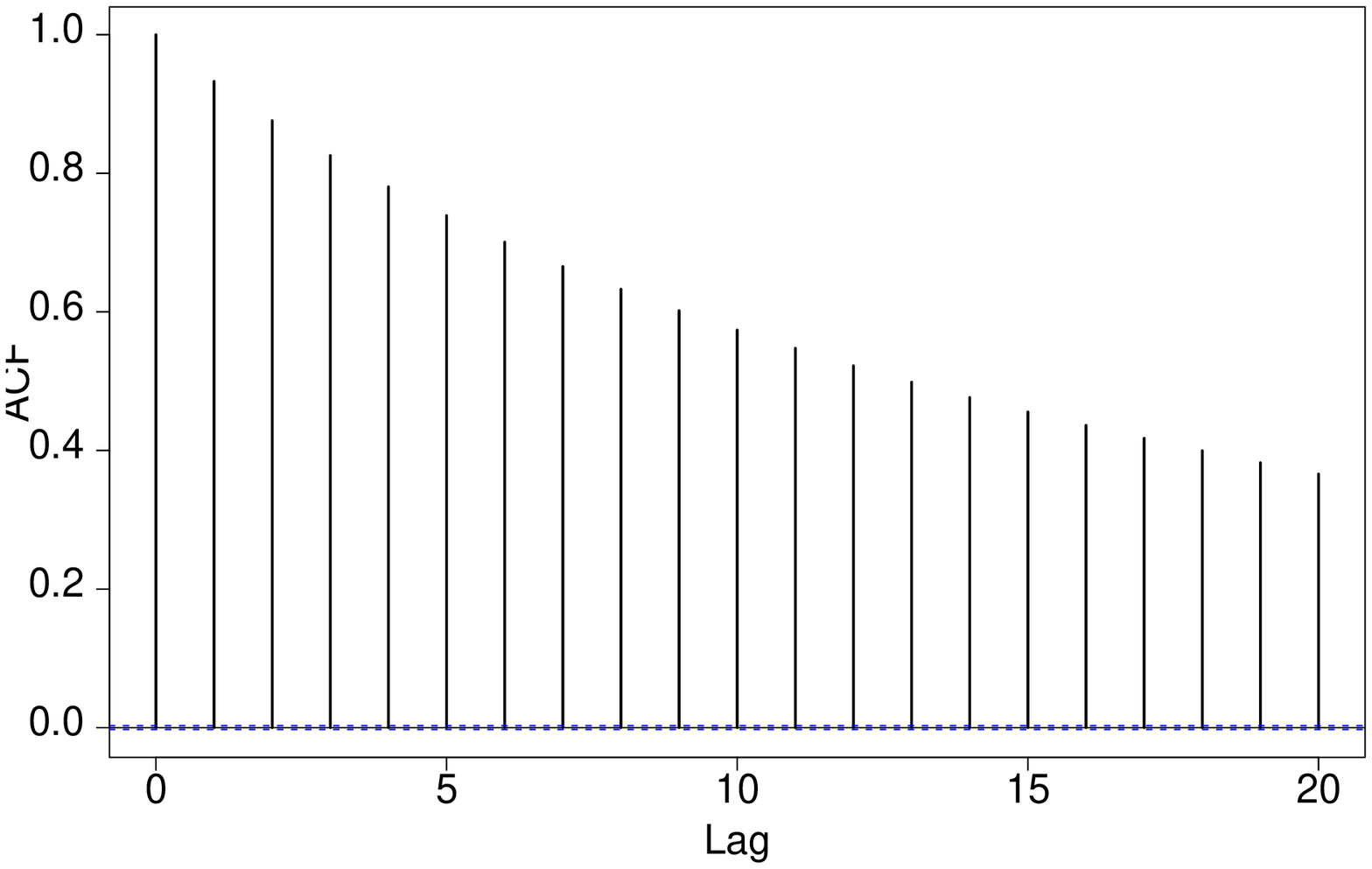}
    \end{minipage}} \\
  \subfigure[$\alpha_{0}$.]{
    \label{fig:ds3-nacfalpha:0}
    \begin{minipage}[b]{0.5\textwidth}
      \centering 
      \includegraphics[width=0.9\textwidth,height=0.15\textheight]{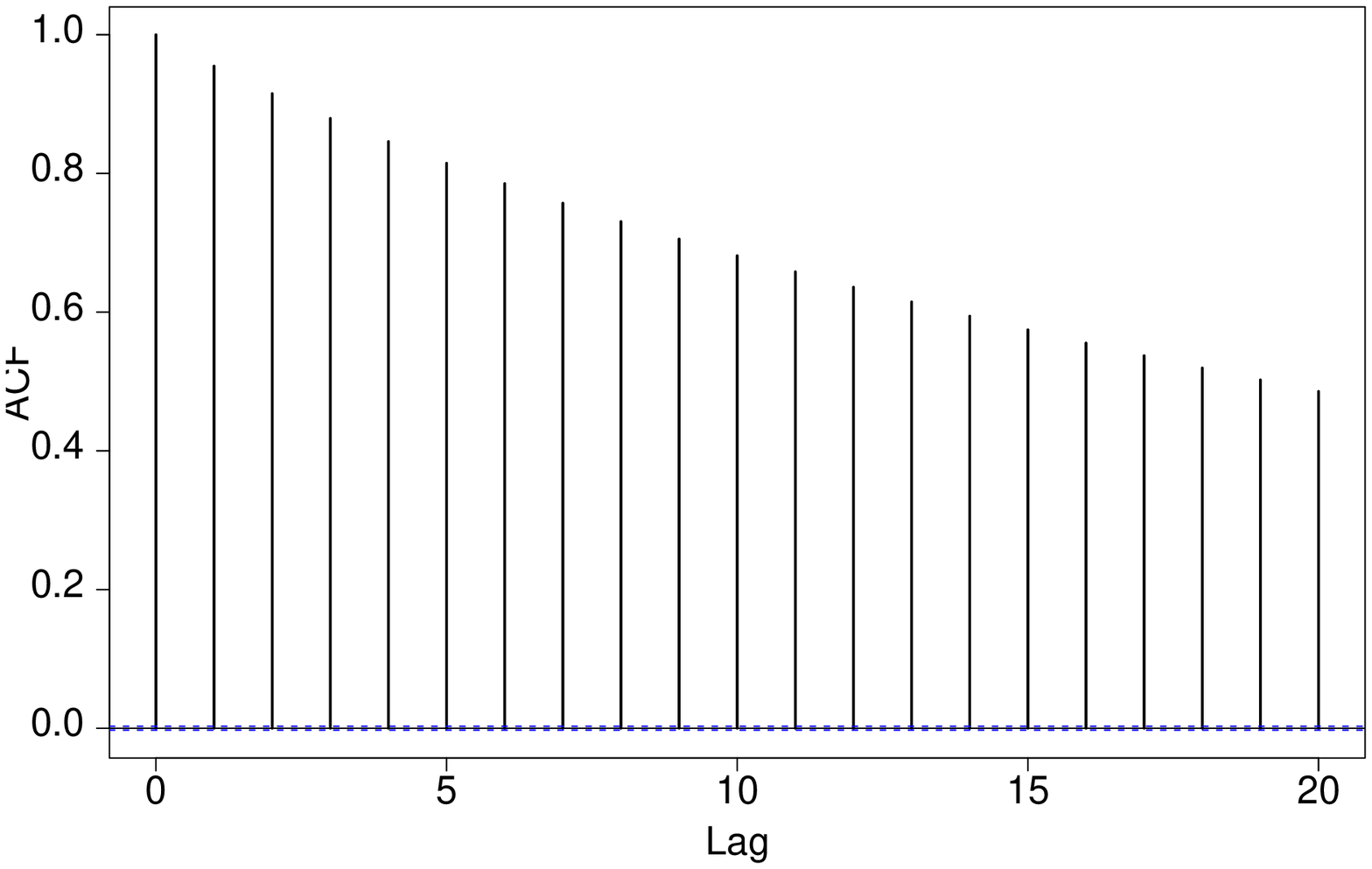}
    \end{minipage}} 
  \subfigure[$\alpha_{1}$.]{
    \label{fig:ds3-nacfalpha:1}
    \begin{minipage}[b]{0.5\textwidth}
      \centering 
      \includegraphics[width=0.9\textwidth,height=0.15\textheight]{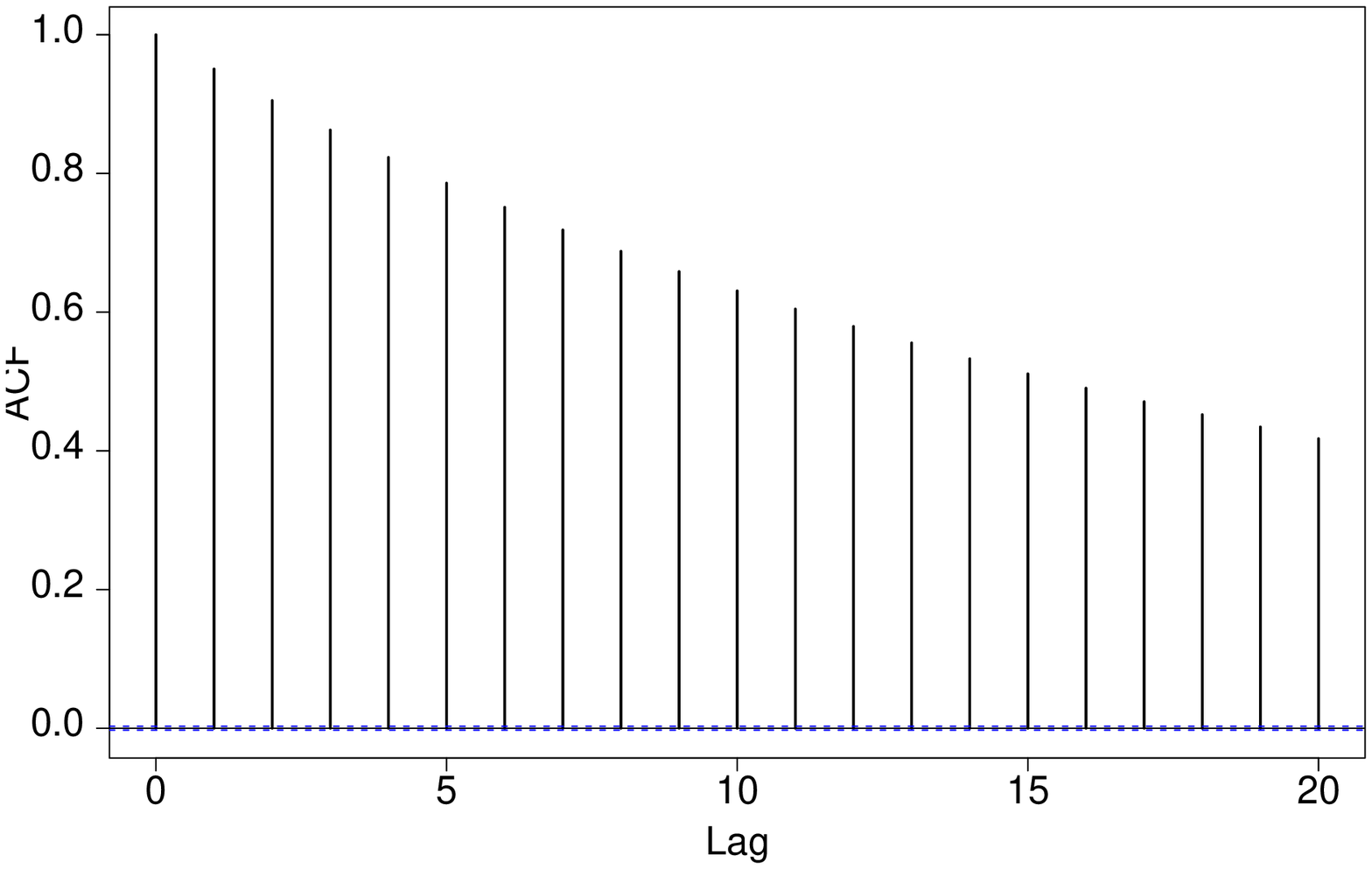}
    \end{minipage}} \\
  \subfigure[$\alpha_{2}$.]{
    \label{fig:ds3-nacfalpha:2}
    \begin{minipage}[b]{0.5\textwidth}
      \centering 
      \includegraphics[width=0.9\textwidth,height=0.15\textheight]{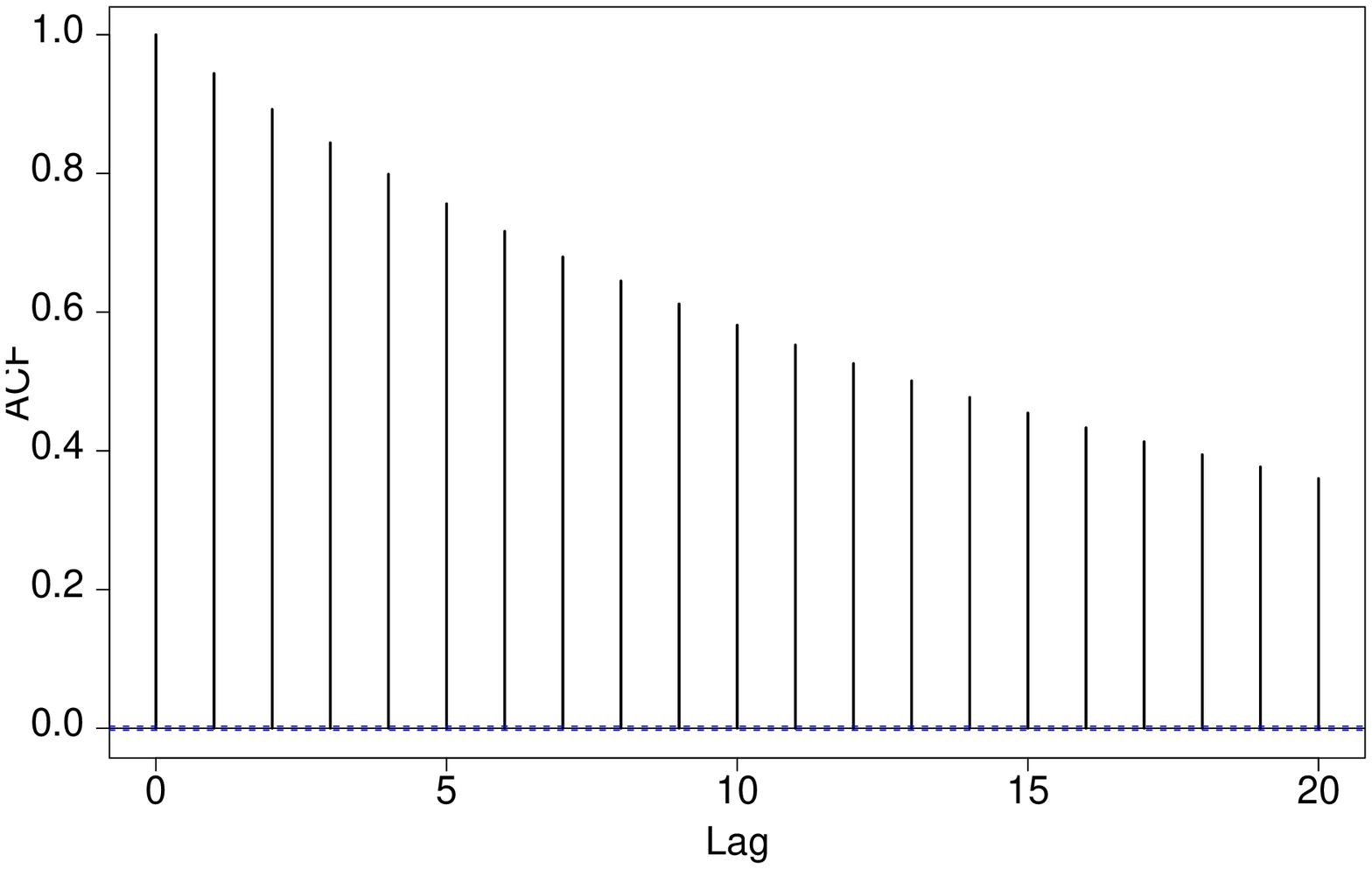}
    \end{minipage}}%
  \subfigure[$\alpha_{3}$.]{
    \label{fig:ds3-nacfalpha:3}
    \begin{minipage}[b]{0.5\textwidth}
      \centering 
      \includegraphics[width=0.9\textwidth,height=0.15\textheight]{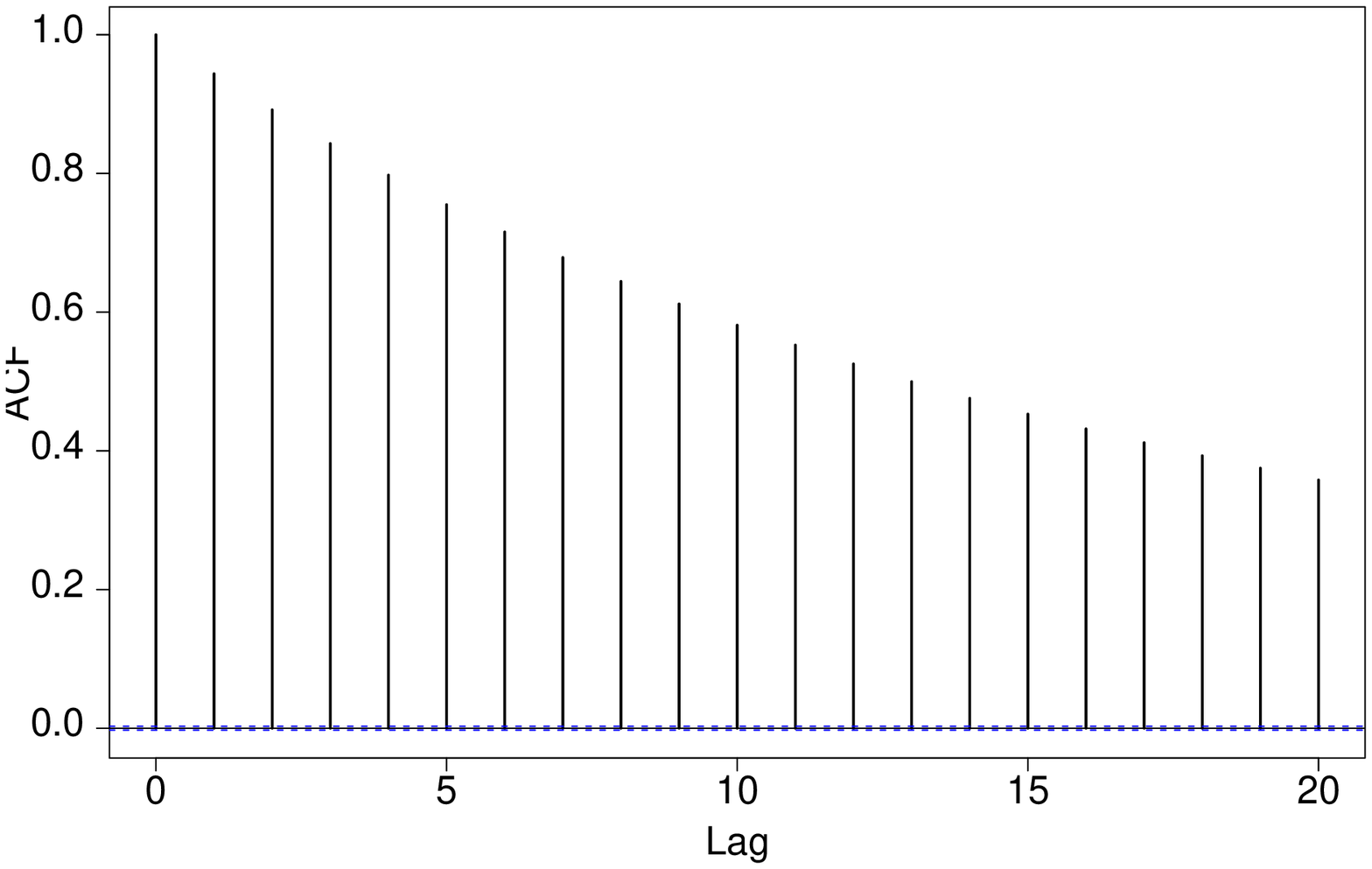}
    \end{minipage}}  \\
  \subfigure[$\alpha_{4}$.]{
    \label{fig:ds3-nacfalpha:4}
    \begin{minipage}[b]{0.5\textwidth}
      \centering 
      \includegraphics[width=0.9\textwidth,height=0.15\textheight]{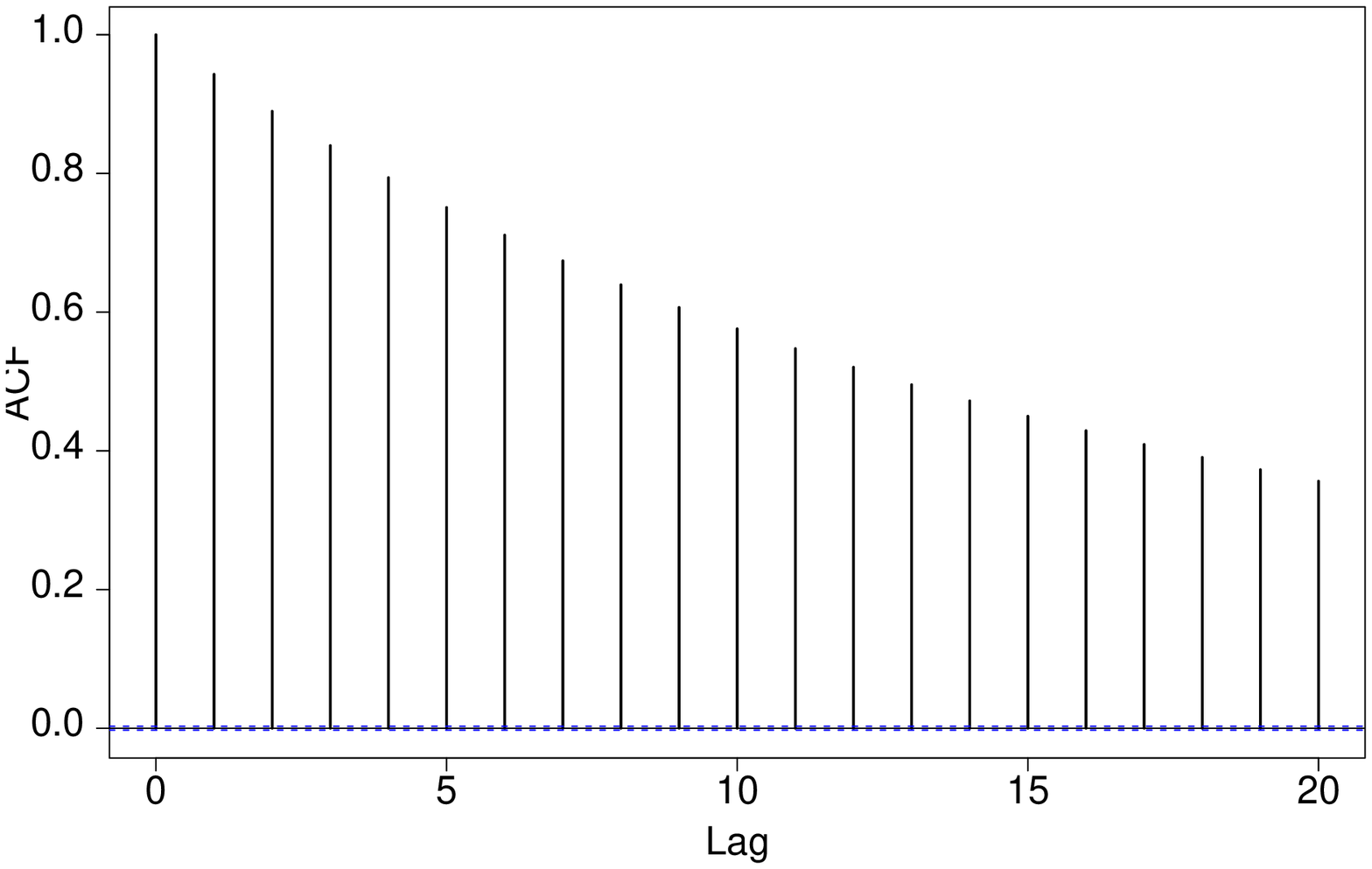}
    \end{minipage}}%
  \subfigure[$\alpha_{5}$.]{
    \label{fig:ds3-nacfalpha:5}
    \begin{minipage}[b]{0.5\textwidth}
      \centering 
      \includegraphics[width=0.9\textwidth,height=0.15\textheight]{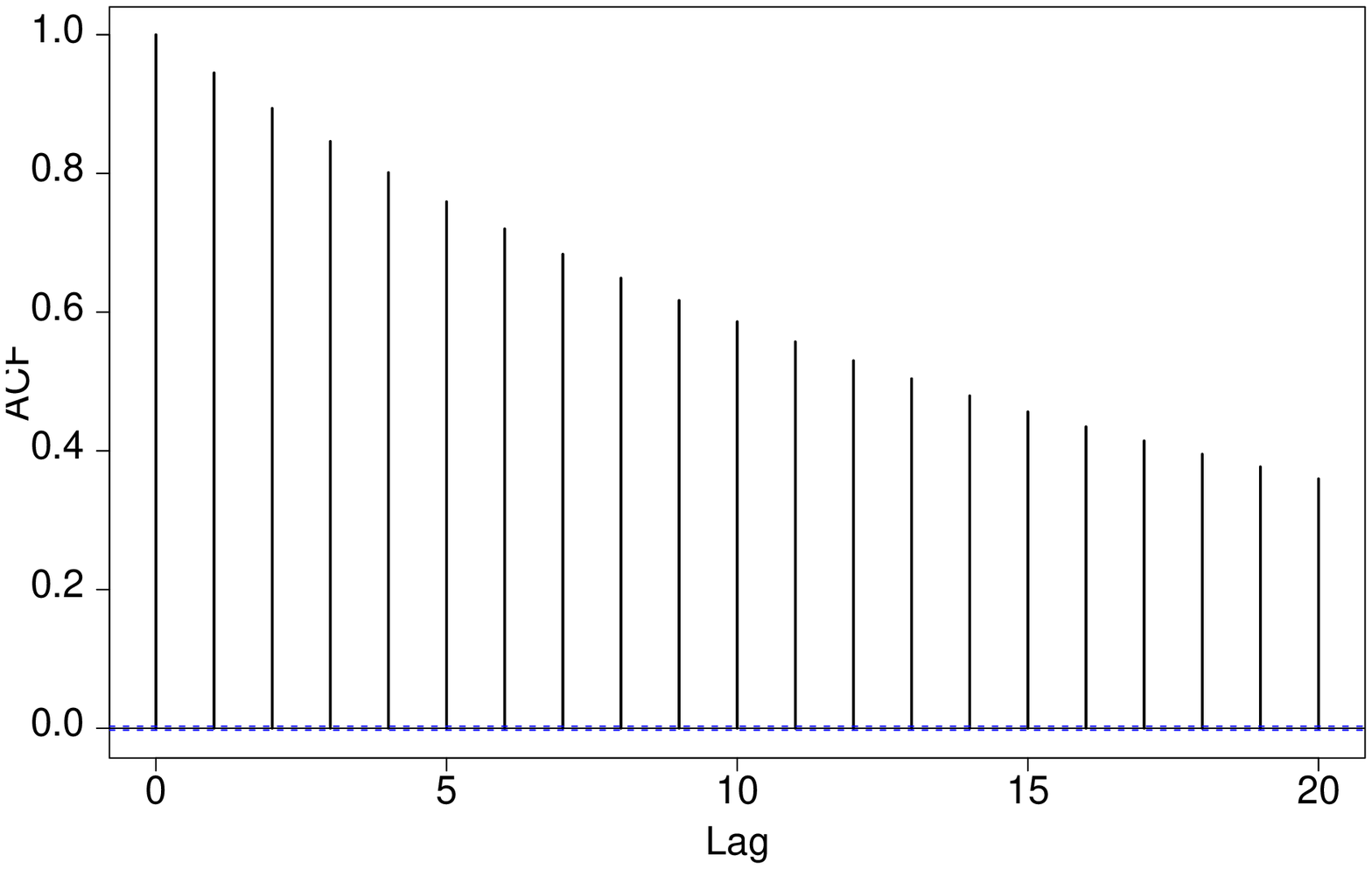}
    \end{minipage}} \\
  \subfigure[$\alpha_{6}$.]{
    \label{fig:ds3-nacfalpha:6}
    \begin{minipage}[b]{0.5\textwidth}
      \centering 
      \includegraphics[width=0.9\textwidth,height=0.15\textheight]{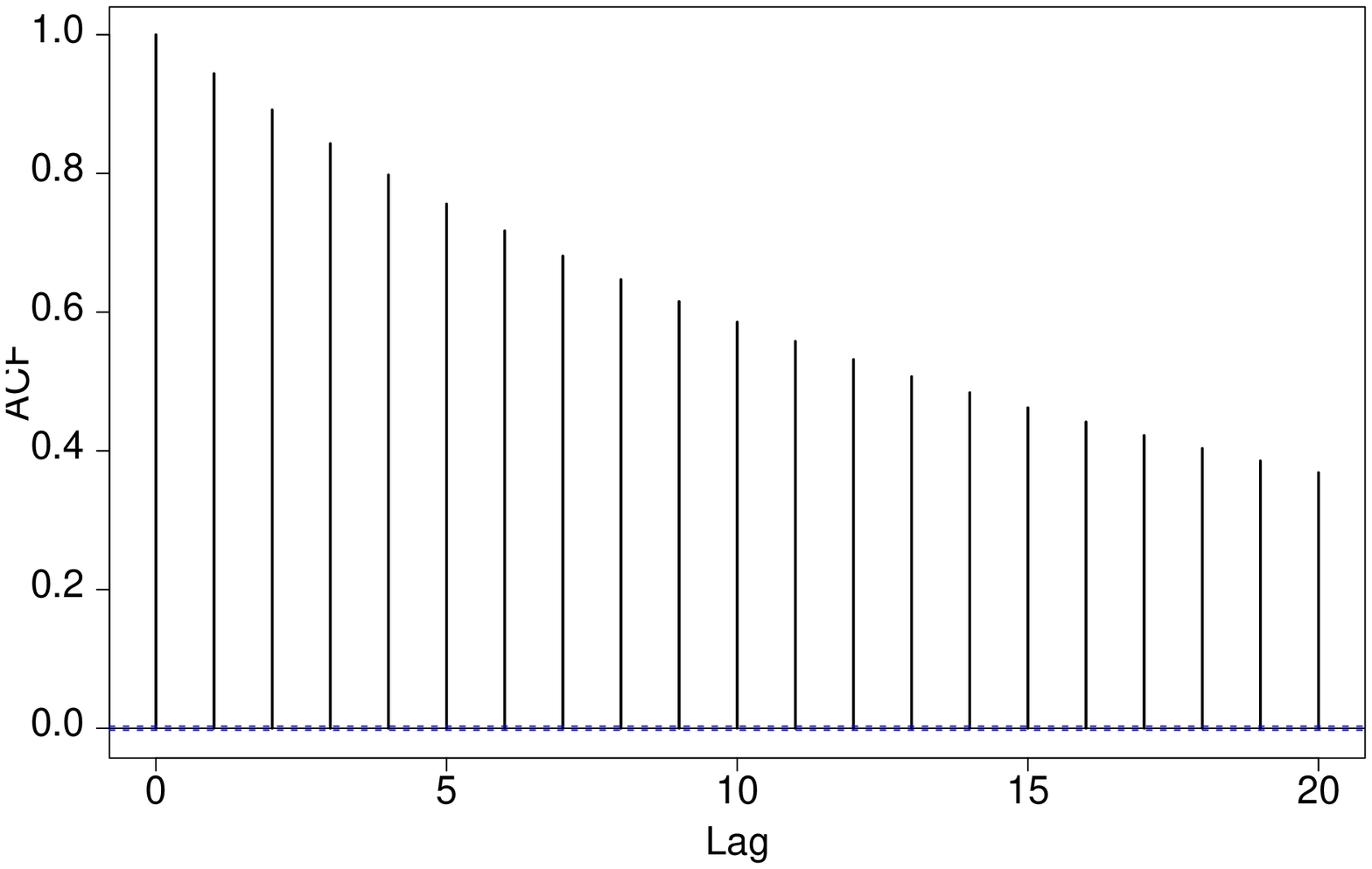}
    \end{minipage}} 
  \subfigure[$\alpha_{7}$.]{
    \label{fig:ds3-nacfalpha:7}
    \begin{minipage}[b]{0.5\textwidth}
      \centering 
      \includegraphics[width=0.9\textwidth,height=0.15\textheight]{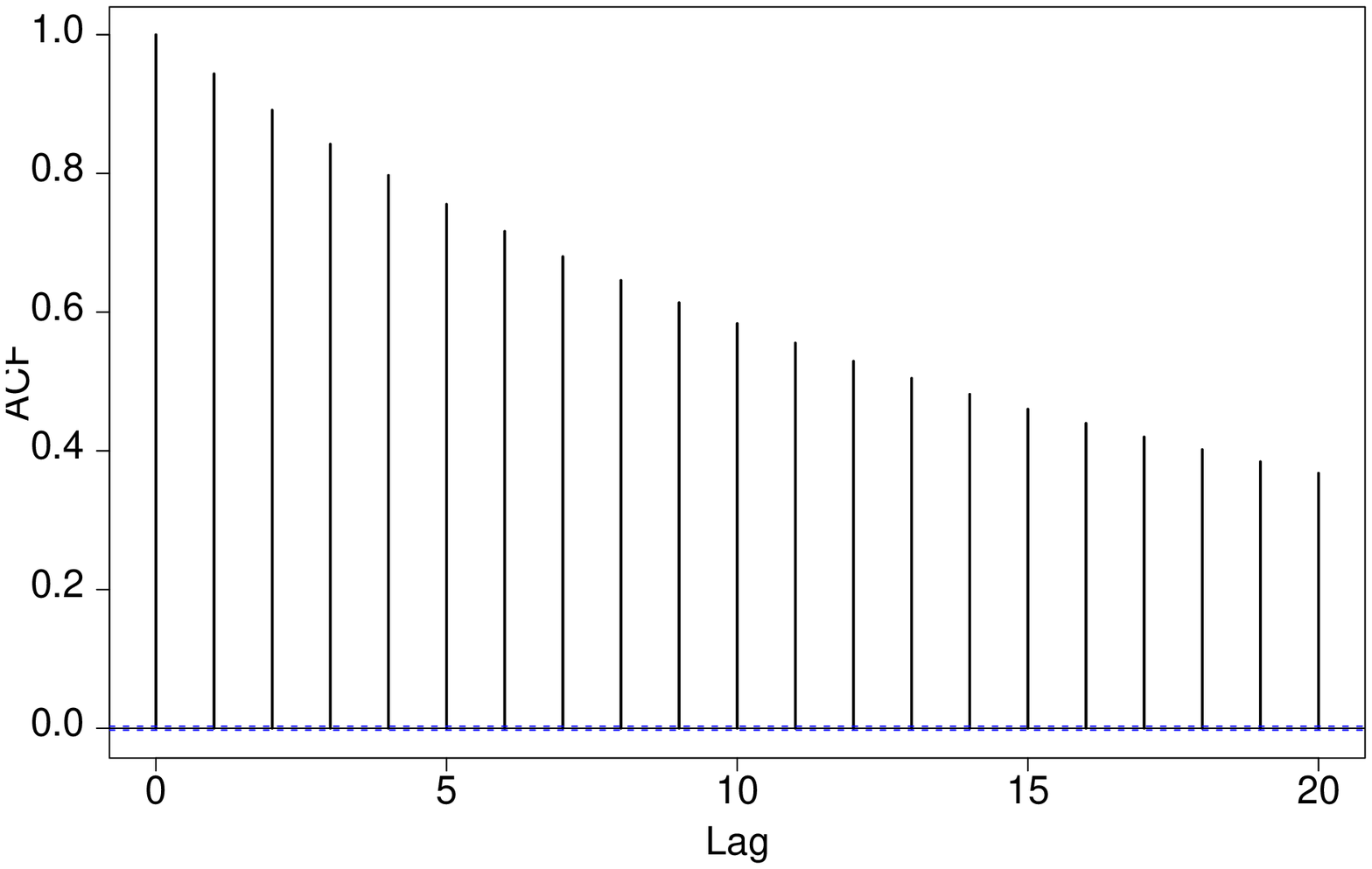}
    \end{minipage}}%
  \caption{Autocorrelation plots for $\rho$, $\eta$ and $\bfalpha$ with 
    inverse-variance parameters $\sigma$ and $\tau$ integrated out.}
  \label{fig:ds3-nacf} %% label for entire figure
\end{figure}

\begin{figure}
  \subfigure[Trace plot of $\rho$.]{
    \label{fig:ds3-nrho}
    \begin{minipage}[b]{0.5\textwidth}
      \centering 
      \includegraphics[width=0.9\textwidth,height=0.15\textheight]{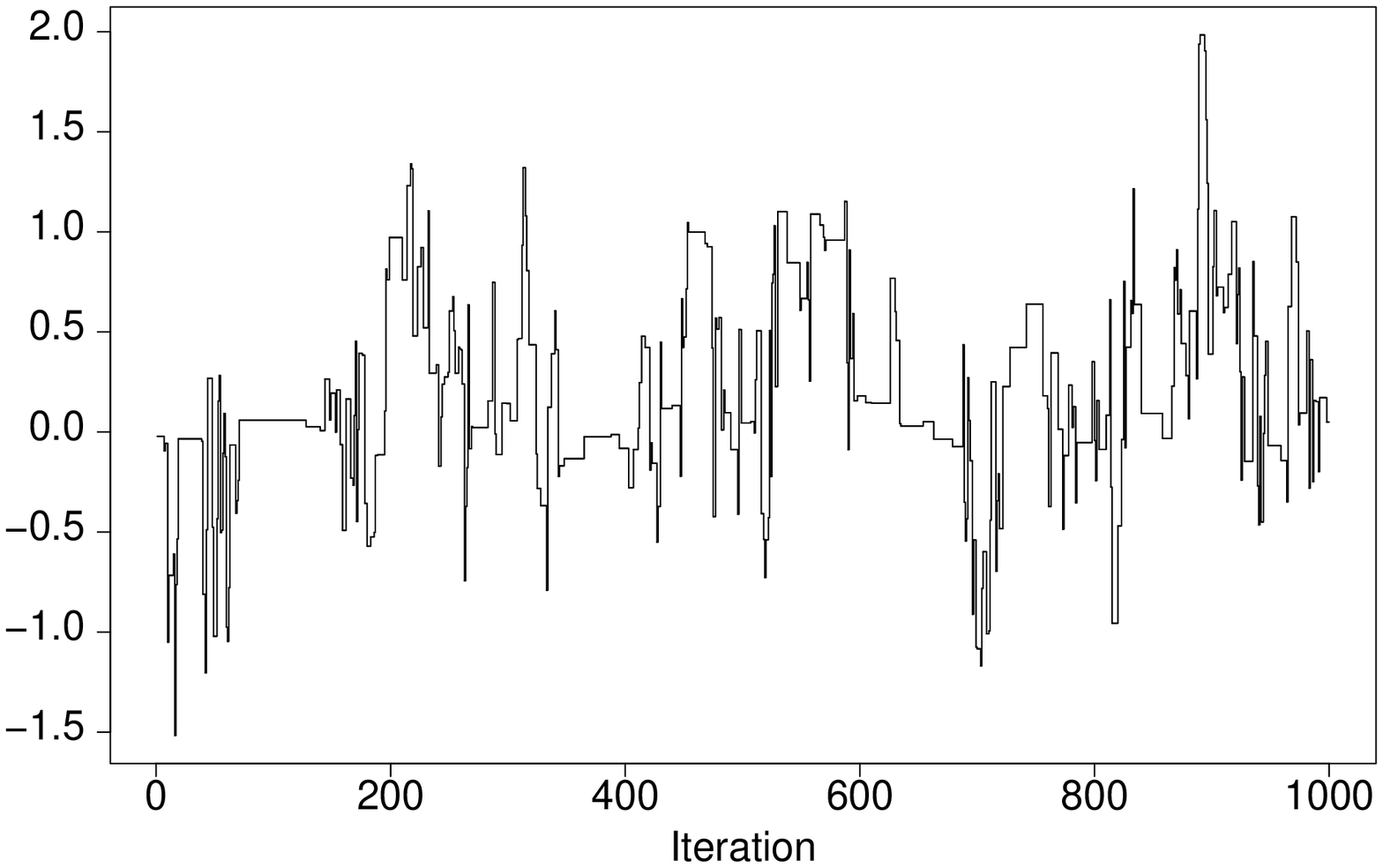}
    \end{minipage}}%
  \subfigure[Trace plot of $\eta$.]{
    \label{fig:ds3-neta}
    \begin{minipage}[b]{0.5\textwidth}
      \centering 
      \includegraphics[width=0.9\textwidth,height=0.15\textheight]{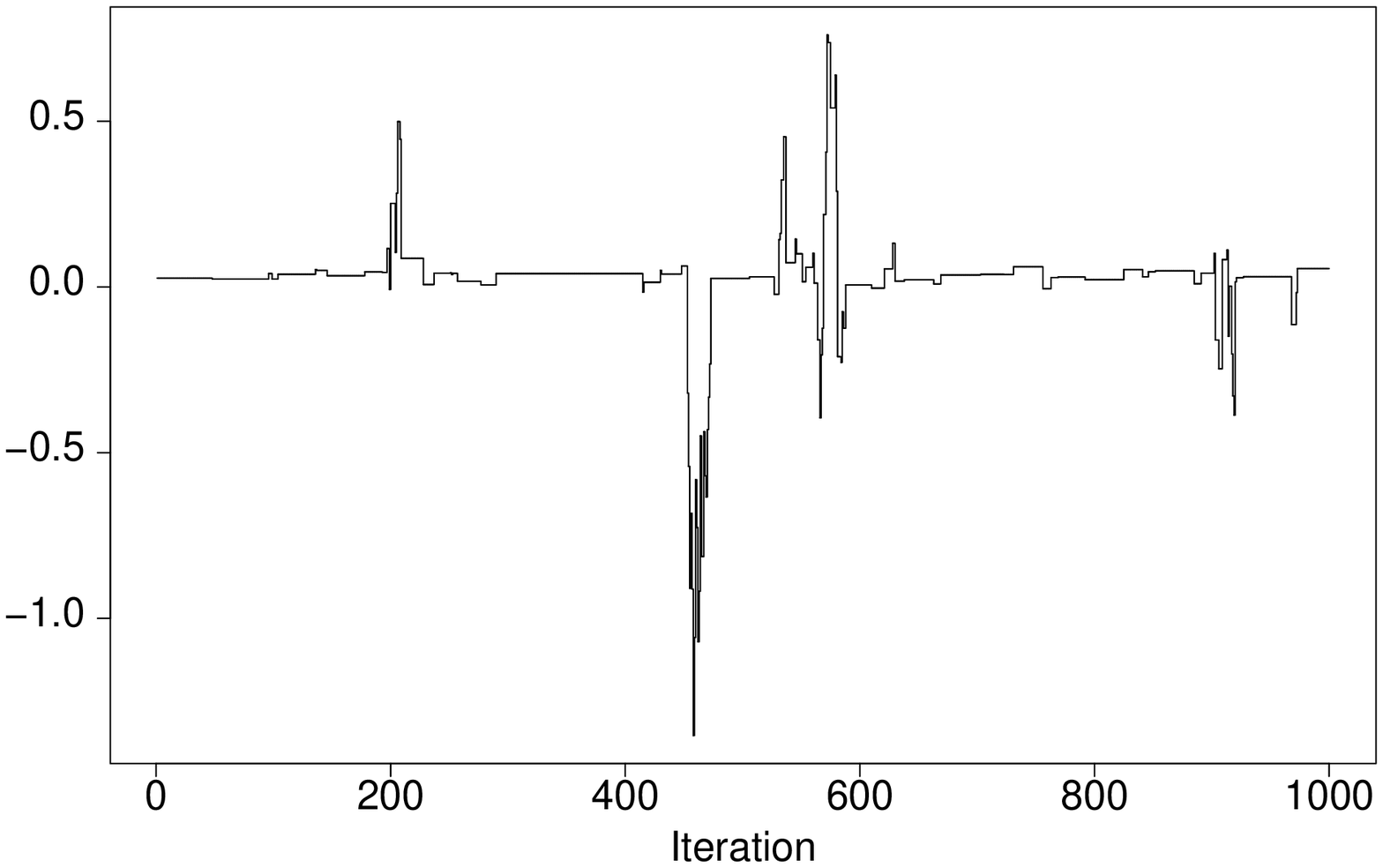}
    \end{minipage}} \\
  \subfigure[Trace plot of $\alpha_{0}$.]{
    \label{fig:ds3-nalpha:0}
    \begin{minipage}[b]{0.5\textwidth}
      \centering 
      \includegraphics[width=0.9\textwidth,height=0.15\textheight]{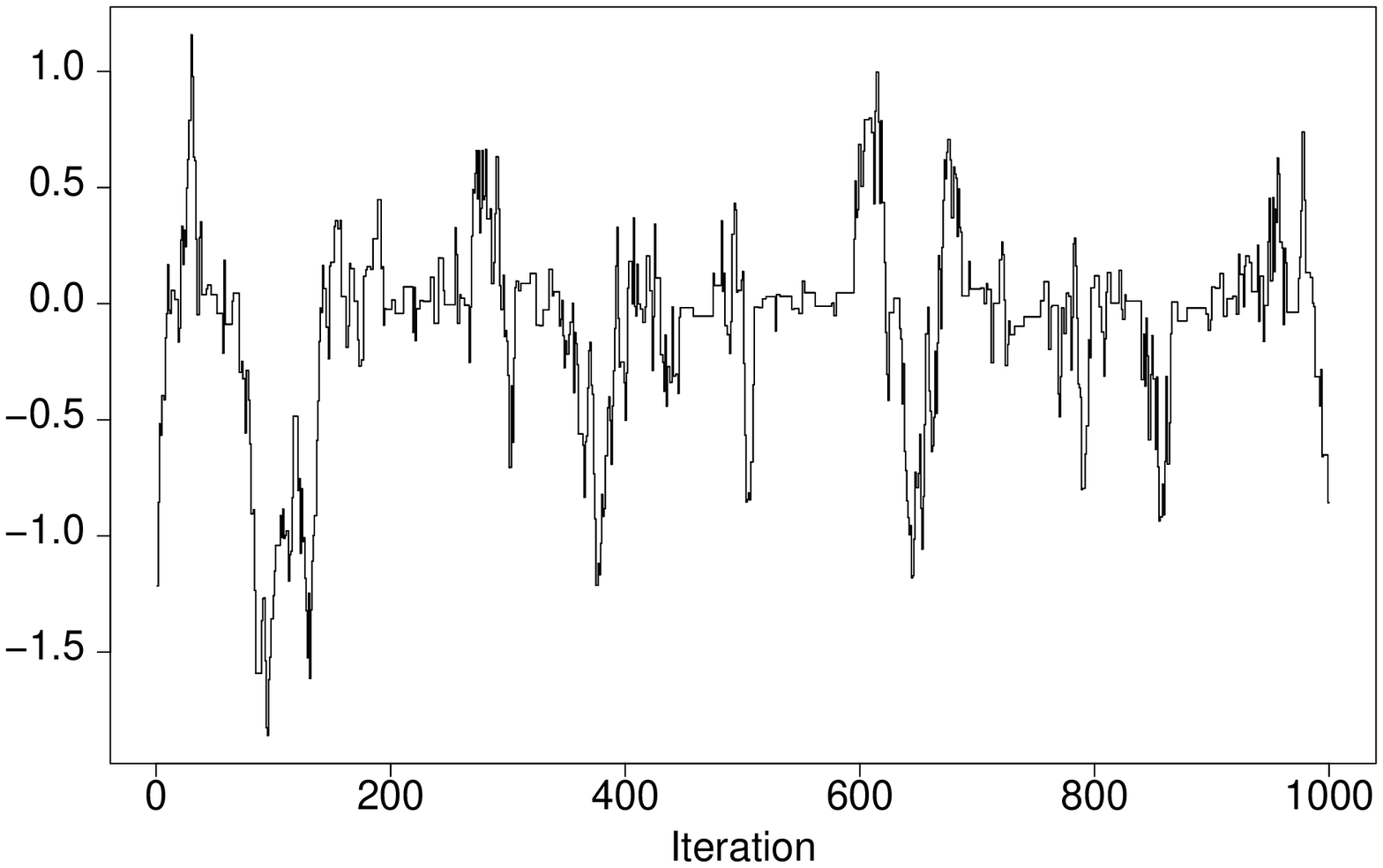}
    \end{minipage}} 
  \subfigure[Trace plot of $\alpha_{1}$.]{
    \label{fig:ds3-nalpha:1}
    \begin{minipage}[b]{0.5\textwidth}
      \centering 
      \includegraphics[width=0.9\textwidth,height=0.15\textheight]{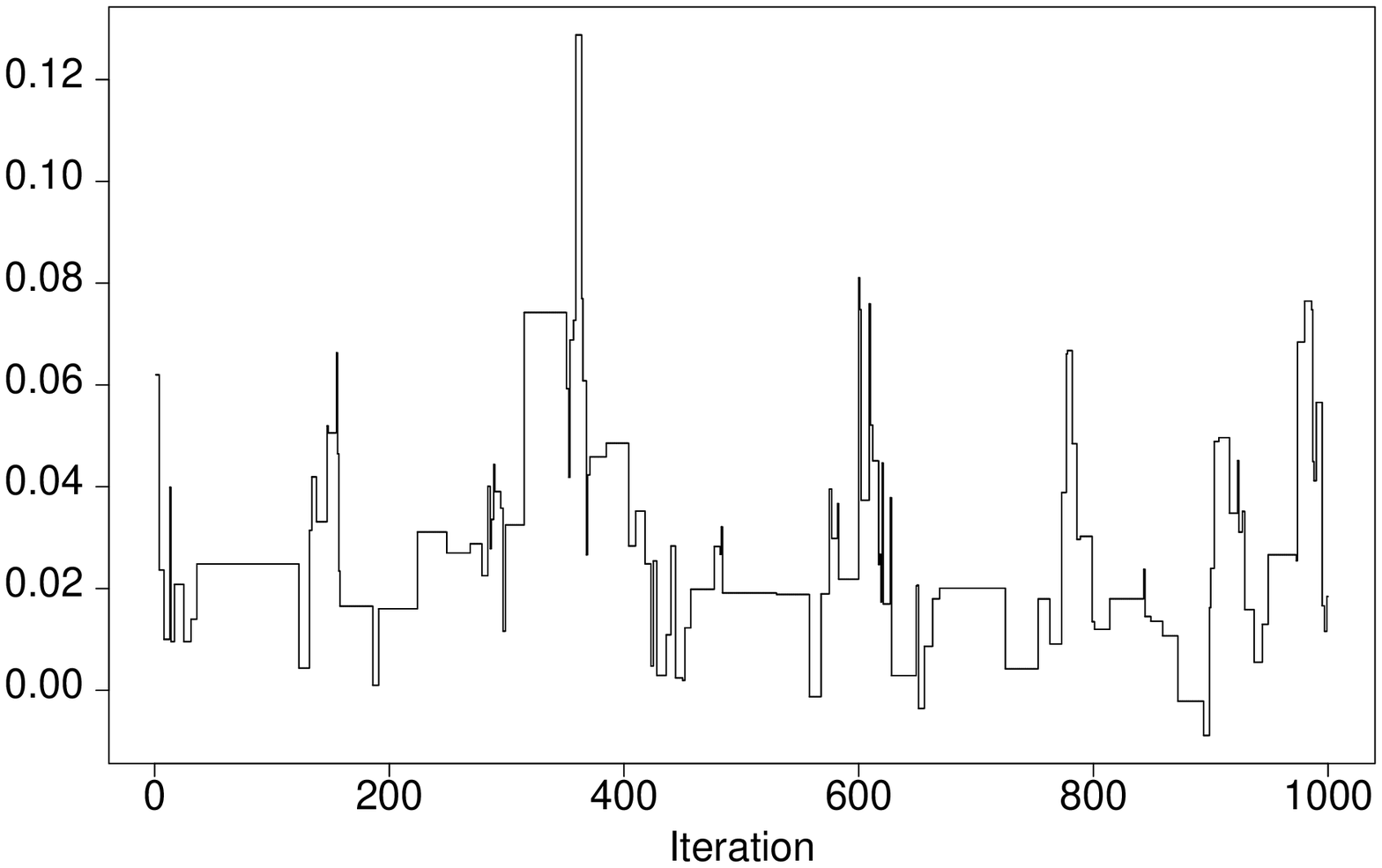}
    \end{minipage}} \\
  \subfigure[Trace plot of $\alpha_{2}$.]{
    \label{fig:ds3-nalpha:2}
    \begin{minipage}[b]{0.5\textwidth}
      \centering 
      \includegraphics[width=0.9\textwidth,height=0.15\textheight]{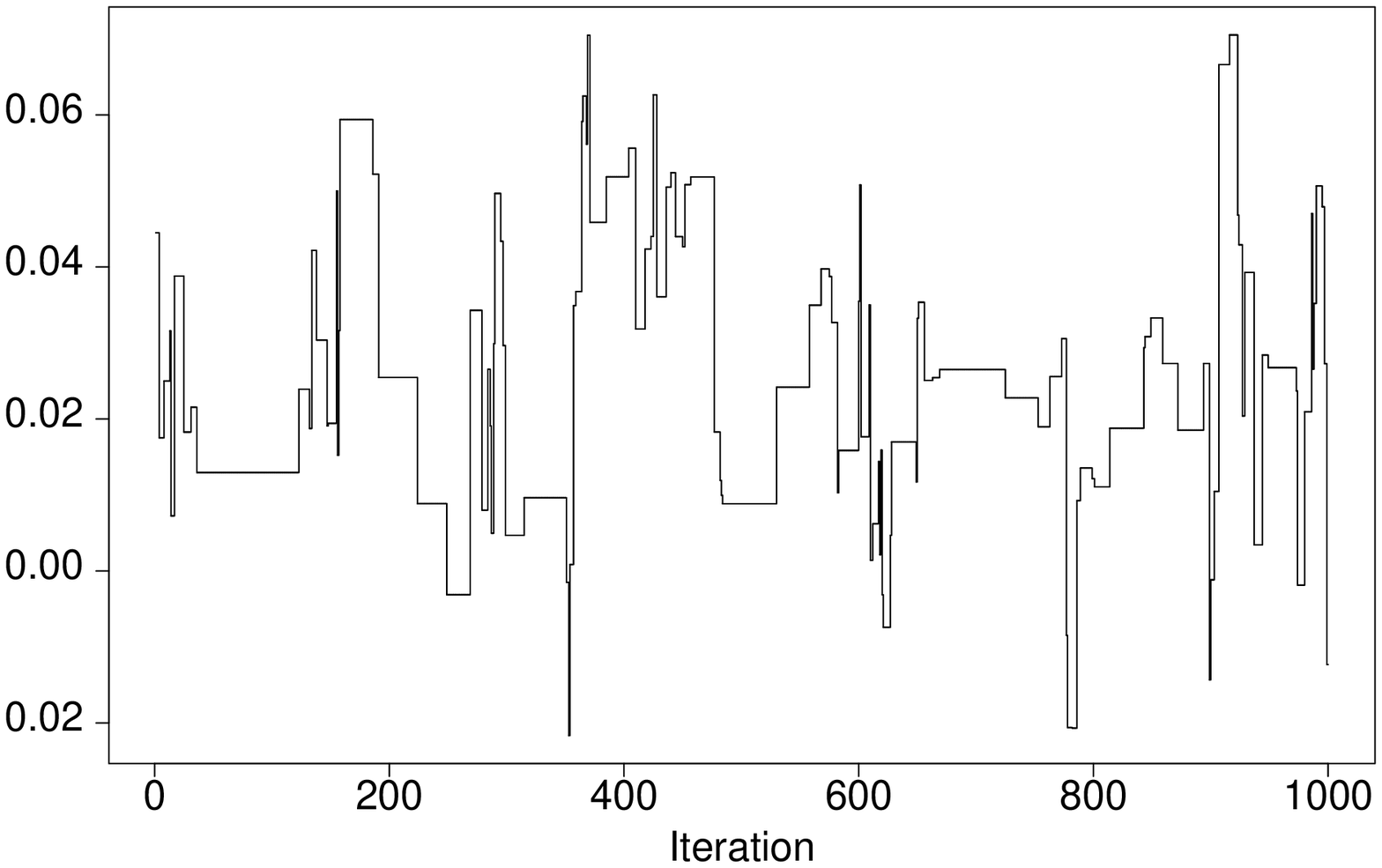}
    \end{minipage}}%
  \subfigure[Trace plot of $\alpha_{3}$.]{
    \label{fig:ds3-nalpha:3}
    \begin{minipage}[b]{0.5\textwidth}
      \centering 
      \includegraphics[width=0.9\textwidth,height=0.15\textheight]{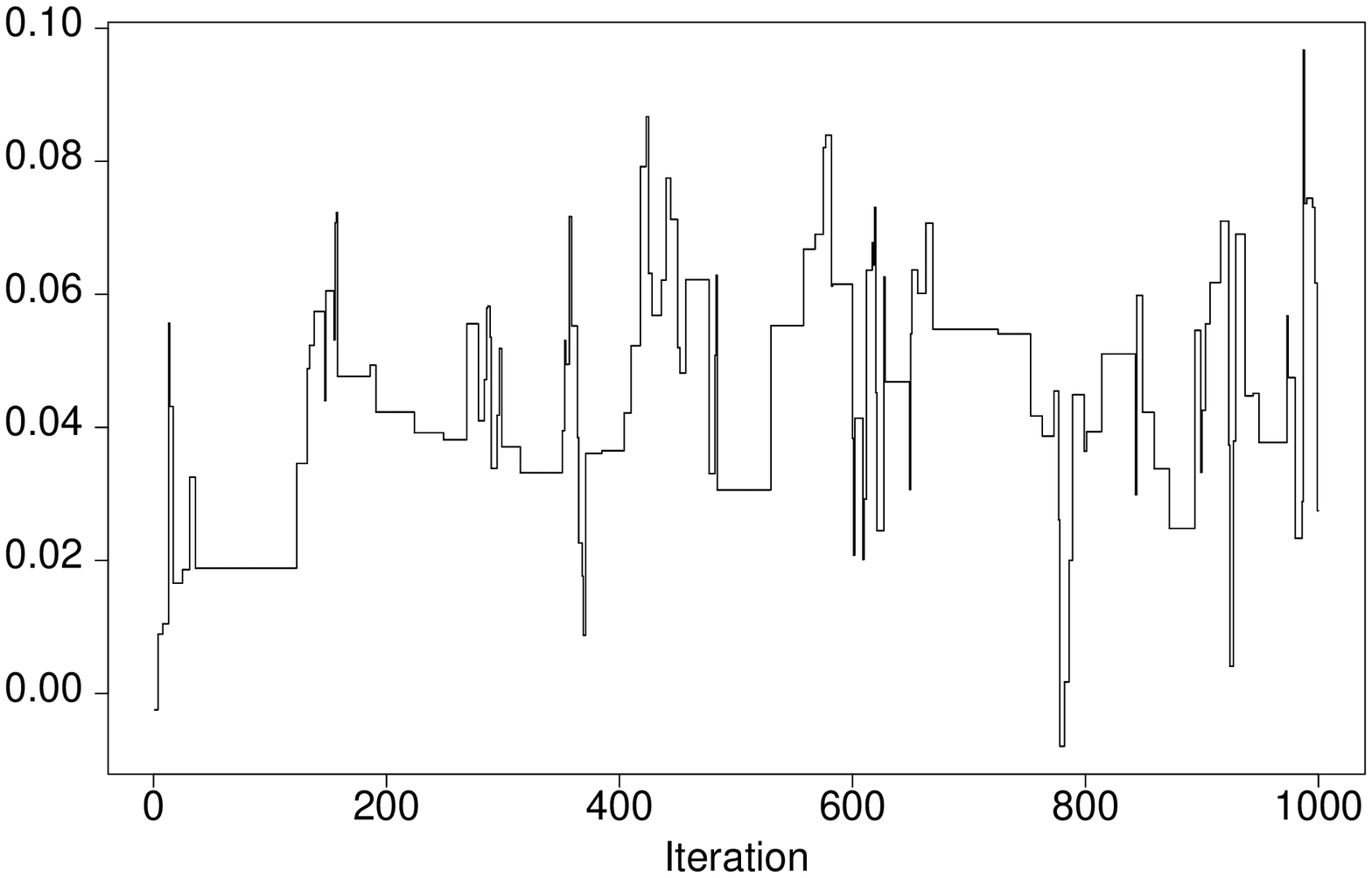}
    \end{minipage}} \\
  \subfigure[Trace plot of $\alpha_{4}$.]{
    \label{fig:ds3-nalpha:4}
    \begin{minipage}[b]{0.5\textwidth}
      \centering 
      \includegraphics[width=0.9\textwidth,height=0.15\textheight]{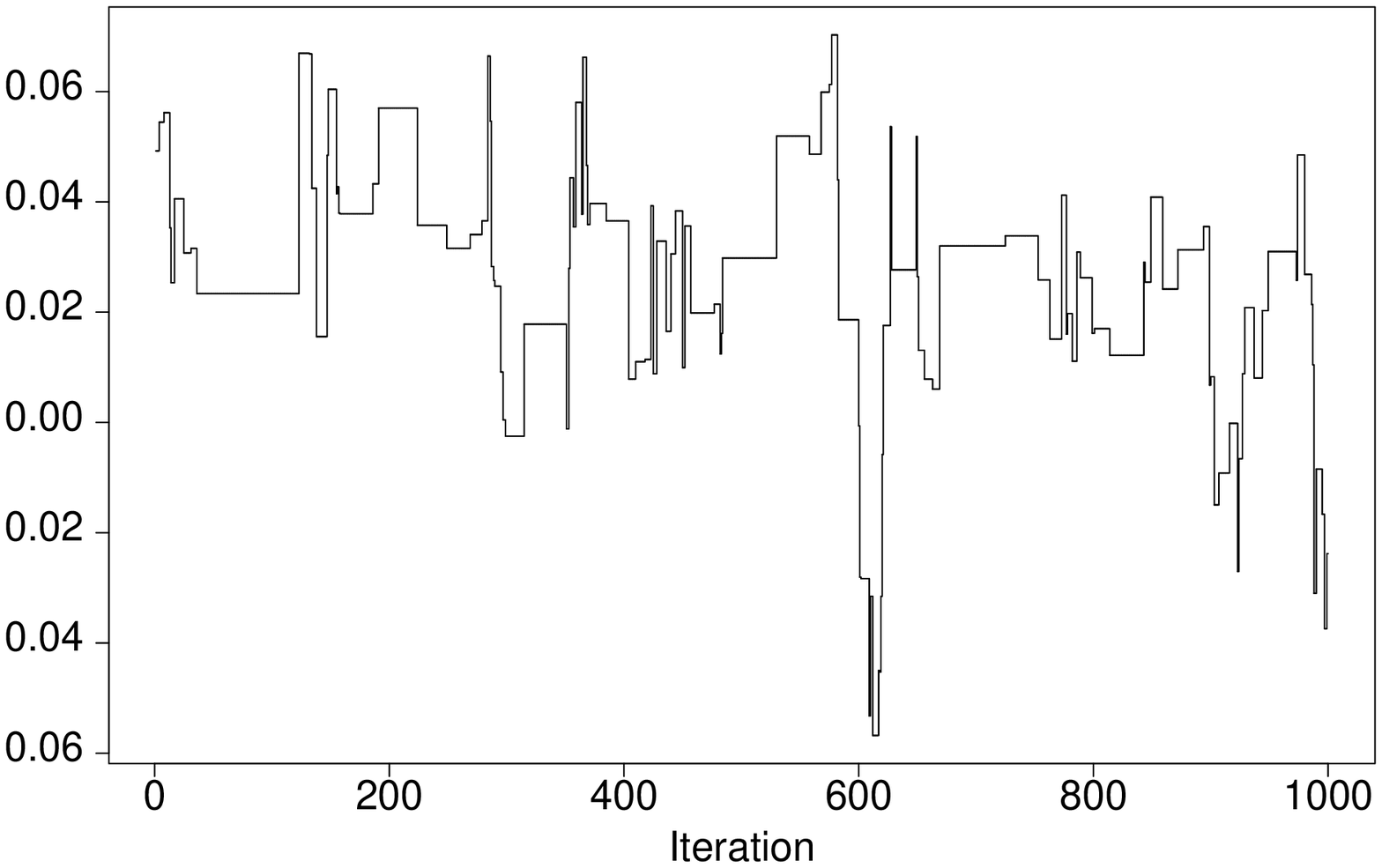}
    \end{minipage}}%
  \subfigure[Trace plot of $\alpha_{5}$.]{
    \label{fig:ds3-nalpha:5}
    \begin{minipage}[b]{0.5\textwidth}
      \centering 
      \includegraphics[width=0.9\textwidth,height=0.15\textheight]{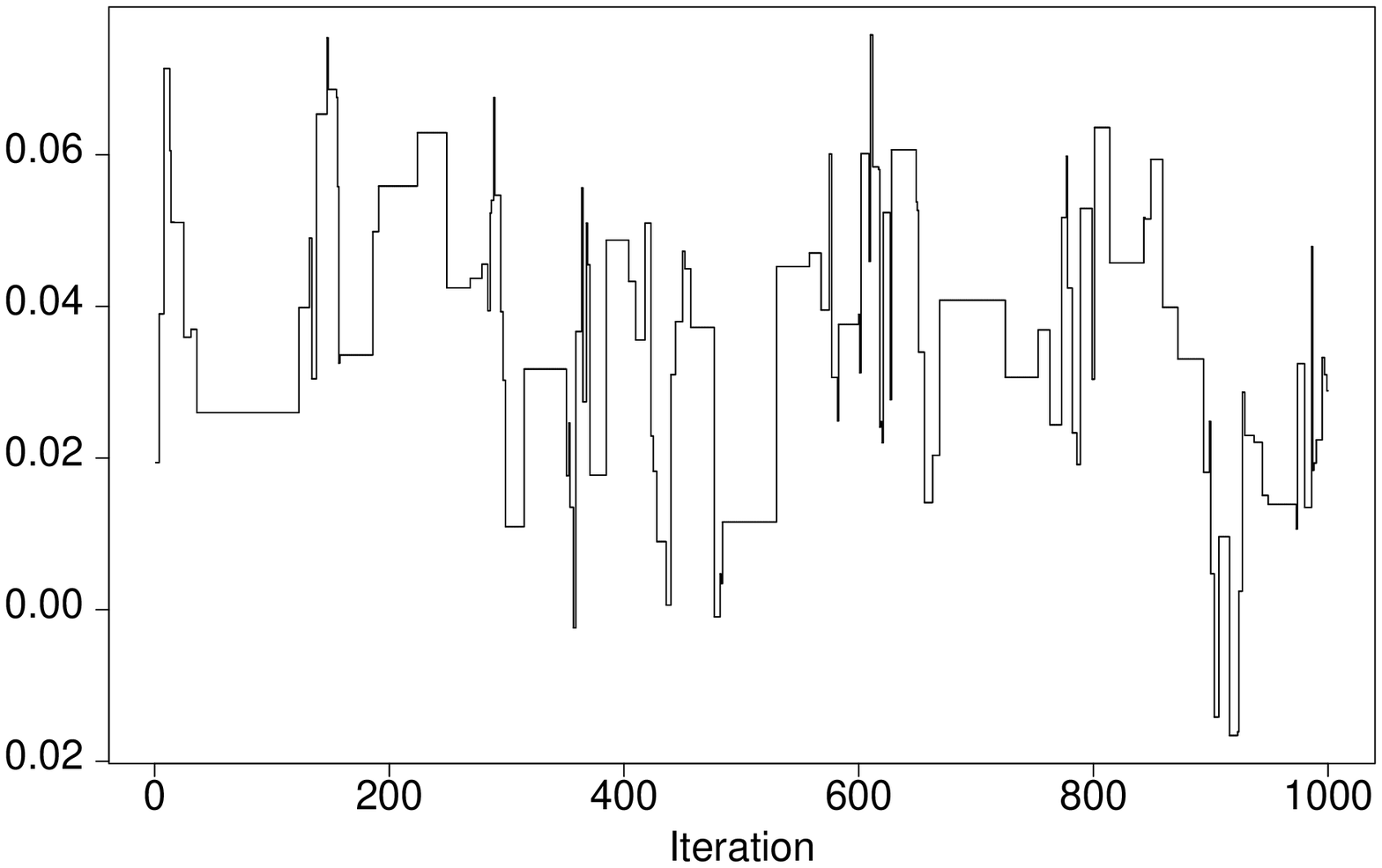}
    \end{minipage}} \\
  \subfigure[Trace plot of $\alpha_{6}$.]{
    \label{fig:ds3-nalpha:6}
    \begin{minipage}[b]{0.5\textwidth}
      \centering 
      \includegraphics[width=0.9\textwidth,height=0.15\textheight]{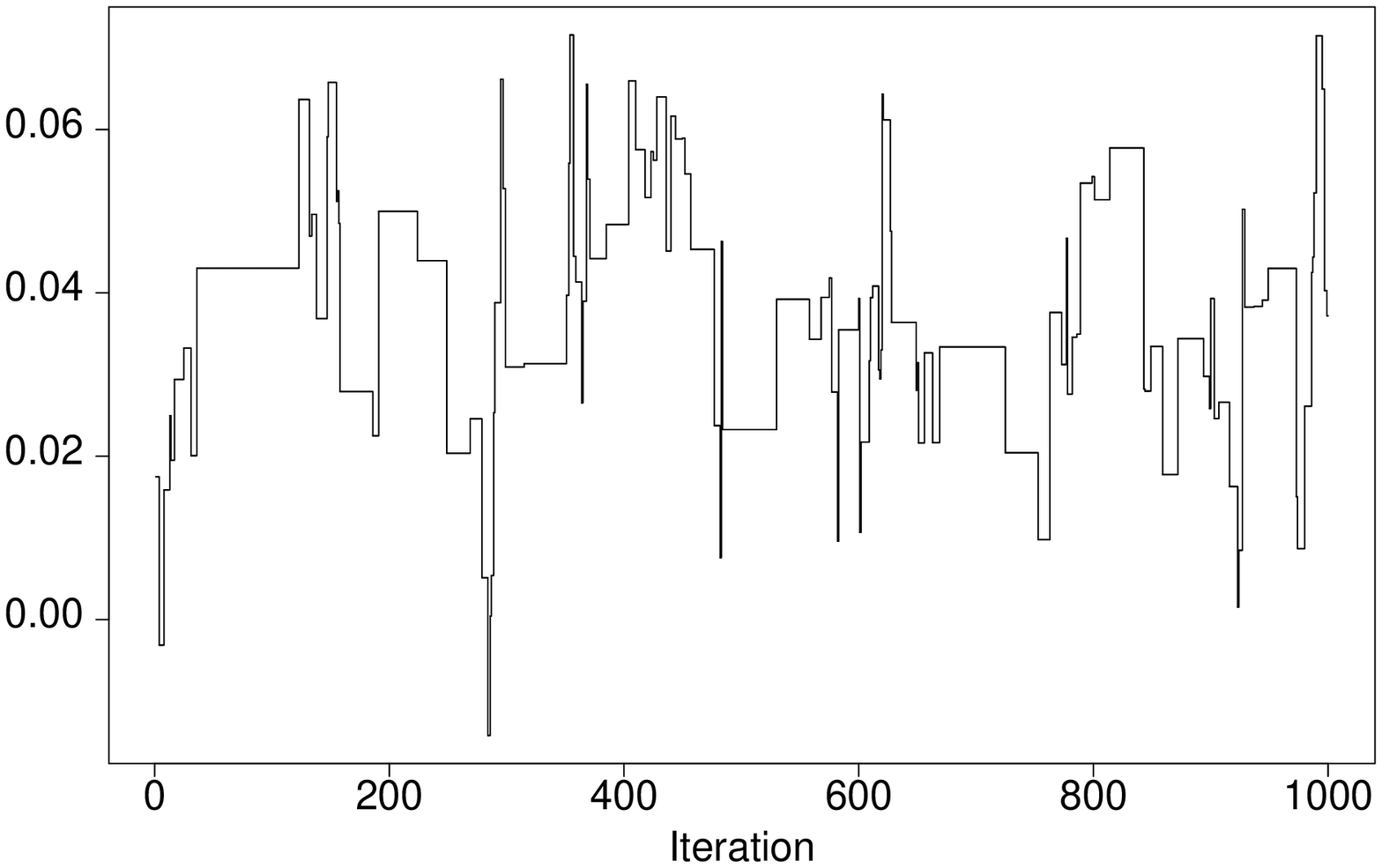}
    \end{minipage}}%
  \subfigure[Trace plot of $\alpha_{7}$.]{
    \label{fig:ds3-nalpha:7}
    \begin{minipage}[b]{0.5\textwidth}
      \centering 
      \includegraphics[width=0.9\textwidth,height=0.15\textheight]{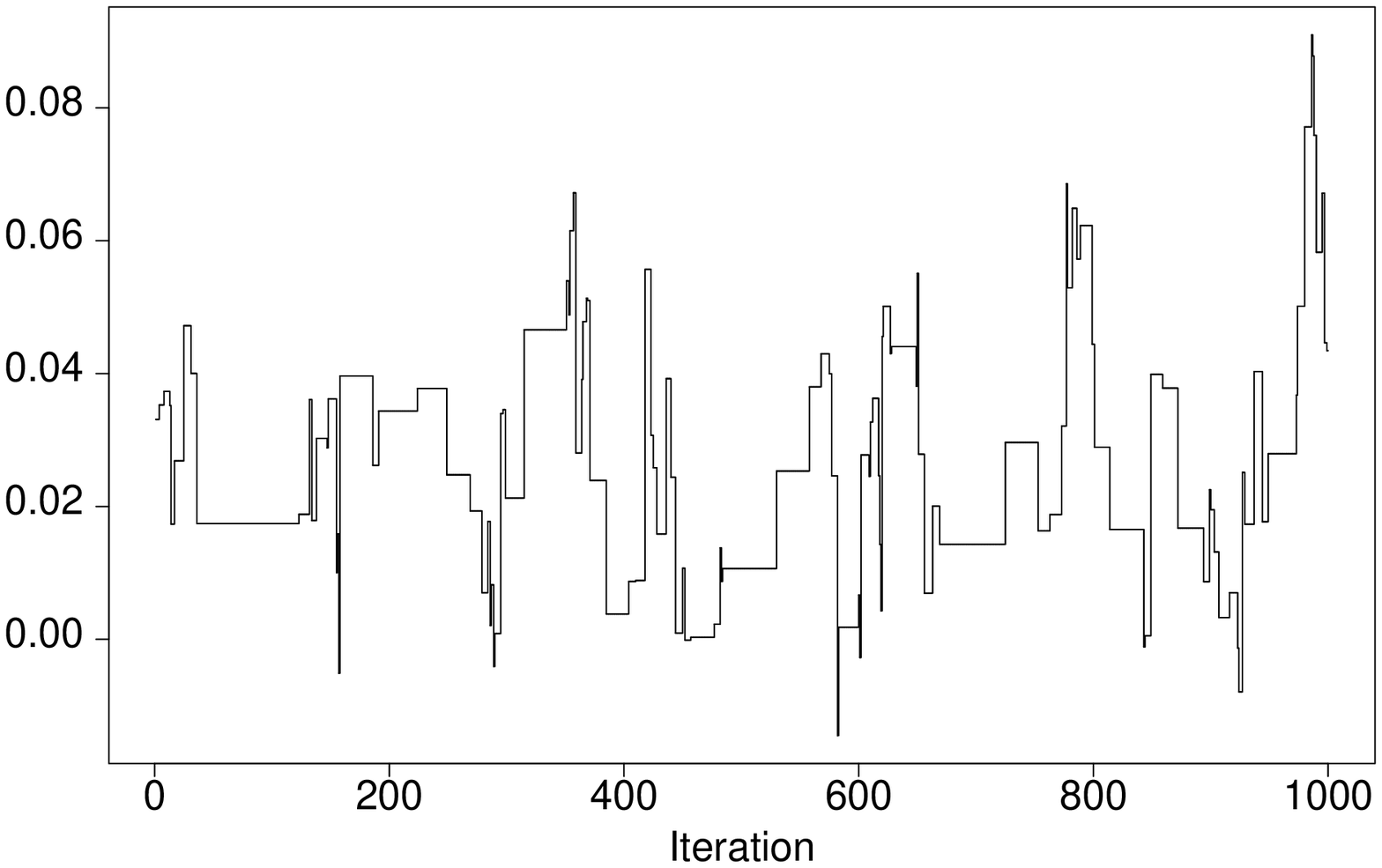}
    \end{minipage}}%
  \caption{Trace plots of model parameters, after integrating out the 
    inverse-variance parameters. }
  \label{fig:ds3-nalpha} %% label for entire figure
\end{figure}

An important diagnostic tool in MCMC modelling is the autocorrelation plot of
the parameter of interest. Figures~\ref{fig:ds3-acf} and \ref{fig:ds3-nacf}
show the autocorrelation functions for the parameters of interest in our
model. For the full implementation the autocorrelation values are essentially
zero at lags greater than $5$, except for $\rho$ where lags up to $25$ are
large. The implementation with the inverse-variance parameters integrated out
does not appear to be better than the full implementation. The
autocorrelations for this implementation are bigger than those of the full
implementation at all lags and are significant up to lag 10, excepting for
$\rho$ which has autocorrelation significant up to lag $15$. The
autocorrelation plots and trace plots for the implementation with the
variance parameters integrated out are shown in Figures~\ref{fig:ds3-nalpha}
and \ref{fig:ds3-nacf}, respectively. This could be indicating poor mixing of
the Metropolis algorithm due to the complexity of the terms in
Equation~\eqref{eq:marg} and small Metropolis acceptance rates.  The small
Metropolis rates could also be indicating that the proposal variances need to
be smaller so that proposed values are closer to the current values and will
have a greater chance of being accepted.

%%\input{ds3-acf.tex}
%% Garfield Brown <gob20@cam.ac.uk>
%% Last change: Time-stamp: <06-Jun-2004 18:22:50 gob20>
%%\endinput

%%\input{ds3-alpha.tex}
%%\endinput

%%\input{ds3-nacf.tex}
%%\endinput

%%\input{ds3-nalpha.tex}
%%%\endinput

These results suggest that fitting models with $\rho=0$ or $\rho=1$ should
give better description of the data. i.e.  $\rho=0$ and $\rho=1$ might be
plausible alternative models. In the remainder of this paper we examine in
more detail the simplified models with $\rho=0$ and $\rho=1$, and how to
discriminate between them. The model with $\rho=0$ means that a simple
variance components model will be enough to describe the data whereas the
model with $\rho=1$ means a simple autoregressive model with Normal errors
will be adequate to describe the data.

\section{The Reversible Jump Algorithm}\label{sec:reversible}
In this section we discuss trans-dimensional algorithms. The algorithms we
have discussed before, notably the Metropolis--Hastings algorithm and the
Gibbs sampling algorithm, cannot be used to simulate Markov chains where the
dimension of the state vector can change at each iteration. This situation
arises particularly in model selection problems where there are competing
models, and where the size of the parameter vector is allowed to vary between
models.

In this context the distribution of interest is defined jointly over both
parameter and model space. Several authors have proposed simulation methods
to construct Markov chains which can explore such state spaces. These include
the product space formulation given in \citet{carlin1995}, the reversible
jump (RJMCMC) algorithm of \citet{green1995}, the jump diffusion method of
~\citet{grenander1994}, and \citet{phillips1996:mcmcip} and the continuous
time birth-death method of \citet{stephens2000}. Also for particular problems
involving the size of the regression vector in regression analysis there is
the stochastic search variable selection method of \citet{george1993}. In the
remainder of this section we describe the reversible jump method of
\citet{green1995}. In practice trans--dimensional algorithms work by updating
model parameters for the current model then proposing to change models with
some specified probability.

The Reversible jump algorithm represents an extension of the
Metropolis--Hastings algorithm.  We assume there is a countable collection of
candidate models, indexed by $M\in\mathcal{M}=\{M_1$, $M_2$,$\ldots$ ,
$M_k\}$. We further assume that for each model $M_i$, there exists an unknown
parameter vector $\bftheta_i \in \mathbb{R}^{n_i}$ where $n_i$, the dimension
of the parameter vector, can vary with $i$.

Typically we are interested in finding which models have the greatest
posterior probabilities and also estimates of the parameters. Thus the
unknowns in this modelling scenario will include the model index $M_i$ as
well as the parameter vector $\bftheta_i $. We assume that the models and
corresponding parameter vectors have a joint density $\pi(M_i,\bftheta_i )$.
The reversible jump algorithm constructs a reversible Markov chain on the
state space $\mathcal{M} \times \bigcup_{M_i\in\mathcal{M}} \mathbb{R}^{n_i}$
which has $\pi$ as its stationary distribution \citep{green1995}.  In many
instances, and in particular for Bayesian problems this joint distribution
is of the form
\begin{equation*}
  \pi(M_i, \bftheta_i) = \pi(M_i, \bftheta_i\vert
  \bfX)\propto
  \bfL(\bfX\vert M_i, \bftheta_i) \; 
  p(M_i, \bftheta_i ) ,
\end{equation*}
where the prior on $(M_i,\bftheta_i)$ is often of the form
\begin{equation*}
  p(M_i, \bftheta_i) = p(\bftheta_i\vert M_i) \;p(M_i)
\end{equation*}
with $p(M_i)$ being the density of some counting distribution.

Suppose now that we are at model $M_i$ and a move to model $M_j$ is proposed
with probability $r_{ij}$. The corresponding move from
$\bftheta_i$ to $\bftheta_j$ is achieved by using a
deterministic transformation $h_{ij}$, such that
\begin{equation}\label{eq:revjumpmapping}
  (\bftheta_j , \bfv) = h_{ij}(\bftheta_i, \bfu ),
\end{equation}
where $\bfu$ and $\bfv$ are random variables introduced to ensure dimension
matching necessary for reversibility. To ensure dimension matching we must
have
\begin{equation*}%%\label{eq:dimmatch}
  \dim(\bftheta_j)+\dim(\bfv)=\dim(\bftheta_i)+\dim(\bfu). 
\end{equation*}
For discussions about possible choices for the function $h_{ij}$ we refer the
reader to \citet{green1995}, and \citet{brooks2003}.  Let %%If we denote the
%%ratio
\begin{equation}\label{eq:acceptratio}
  A(\bftheta_i, \bftheta_j ) = 
  \frac{\pi(M_j, \bftheta_j)}{\pi(M_i, \bftheta_i )}  
  \frac{q(\bfv)}{q(\bfu)} 
  \frac{r_{ji}}{r_{ij}}\hskip .15cm
  \biggl\lvert\frac{\partial h_{ij}(\bftheta_i, \bfu)}
  {\partial (\bftheta_i,\bfu)} \biggr\rvert
\end{equation}
then the acceptance probability for a proposed move from model $(M_i,
\bftheta_i)$ to model $(M_j, \bftheta_j)$ is
\begin{equation*}
  \min\left\{1, A(\bftheta_i, \bftheta_j )  \right\}
\end{equation*} 
where $q(\bfu)$ and $q(\bfv)$ are the respective proposal densities for
$\bfu$ and $\bfv$, and $\lvert \partial h_{ij}(\bftheta_i,\bfu) / \partial
(\bftheta_i,\bfu) \rvert$ is the Jacobian of the transformation induced by
$h_{ij}$. \citet{green1995} shows that the algorithm with acceptance
probability given above simulates a Markov chain which is reversible and
follows from the detailed balance equation
\begin{equation*}
  \pi(M_i, \bftheta_i) q(\bfu) r_{ij} = 
  \pi(M_j, \bftheta_j) q(\bfv) r_{ji} \hskip .1cm \biggl\lvert
  \frac{\partial h_{ij}(\bftheta_i,\bfu )} 
  {\partial (\bftheta_i, \bfu)}
  \biggr\rvert .
\end{equation*}
Detailed balance is necessary to ensure reversibility and is a sufficient
condition for the existence of a unique stationary distribution.
%%If the jump is accepted then the updating equations for model $j$ are used to
%%update the parameters. Otherwise the algorithm stays in the current model.
For the reverse move from model $M_j$ to model $M_i$ it is easy to see that
the transformation used is $(\bftheta_i, \bfu) =
h_{ij}^{-1}(\bftheta_j, \bfv)$ and the acceptance probability for
such a move is
\begin{equation*}
  \min\left\{1, 
    \frac{\pi(M_i, \bftheta_i)}{\pi(M_j, \bftheta_j)}  
    \frac{q(\bfu)}{q(\bfv)} 
    \frac{r_{ij}}{r_{ji}}\hskip .15cm
    \biggl\lvert\frac{\partial h_{ij}(\bftheta_i,\bfu)}
    {\partial (\bftheta_i,\bfu)} \biggr\rvert^{-1} \right\} 
  =\min\left\{1, A(\bftheta_i, \bftheta_j)^{-1} \right\}.
\end{equation*} 
For inference regarding which model has the greater posterior probability we
can base our analysis on a realisation of the Markov chain constructed above.
The marginal posterior probability of model $M_i$ 
\begin{equation*}
  \pi(M_i\vert \bfX) = \frac{p(M_i) f(\bfX\vert M_i)}
  {\sum_{M_j\in\mathcal{M}} p(M_j) f(\bfX\vert M_j) },
\end{equation*}
where 
\begin{equation*}
  f(\bfX\vert M_i) =\int \bfL(\bfX\vert 
  M_i,\bftheta_i)
  p(\bftheta_i|M_i)\, d\,\bftheta_i
\end{equation*} 
is the marginal density of the data after integrating over the unknown
parameters $\bftheta$. In practice we estimate $\pi(M_i| \bfX)$ by counting
the number of times the Markov chain visits model $M_i$ in a single long run
after reaching stationarity.  These between model moves described in this
section are also augmented with within model Gibbs updates as given in
Section~\ref{sec:postcond} to update model parameters.

\subsection{Efficient Proposals} \label{sec:efficientprop}
In practice the between model moves can be small resulting in poor mixing of
the resulting Markov chain. In this section we discuss recent attempts at
improving between model moves by increasing the acceptance probabilities for
such moves. Several authors have addressed this problem including
\citet{troughton1997}, \citet{giudici1998}, \citet{godsill2001},
\citet{rotondi2002}, and \citet{alawadhi2004}. \citet{green2001:mira}
proposes an algorithm so that when between model moves are first rejected, a
second attempt is made. This algorithm allows for a different proposal to
generated from a new distribution, that is allowed to depend on the
previously rejected proposal. Methods to improve mixing of reversible jump
chains have also been proposed by \citet{green2002} and
\citet{brooks2003}, which has been extended by \citet{ehlers2002}.

A general  strategy proposed by \citet{brooks2003} and extended to more general
cases by \citet{ehlers2002} is based on making the term
$A_{ij}(\bftheta_i,\bftheta_j)$ in the acceptance probability for between
model moves given in Equation~\eqref{eq:acceptratio} as close as possible to
1. The motivating reason for this is that if we make this term as close as
possible to 1 the the reverse move acceptance governed by $1 /
A_{ij}(\bftheta_i,\bftheta_j)$ will also be maximised resulting in easier
between model moves. In general, if the move from $(M_i,
\bftheta_i)\Rightarrow(M_j,\bftheta_j)$ involves a change in dimension, the
best values of the parameters for the densities $q(\bfu)$ and $q(\bfv)$ in
Equation~\eqref{eq:acceptratio} will generally be unknown, even if their
structural forms are known. Using some known point $(\widetilde \bfu,
\widetilde \bfv)$, which we call the centering point, we can solve
$A_{ij}(\bftheta_i,\bftheta_j)=1$ to get the parameter values for these
densities. Setting $A_{ij}=1$ at some chosen centering point is called the
zeroth-order method. Where more degrees of freedom are required we can expand
$A_{ij}$ as a Taylor series about $(\widetilde\bfu, \widetilde \bfv)$ and
solve for the proposal parameters. For the methods we use in this paper the
new parameters are proposed so that the mapping function in
Equation~\eqref{eq:revjumpmapping} is the identity function, i.e.,
\begin{equation*}
  (\bftheta_j, \bfv) = h_{ij}(\bftheta_i, \bfu) = (\bfu, \bftheta_i)
\end{equation*}
and the acceptance ratio term $A_{ij}(\bftheta_i, \bftheta_j)$ probability in
Equation~\eqref{eq:acceptratio} becomes
\begin{align*}
  A_{ij}(\bftheta_i, \bftheta_j) &=
  \frac{\pi(M_j, \bftheta_j)}{\pi(M_i, \bftheta_i)} 
  \frac{r_{ji}}{r_{ij}} 
  \frac{q (\bfv)}{ q(\bfu)} \\
  &=
  \frac{\pi(M_j, \bftheta_j)}{\pi(M_i, \bftheta_i)} 
  \frac{r_{ji}}{r_{ij}} 
  \frac{q (\bftheta_i)}{ q(\bftheta_j)} .
\end{align*}

\subsection{Convergence Assessment}\label{sec:conv}
% In this section we discuss various methods for assessing the convergence, to
% the stationary distribution, of the fixed dimensional Markov chains generated
% by the simulation algorithms of Section~\ref{sec:computation}, and the
% trans--dimensional algorithms of Section~\ref{sec:reversible}.

Convergence assessment for trans-dimensional algorithms are still in their
infancy. ~\citet{brooks1999:giudici} propose to run $I\geq 2$ chains in
parallel and base their convergence diagnostic on splitting the total
variation not just between chains but also between models. Their method was
extended by ~\citet{brooks2003:jcgs} to include non-parametric techniques,
including chi-square tests, Kolmogorov--Smirnov tests and direct convergence
rate estimation. The latter being similar to the ideas of
~\citet{raftery1992:lewis} for the fixed dimensional Metropolis--Hastings or
Gibbs algorithms. \citet{castelloe2002} also develop methods based on the
ideas of ~\citet{brooks2003:jcgs} which can be used only where the parameters
have the same interpretation across all models.

\citet{brooks2003:jcgs} suggest several methods for assessing
convergence within the context of model selection problems. In particular for
reversible jump algorithms we can have some idea of how fast the simulations
approach stationarity by comparing the empirical stationary distribution on
the observed model orders. They propose specific test statistics based on the
$\chi$--square distribution and also a Kolmogorov--Smirnov test for goodness
of fit.  The $\chi$--square and Kolmogorov--Smirnov compare the stationary
distribution of each chain and computes $p$--values for the computed test
statistics. A critical value of $5\%$ is used so that if the $\chi$--square
or Kolmogorov--Smirnov statistic is above this significance level there is no
reason to reject the chains as not being from the same stationary
distribution. See \citet{brooks2003:jcgs} for further details.

\section{Model Selection Using Reversible Jump Algorithms}
\label{sec:ds3:revjump}
In this section we introduce two additional models and describe a reversible
jump model selection technique to discriminate between them. Denote the full
model in Equations~\eqref{eq:model} and \eqref{eq:hyper} with $M_1$, we
introduce two additional models, which are sub-models of $M_1$. The second
model, $M_2$, has $\rho$ fixed at 1. For this model there is no $\eta$ and
the first two levels are
\begin{align*}
  R_j\vert\alpha'_j,\sigma &\sim \dnorm{\alpha'_j}{(\sigma' E_j)^{-1}} \\
  \alpha'_j\vert\alpha'_{j-1},\tau' &\sim \dnorm{\alpha'_{j-1}}{{\tau'}^{-1}}.
\end{align*}
The prior distribution on $\alpha_0$, $\sigma$ and $\tau$ remain as in $M_1$.
The posterior conditionals are exactly the same as in
Section~\ref{sec:postcond}, simplified with $\rho=1$ where necessary. The
third model, $M_3$, has $\rho$ fixed as well, however this time at 0, which
results in a simple random effects model:
\begin{align*}
  R_j\vert\alpha''_j,\sigma'' &\sim \dnorm{\alpha''_j}{(\sigma'' E_j)^{-1}}\\
  \alpha''_j\vert\eta'',\tau''&\sim \dnorm{\eta''}{{\tau''}^{-1}} .
\end{align*}
The prior distributions on $\eta$, $\sigma$ and $\tau$ remain as in $M_1$.
Again the posterior conditionals are as those in Section~\ref{sec:postcond}
with $\rho=0$ where necessary.

The computation here is a simple extension to the Bayesian posterior
distribution described in Section~\ref{sec:postcond} above. Here we have
model space $\mathcal{M}$=$\{M_1$, $M_2$, $M_3\}$ with three models, where
$M_1$ is the original model described in Equations~\eqref{eq:model} and
\eqref{eq:hyper}.  Models $M_2$ and $M_3$ correspond to the simplified
models with $\rho$ fixed at 1 and 0, respectively. We can extend our
posterior distribution to consider both parameter and model space by taking
as our posterior for model $M_1$
\begin{equation*}
  \pi(M_1, \bfalpha, \tau, \sigma, \rho, \eta | R^n) \propto
  \bfL(R^n | \bfalpha, \sigma) p(\bfalpha | \rho,\eta,\tau,\alpha_0)
  p(\sigma)p(\tau)p(\rho)p(\eta)p(\alpha_0) p(M_1).
\end{equation*}
For the simplified models $M_2$ and $M_3$ the posteriors defined up to the
constant of proportionality are
\begin{equation*}
  \pi(M_2, \bfalpha', \tau', \sigma' | R^n) \propto
  \bfL(R^n | \bfalpha', \sigma') p(\bfalpha' |\tau',\alpha'_0)
  p(\sigma')p(\tau')p(\alpha'_0) p(M_2),
\end{equation*}
and
\begin{equation*}
  \pi(M_3, \bfalpha'', \tau'', \sigma'', \eta'' | R^n) \propto
  \bfL(R^n | \bfalpha'', \sigma'') p(\bfalpha'' | \eta'',\tau'')
  p(\sigma'')p(\tau'')p(\eta'')p(M_3),
\end{equation*}
respectively, where $p(M_i)$ is some discrete prior distribution on the model
space $\mathcal{M}$.  Posterior model probabilities may then be obtained by
marginalisation i.e., integrating out $\bfalpha$, $\alpha_0$, $\rho$, $\eta$,
$\sigma$ and $\tau$ to obtain the posterior marginal for $M_i$ given the
data.
%%The algorithm works as described in Section~\ref{sec:reversible}.
%% These probabilities can then be used either to discriminate between
%% competing models or to model-average predictive inference for example.
%% Essentially the reversible jump method works by proposing a new model $M_j$
%% with probability $r_{ij}$ when at model $M_i$ and between model moves are
%% supplemented by within model moves which update the parameters for the given
%% model.
For the implementation we start with each model having equal prior
probability
\begin{equation*}
  p(M_1) = p(M_2) = p(M_3)  =\tfrac1{3},
\end{equation*}
and $r_{ij}$ the probability of proposing a move to model $M_j$ when at model
$M_i$ taken to be $\tfrac1{2}$ for $i,j = 1,2,3$ and $i\neq j$.

In the discussion that follows for ease of notation we suppress the
dependence of the densities on the parameters $\bfalpha=(\alpha_1,
\ldots,\alpha_n)$, $\sigma$, and $\tau$ since these parameters are common to
all models.  In addition for our reversible jump moves these common
parameters are kept fixed between models.
\subsection{Pilot Tuned Methods}
Consider a proposed move from $(M_2, \alpha'_0)$ to $(M_1,
\alpha_0,\rho,\eta) $, we need to increase the dimensionality of the
parameter vector by adding three components $\alpha_0$, $\rho$ and $\eta$ and
removing $\alpha'_0$.  To achieve this we simulate $u_1$, $u_2$ and $u_3$
from densities $q(u_1)$, $q(u_2)$ and $q(u_3)$, respectively, and set
\begin{equation*}
  (\alpha_0, \rho, \eta, v)
  = h_{21}(\alpha'_0, u_1, u_2, u_3)
  = (u_1, u_2, u_3, \alpha'_0),
\end{equation*}
where the variable $v$ is needed to ensure dimension matching and
reversibility. We further assume $v$ has some density $q(v)$, which we use to
simulate values of $v$ for the reverse move from $M_1$ to $M_2$.
The acceptance probability for such a move is then $\min\{1, A_{21}\}$ where
\begin{align}\label{eq:efficient21}
  A_{21} & = 
  \frac{\pi(M_1,\alpha_0,\rho,\eta)}
  {\pi(M_2, \alpha'_0)} \times
  \frac{q(v)}{q(u_1)q(u_2)q(u_3)} \times
  \Biggl\lvert
  \frac{\partial h_{21}(\alpha'_0, u_1, u_2, u_3 )}
  {\partial(\alpha'_0, u_1, u_2, u_3)}\Biggr\rvert \nonumber\\
  & =  \frac{\pi(M_1,\alpha_0,\rho,\eta)}
  {\pi(M_2, \alpha'_0)} \times
  \frac{q(v)}{q(u_1)q(u_2)q(u_3)} \nonumber \\
  & =  \frac{\pi(M_1,\alpha_0,\rho,\eta)}
  {\pi(M_2, \alpha'_0)} \times
  \frac{q( \alpha'_0)}{q(\alpha_0)q(\rho)q(\eta)},
\end{align}
since the Jacobian term
%%\begin{equation*}
$
  \Biggl\lvert
  \frac{\partial h_{21}(\alpha'_0, u_1, u_2, u_3 )}
  {\partial(\alpha'_0, u_1, u_2, u_3)}\Biggr\rvert $
%%\end{equation*}
evaluates to 1. 

The densities $q(\alpha_0)$, $q(\rho)$, $q(\eta)$ and $q(\alpha'_0)$ are all
assumed to be Gaussian densities, with respective parameters $(m_1,
\sigma_1)$, $(m_2, \sigma_2)$, $(m_3, \sigma_3)$ and $(m_v,
\sigma_v)$. Theoretically, we can choose arbitrary values for the location
parameters $m_1$, $m_2$, $m_3$ and $m_v$ and for the scale parameters
$\sigma_1$, $\sigma_2$, $\sigma_3$ and $\sigma_v$. However, some choices will
result in an algorithm which takes longer to reach stationarity, since poor
choices will result in low acceptance rates for between model moves. We
fine-tuned the between model transitions by trying several different choices
for these quantities and all resulted in the same posterior model
probabilities. %%
%%, however some values took longer and resulted in very low acceptance rates.  
Generally, picking $m_1$ and $\sigma_1$ close to the
posterior marginal mean and variance for $\alpha_0$; $m_2$ and $\sigma_2$
close to the marginal posterior mean and variance of $\rho$; $m_3$ and
$\sigma_3$ close to the marginal mean and variance of $\eta$ ; $m_v$ and
$\sigma_v$ close to the posterior marginal mean and variance of $\alpha'_0$
results in an algorithm where between model jumps are easier. We determined
these posterior values by running each model in turn and recording posterior
estimates of the mean and variance of the model parameters. These estimates
are then used as proposal parameters in the reversible jump
implementation. This scheme can only be used when there are a small number of
candidate models as it becomes infeasible when the number of candidate models
is large. In Section ~\ref{sec:efficient} we propose to use an automatic
sampler which can choose location and scale parameters to maximise between
model transitions based on methods presented in \citet{brooks2003}.

The reverse move from $(M_1, \alpha_0,\rho,\eta) $ to $(M_2, \alpha'_0)$ is
achieved by simulating $v$ from density $q(v)$ then setting
\begin{equation*}
  (\alpha'_0, u_1, u_2, u_3)
  = h_{21}^{-1}(\alpha_0, \rho,\eta, v)
  = (v, \alpha_0, \rho, \eta)
\end{equation*}
for which the acceptance probability of accepting this dimension changing
move is then $\min\{ 1, A_{21}^{-1}\}$ where $A_{21}$ is given in
Equation~\eqref{eq:efficient21}.

The description is similar for moves between models $M_1$ and $M_3$. Assume
we are at model $(M_3, \eta'')$ and a move to model $M_1$ is proposed. We
simulate $u_1$, $u_2$, $u_3$ from densities $q(u_1)$, $q(u_2)$ and $q(u_3)$
respectively and set
\begin{equation*}
  (\alpha_0, \rho, \eta, w)
  = h_{31}(\eta'', u_1, u_2, u_3)
  =(u_1, u_2, u_3, \eta''),
\end{equation*}
where $w$ is introduced to ensure dimension matching.

The probability of accepting this move is then $\min\{1, A_{31}\}$ where
\begin{equation}\label{eq:efficient13}
  A_{31} = \frac{\pi(M_1, \alpha_0, \rho, \eta) }{\pi(M_3, \eta'')} \times
  \frac{q(w) }{q(u_1) q(u_2) q(u_3)} \times
  \Biggl\lvert
  \frac{\partial h_{31}(\eta'',u_1,u_2,u_3 )}{\partial(\eta'',u_1,u_2,u_3 )}
  \Biggr\rvert.
\end{equation}
For reasons similar to those given above $q(u_1)$, $q(u_2)$, $q(u_3)$ and
$q(w)$ are densities approximating the posterior marginals of $\alpha_0$,
$\rho$, $\eta$ and $\eta''$, respectively. The reverse move from $(M_1,
\alpha_0, \rho, \eta)\Rightarrow (M_3, \eta'')$ is achieved by simulating $w$
with density $q(w)$ and setting
\begin{equation*}
  (\eta'', u_1, u_2, u_3) = h_{31}^{-1}(\alpha_0, \rho, \eta, w)
  =(w, \alpha_0, \rho,\eta)
\end{equation*}
for which the acceptance probability is the $\min\{1, A_{31}^{-1}\}$ where
$A_{31}$ is given in Equation~\ref{eq:efficient13}.

For a proposed move from $(M_2,\alpha'_0)$ to $(M_3,\eta'')$, we simulate $w$
with density $q(w)$ and set 
$(\eta'',v )= h_{23}(\alpha'_0, w)=(w, \alpha'_0,)  $. 
For such a proposal the acceptance probability is $\min\{1, A_{23} \}$
where
\begin{equation*}
  A_{23} =
  \frac{\pi(M_3,\eta'')}{\pi(M_2,\alpha'_0)} \times
    \frac{q(v)}{q(w)} \times
    \Biggl|
    \frac{\partial h_{32}(\alpha'_0, w)}{\partial(\alpha'_0, w)}
    \Biggr|
\end{equation*}
where $q(w)$ and $q(v)$ are the densities discussed above. Again the Jacobian
for this proposed move is $1$ since the transformation from $(M_2,\alpha'_0)$
to $(M_3,\eta'')$ is the identity function.  Notice that with this move we
are not changing the number of parameters, but swapping $\alpha'_0$ for
$\eta''$. The acceptance probability for the reverse move is then $\min\{1,
A_{23}^{-1} \}$.

\begin{table}
  \caption{\label{tab:ds3results}Parameter Estimates and 95\% HPD Intervals.
    The corresponding results for the full model are given in 
    Table~\ref{tab:ds3Fullresults}.}
  \centering
  \fbox{%
    \begin{tabular}{rrrrr}
      \hline\hline
      & $M_2$& 95\% HPD Interval & $M_3$& 95\% HPD Interval\\
      \hline
      $\alpha_0$ &0.0252&(-0.0586, 0.1092) & -   &   \\
      $\alpha_1$ &0.0253&(-0.0185, 0.0700) &0.0275&(-0.0145, 0.0697)\\
      $\alpha_2$ &0.0253&(-0.0129, 0.0648) &0.0244&(-0.0170, 0.0666)  \\
      $\alpha_3$ &0.0368&(-0.0015, 0.0744) &0.0403&(-0.0024, 0.0818)  \\
      $\alpha_4$ &0.0292&(-0.0073, 0.0664) &0.0261&(-0.0143, 0.0670)  \\
      $\alpha_5$ &0.0358&(-0.0004, 0.0720) &0.0359&(-0.0048, 0.0754)  \\
      $\alpha_6$ &0.0362&(-0.0003, 0.0726) &0.0361&(-0.0032, 0.0747)  \\
      $\alpha_7$ &0.0304&(-0.0127, 0.0740) &0.0288&(-0.0127, 0.0706)  \\
      $\eta$     & -  &   &0.0313&(-0.0014, 0.0636)  \\
      $\sigma$   &1145.7&(0.18, 2884.4) &1115.2&(1.46, 2695.4)  \\
      $\tau$     &1460.8&(35.6, 3359.3) &1617.1&(53.7, 3783.7)  \\
    \end{tabular}}
\end{table}

\begin{figure}
  \centering
  \makebox{
  \includegraphics[width=0.8\textwidth]{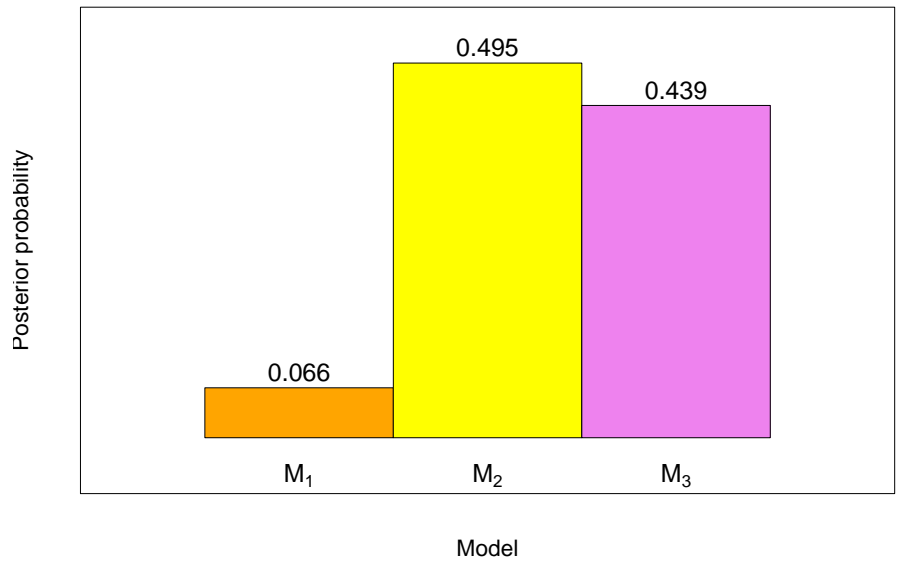}}
  \caption{\label{fig:ds3rj}Posterior model probabilities for 
    models $M_1$, $M_2$, and $M_3$.}
\end{figure}

\begin{figure}
  \centering
  \makebox{
  \includegraphics[width=0.8\textwidth]{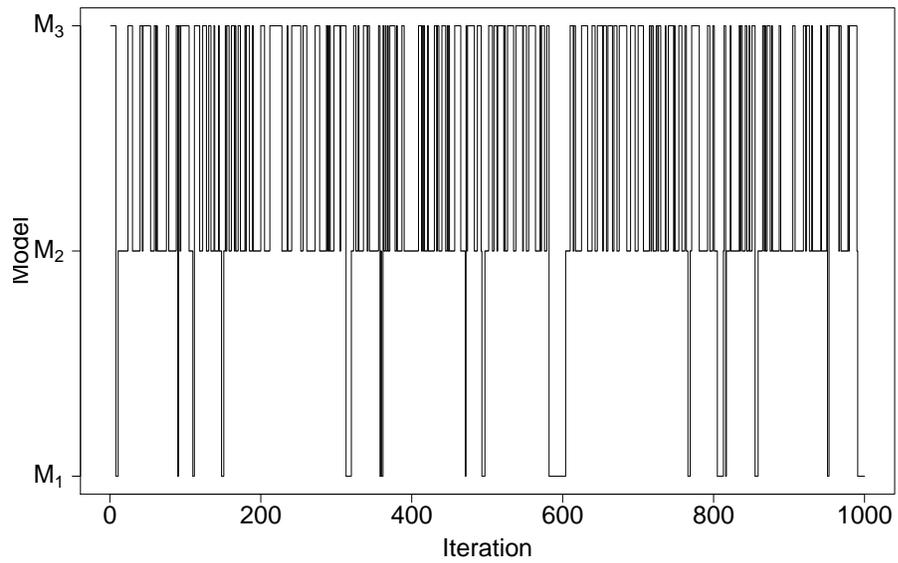}}
  \caption{\label{fig:ds3-model}Trace plot of the model indicator.}
\end{figure}

\subsubsection{Simulation Study}
To test how well the model discrimination scheme works we simulated
several datasets and applied the algorithm to them. In all cases
where data were simulated from model $M_2$ the algorithm placed the
largest posterior probability on that model, like with data
simulated from model $M_3$ the algorithm placed the highest
posterior probability on that model. For data simulated from model
$M_1$ in some instances the highest posterior probability is placed
on either model $M_2$ or model $M_3$. As the value of $n$ increases,
it appears as though the algorithm will place most of the posterior
probabilities on either $M_2$ or $M_3$ since for large values of $n$
the values of $R_j$ simulated approach $\eta$ asymptotically, hence
the smaller models $M_2$ and $M_3$ offer a better fit to the data.

\subsubsection{Model Averaged Results}
The posterior parameter estimates with 95\% HPD intervals for each of the
three models are given in Table~\ref{tab:ds3results}. The posterior model
probabilities are shown in Figure~\ref{fig:ds3rj}, this shows that model
$M_2$ has the greatest posterior probability of $0.495$, followed by $M_3$
with probability $0.439$.  The full model $M_1$ has the least posterior
probability, $0.066$. The posterior model probabilities of $M_2$ and $M_3$
seem to contradict the results if we consider the posterior distribution of
$\rho$.  Figure~\ref{fig:ds3-rho-dens_M1} shows that the posterior density of
$\rho$ clearly has most mass around the node $\rho=0$, so we might expect
model $M_3$ to have the greater posterior probability.

% For each model we computed the $DIC$. These values are shown in
% Table~\ref{tab:ds3-dic} along with the $p_D$ value for each model. The value
% for model $M_1$ is $-27.14$, while for models $M_2$ and $M_3$ the computed
% values are $-29.45$ and $-28.75$ respectively. This adds support to $M_2$
% being the most likely model with $M_3$ being likely also, and $M_1$ not being
% a very good model for describing the data.
% \begin{table}
%   \caption{\label{tab:ds3-dic}Complexity Measures.}
%   \centering
%   \fbox{
%     \begin{tabular}{rrrrr}
%       \hline
%       Model& $\overline{D(\bftheta)}$& $D(\bar{\bftheta})$& $p_D$& $DIC$\\
%       \hline
%       $M_1$ & -34.27 & -41.41 & 7.13 & -27.14 \\
%       $M_2$ & -35.71 & -41.97 & 6.26 & -29.45 \\
%       $M_3$ & -35.42 & -42.10 & 6.67 & -28.75 \\
%     \end{tabular}}
% \end{table}

It is interesting to note that many of the parameter estimates are similar
under all three models. In particular the error variances seem to take very
similar values under all three models. Thus model-averaged estimates look
very similar from those derived from just a single model for this example.
Note also the posterior distribution for $\rho$ in the full model has a
posterior mean of $0.220$. This might naively be interpreted as suggesting
that the ratio of model probabilities between Model $M_2$ and $M_3$ should be
roughly $1:4$ rather than the $1:1$ ratio observed. The posterior density of
rho is shown in Figure~\ref{fig:posteriordensrho}.

Figure~\ref{fig:ds3-model} shows the mixing of the deterministic proposal
reversible jump algorithm. It is noticeable that even though models $M_2$ and
$M_3$ have approximately equal posterior probabilities the algorithm does not
mix very well. In the next section we set try to improve the mixing of the
reversible jump algorithm.

\subsection{Automatic Proposal Choices}\label{sec:efficient}
The choice of proposal densities in the reversible jump MCMC implementations
given in Section~\ref{sec:ds3:revjump} are determined by doing a pilot run to
determine good parameter choices to describe these densities. In this section
we show how this process can be made more automatic by proposing an adaptive
scheme where the proposals are chosen as to maximise the probability of
between model moves.  Automatic proposals are desirable for a number of
reasons, mainly because they reduce the need to do trial runs in order to get
parameter estimates for proposal densities. The method we use is based on
\citet{brooks2003} and uses the idea of so-called weak non-identifiability
and centering to determine the choice of proposal densities which maximises
the probability of between model moves. The weak non-identifiability
centering point is a choice of parameter values which essentially reduces the
more complex model to the simpler model. We refer to this new implementation
as the efficient proposals method and the previous implementation in
Section~\ref{sec:ds3:revjump} as the vanilla implementation. In the remainder
of this section we show the details of how the between model moves are
implemented. 
\subsubsection{Moving between Models $M_1$ and $M_2$}
Consider a move from model $M_2$ to model $M_1$. The acceptance probability
for such a move is $\min\{1, A_{21}\}$, where
\begin{equation}\label{eq:accept21}
  A_{21} =\frac{\pi(M_1,\alpha_0,\rho,\eta)}{\pi(M_2,\alpha'_0)}
  \frac{q(\alpha'_0)}{ q(\alpha_0,\rho,\eta)}.
\end{equation}
An ideal choice for $q(\alpha_0,\rho,\eta)$ would be
$\pi(\alpha_0,\rho,\eta\vert M_1)$, the conditional posterior for $(\alpha_0,
\rho, \eta)$ given $M=M_1$. This density is non-standard, furthermore we
would also need to know its normalising constant to compute the ratio
$A_{21}$.  We cannot sample directly from this density, but instead we
approximate $q(\alpha_0,\rho,\eta)$ with a trivariate normal density. We
approximate $q(\alpha'_0)$ using a Gaussian density whose parameters we
derive below.  Similar methods have been proposed
\citep{carlin1995,madigan1995}.

The best approximating density for $q(\alpha_0,\rho,\eta)$ in our case is one
that will maximise $A_{21}$. To do this we use the $k^{th}$-order method of
\citet{brooks2003} and expand $A_{21}$ as a Taylor series around some
point $(\tilde\alpha_0, \tilde\rho,\tilde\eta)$ which they call the centering
point.  Since we need only to estimate the mean and variance of this
trivariate normal density, partial derivatives of order 1 and 2 will suffice.
Essentially this means solving
\begin{equation*}
  \frac{\partial^k}{\partial(\alpha_0,\rho,\eta)^k}
  A_{21}\Bigl\vert_{(\tilde\alpha_0,\tilde\rho,\tilde\eta)} = 0,\quad k=1,2\,
\end{equation*}
for the mean vector and covariance matrix for the density
$q(\alpha_0,\rho,\eta)$, where $(\tilde\alpha_0,\tilde\rho,\tilde\eta)$ is
our chosen centering point. However it is usually much easier to do
computations with the $\log$ of $A_{21}$, in which case we solve
\begin{equation}\label{eq:kthorder}
  \frac{\partial^k}{\partial(\alpha_0,\rho,\eta)^k} \log
  A_{21}\Bigl\vert_{(\tilde\alpha_0,\tilde\rho,\tilde\eta)} =0,\quad k=1,2.
\end{equation}
With $A_{21}$ as given in \eqref{eq:accept21} it is not very difficult to see
that when we take derivatives of $A_{21}$ (or $\log A_{21}$) with respect to
$(\alpha_0,\rho,\eta)$ the terms involving $\pi(M_2,\alpha'_0)$ and
$q(\alpha'_0)$ will contribute nothing to that derivative and likewise when
we take the derivative of $A_{21}$ or ($\log A_{21}$) with respect to
$\alpha'_0$ the terms $\pi(M_1,\alpha_0,\rho,\eta)$ and
$q(\alpha_0,\rho,\eta)$ will contribute nothing to that derivative. In what
follows we will ignore terms where appropriate. Thus we can compute the
first and second partial derivatives of $\log A_{21}$ as
\begin{equation}\label{eq:firstpartial}
  \frac{\partial\log A_{21}}{\partial(\alpha_0,\rho,\eta)}=
  \frac{\partial}{\partial(\alpha_0,\rho,\eta)}
  \Bigl(\log \pi(M_1,\alpha_0,\rho,\eta)- \log q(\alpha_0,\rho,\eta)
  + K_{2}\Bigr)
\end{equation}
and
\begin{equation}\label{eq:secondpartial}
  \frac{\partial^2\log A_{21}}{\partial(\alpha_0,\rho,\eta)^2}=
  \frac{\partial^2}{\partial(\alpha_0,\rho,\eta)^2}
  \Bigl(\log \pi(M_1,\alpha_0,\rho,\eta)- \log q(\alpha_0,\rho,\eta)
  +K_{2}\Bigr),
\end{equation}
where the term $K_{2}= - \log\pi(M_2,\alpha'_0) + \log q(\alpha'_0)$ is
independent of $(\alpha_0,\rho,\eta)$. Also we can expand
the posterior density of $(M_1,\alpha_0,\rho,\eta)$
\begin{equation*}
  \pi(M_1,\alpha_0,\rho,\eta)\propto
  \bfL(R^n\vert\bfalpha,\sigma)
  p(\bfalpha|\rho,\eta,\alpha_0,\tau)
  p(\eta)p(\rho)p(\alpha_0)p(\sigma)p(\tau)p(M_1)
\end{equation*}
and the proposal density for $(\alpha_0,\rho,\eta)$
\begin{equation*}
  q(\alpha_0,\rho,\eta) \propto |\boldsymbol{\varSigma}|^{-3/2}
  \exp\left\lbrace -\tfrac1{2}
    \left(\left(\begin{smallmatrix}\alpha_0\\ \rho \\ \eta
        \end{smallmatrix}\right)-\bfmu\right)'
    \boldsymbol{\varSigma}^{-1}
    \left(\left(\begin{smallmatrix}\alpha_0\\ \rho \\ \eta
        \end{smallmatrix}\right)-\bfmu\right)
  \right\rbrace ,
\end{equation*}
the density of a trivariate normal distribution with mean vector $\bfmu$ and
covariance matrix $\boldsymbol{\varSigma}$.  Setting \eqref{eq:firstpartial}
and \eqref{eq:secondpartial} equal to zero at the point
$(\tilde\alpha_0,\tilde\rho,\tilde\eta)$ we get two equations which can be
solved simultaneously for the variance matrix $\boldsymbol{\varSigma}$ and
mean vector $\bfmu$. Solving simultaneously we can easily see that the
variance matrix $\boldsymbol{\varSigma}$ is
\begin{multline}\label{eq:mvSigma}
  \boldsymbol{\varSigma}^{-1}=\\
  \left(\begin{smallmatrix}
      1 + \tau\tilde\rho^2
      & -\tau(\alpha_1-\tilde\eta+2\tilde\rho(\tilde\eta-\tilde\alpha_0))
      & -\tau\tilde\rho(1-\tilde\rho) \\
      -\tau(\alpha_1-\tilde\eta+2\tilde\rho(\tilde\eta-\tilde\alpha_0))
      & 1+\tau\sum_{j=1}^n(\tilde\eta-\alpha_{j-1})^2
      &-\tau\sum_{j=1}^n[(1-2\tilde\rho)(\tilde\eta-\alpha_{j-1})+\tilde\eta-\alpha_j ]\\
      -\tau\tilde\rho(1-\tilde\rho)
      & -\tau\sum_{j=1}^n[(1-2\tilde\rho)(\tilde\eta-\alpha_{j-1})+\tilde\eta-\alpha_j]
      & 1 + n\tau(1-\tilde\rho)^2 \\
    \end{smallmatrix}\right)
\end{multline}
and the mean vector $\bfmu$ satisfies
\begin{equation*}
  \boldsymbol{\varSigma}^{-1}\left( \left( \begin{smallmatrix}
        \tilde\alpha_0 \\ \tilde\rho\\ \tilde\eta \end{smallmatrix}\right) -
    \bfmu\right)=
  \left(
    \begin{smallmatrix}
      \tilde\alpha_0 - \tilde\rho\tau(\alpha_1 -\tilde\rho\tilde\alpha_0-(1-\tilde\rho)\tilde\eta )\\
      \tilde\rho+\tau\sum_{j=1}^n[\tilde\eta-\alpha_{j-1}]
      [(\tilde\eta-\alpha_{j-1})\tilde\rho+\alpha_j-\tilde\eta ]\\
      \tilde\eta-\tau(1-\tilde\rho)\sum_{j=1}^n[\alpha_j -\tilde\rho\alpha_{j-1}-(1-\tilde\rho)\tilde\eta ]
    \end{smallmatrix}\right)
\end{equation*}
which results in the estimate
\begin{equation}\label{eq:mvMean}
  \bfmu =
  \left( \begin{smallmatrix}
      \tilde\alpha_0 \\ \tilde\rho\\ \tilde\eta \end{smallmatrix}\right) -
  \boldsymbol{\varSigma}
  \left(
    \begin{smallmatrix}
      \tilde\alpha_0 - \tilde\rho\tau(\alpha_1 -\tilde\rho\tilde\alpha_0-(1-\tilde\rho)\tilde\eta )\\
      \tilde\rho+\tau\sum_{j=1}^n[\tilde\eta-\alpha_{j-1}]
      [(\tilde\eta-\alpha_{j-1})\tilde\rho+\alpha_j-\tilde\eta ]\\
      \tilde\eta-\tau(1-\tilde\rho)\sum_{j=1}^n[\alpha_j-\tilde\rho\alpha_{j-1}-(1-\tilde\rho)\tilde\eta ]
    \end{smallmatrix}\right).
\end{equation}
A difficulty arises however, since the above inverse variance matrix is not
guaranteed to be positive definite (symmetric yes!) as the elements are
random. Essentially, this means that the derivatives are not zero within the
range of positive definite matrices, $\boldsymbol{\varSigma}$.  On average in
this implementation $\boldsymbol{\varSigma}$ fails to be positive definite
every 16 iterations. Our approach will be to use \eqref{eq:mvSigma} when it
is positive definite.

%% \note{I think this is a fairly easy result. If the proposal variance matrix
%% depends on first or higher level model parameters only then the matrix will
%%   not necessarily be positive definite. If the matrix depends on the
%%   \emph{data} and first level parameters as in Chapter~\ref{ch:twoway} the
%%   variance matrix will always be positive definite}
In cases where \eqref{eq:mvSigma} is not positive definite we force the
off-diagonal elements to be zero. Note that forcing the off-diagonal elements
to being identically zero reduces our proposal from being a trivariate normal
to being a product of three univariate normals.  There are two possible
centering points if the off-diagonal elements are set to 0, corresponding to
$\tilde\rho=0$ or $\tilde\rho=1$. We pick the one corresponding to
$\tilde\rho=1$ since $M_2$ is a sub-model of $M_1$ with $\rho$ identically
equal to 1. Also with $\tilde\rho=1$ fixing the off-diagonal elements at 0
dictates that $\tilde\alpha_0=\alpha_7$ and $\tilde\eta=2\alpha_7-\alpha_1$.

To get the parameters for the density $q(\alpha'_0)$ we simply use the
conditional posterior of $\alpha_0$ given $M=M_2$. This density has mean
$(1+\tau)^{-1}(\tau\alpha_1)$ and variance $(1+\tau)^{-1}$.  This choice can
be shown to be optimal in terms of maximising the acceptance probability for
proposed moves and also satisfies the $k^{th}$-order equations
\eqref{eq:kthorder}. To see this, we expand
\begin{equation*}
  \pi(M_2,\alpha'_0)\propto
  \bfL(R^n\vert\bfalpha',\sigma')
  p(\bfalpha'\vert\alpha'_0)
  p(\alpha'_0)p(\sigma')p(\tau') p(M_2)
\end{equation*}
and supposing that $q(\alpha'_0)\sim\dnorm{\mu'_0}{v'_0}$, we compute the
equations
\begin{equation*}
  \frac{\partial}{\partial \alpha'_0} \log A_{21}
  \biggl\vert_{\tilde{\alpha}'_0} =
  \frac{\partial}{\partial \alpha'_0}
  \Bigl(\log q(\alpha'_0)-\log \pi(M_2,\alpha'_0)+ K_1\Bigr)
  \biggl\vert_{\tilde{\alpha}'_0} =0 ,
\end{equation*}
\begin{equation*}
  \frac{\partial^2}{\partial (\alpha'_0)^2}\log A_{21}
  \biggl\vert_{\tilde{\alpha}'_0} =
  \frac{\partial^2}{\partial (\alpha'_0)^2}
  \Bigl(\log q(\alpha'_0)-\log\pi(M_2,\alpha'_0)+ K_1 \Bigr)
  \biggl\vert_{\tilde{\alpha}'_0}=0.
\end{equation*}
The term $K_1 = - \log\pi(M_1,\rho,\alpha_0,\eta) + \log
q(\rho,\alpha_0,\eta)$ is independent of the parameter of interest
$\alpha'_0$. Solving simultaneously leads to the estimates
$\mu'_0=(1+\tau)^{-1}(\tau\alpha_1)$ and $v'_0 = (1+\tau)^{-1}$ for the mean
and variance of the proposal distribution. These values are
independent of the centering point $\tilde{\alpha}'_0$ chosen.

Note that when $\tilde \rho = 1$ the new value of $\eta$ is simulated from
the prior density of $\eta$, likewise when $\tilde \rho = 0$ $\alpha_0$ is
simulated from the prior density on $\alpha_0$. This is a form of the birth
death method for reversible jump algorithm. See \citet{green1995} and
\citet[Chapter 8]{brown2004:phd}.

\subsubsection{Moving between Models $M_1$ and $M_3$}
Consider the ratio
\begin{equation*}
  A_{31} =\frac{\pi(M_1,\alpha_0,\rho,\eta)}{\pi(M_3,\eta'')}
  \frac{q(\eta'')}{ q(\alpha_0,\rho,\eta)},
\end{equation*}
notice that in taking logs and differentiating with respect to
$(\alpha_0,\rho,\eta)$ we remove all terms involving $M_2$ and $\eta'$. For
this reason the expressions given for the inverse variance matrix and mean
vector for a proposed move of type $M_3$ to $M_1$ are exactly the same as
those given in Equations~\eqref{eq:mvSigma} and \eqref{eq:mvMean}.  The
principal difference is that since model $M_3$ is a sub-model of $M_1$ with
$\rho$ identically equal to 0, we choose a centering point with
$\tilde\rho=0$. Also whenever the proposed variance matrix is not positive
definite we again force the off-diagonal elements to be zero which forces
$\tilde\alpha_0=2n\alpha_1-2\sum_{j=1}^n\alpha_j+\alpha_7$ and
$\tilde\eta=\alpha_1$.  Likewise the parameters for the proposal density
$q(\eta'')$ which maximises $A_{31}$ can be shown to be the posterior
conditional mean of $\eta''$ and the posterior conditional variance of
$\eta''$ given that $M=M_3$. This density has mean
$(1+n\tau)^{-1}(\tau\sum_{j=1}^n\alpha_j)$ and variance $(1+n\tau)^{-1}$.

\subsubsection{Moving between Models $M_2$ and $M_3$}
For a move between models $M_2$ and $M_3$ there is no change in the size of the
parameter vector. The acceptance probability for such a move is
$\min\lbrace 1, A_{32}\rbrace$ where
\begin{equation*}
  A_{32} =\frac{\pi(M_2,\alpha'_0)}{\pi(M_3,\eta'')}
  \frac{q(\eta'')}{ q(\alpha'_0)}.
\end{equation*}
We use Gaussian densities for the proposals $q(\eta'')$ and $q(\alpha'_0)$.
Solving
\begin{equation*}
\frac{\partial}{\partial \eta''} \log A_{32}\Bigl\vert_{\tilde\eta''}=0
\hbox{ and }
\frac{\partial^2}{\partial (\eta'')^2} \log A_{32}\Bigl\vert_{\tilde\eta''} =0
\end{equation*}
simultaneously shows that $q(\eta'')$ has mean
$(1+n\tau)^{-1}(\tau\sum_{j=1}^n\alpha_j)$ and variance $(1+n\tau)^{-1}$.
The reader will notice at once that these quantities are the conditional
posterior mean and variance of $\eta''$ given $M=M_3$.
Similarly solving
\begin{equation*}
\frac{\partial}{\partial \alpha'_0} \log A_{32}\Bigl\vert_{\tilde\alpha'_0}=0
\hbox{ and }
\frac{\partial^2}{\partial (\alpha'_0)^2} \log A_{32}
\Bigl\vert_{\tilde\alpha'_0} =0
\end{equation*}
simultaneously shows that $q(\alpha'_0)$ has mean
$(1+\tau)^{-1}(\tau\alpha_1)$ and variance $(1+\tau)^{-1}$, which are the
conditional posterior mean and variance of $\alpha'_0$ given $M=M_2$.

We can summarise this by saying that $q(\eta'') = \pi(\eta''\vert M_3)$ and
$q(\alpha'_0) = \pi(\alpha'_0\vert M_2)$ are the proposals which will
maximise the acceptance probability for proposed moves between models $M_2$
and $M_3$, and that these choices are independent of the centering point
chosen. In this case the ratio $A_{32}$ reduces to
\begin{align*}
  A_{32} &= \frac{\pi(M_2, \alpha'_0)}{\pi(M_3, \eta'')}
  \frac{q(\eta'')}{ q(\alpha'_0) }\\
  &=  \frac{\pi(M_2, \alpha'_0)}{\pi(M_3, \eta'')}
  \frac{\pi(\eta''\vert M_3)}{\pi(\alpha'_0\vert M_2)}.
%%  =&\frac{\pi(M_2)\pi(\alpha'_0\vert M_2)}{ \pi(M_3)\pi(\eta''\vert M_3)}
%%  \frac{\pi(\eta''\vert M_3)}{\pi(\alpha'_0\vert M_2)}
%%  \text{ (Bayes' Theorem)} \\
%%  =&\frac{\pi(M_2)}{\pi(M_3)} .
\end{align*}
In our simulations using this term should increase the between model moves.
This was observed in our simulations as all proposed moved from model $M_3$
to model $M_2$ were accepted, whereas for the vanilla implementation such
moves were accepted with probability $0.498$. Similarly a proposed move from
model $M_2$ to model $M_3$ is accepted with probability $0.930$ when the
posterior conditionals are used as proposals, improving upon the $0.440$
probability obtained with the vanilla implementation. The empirical results
observed here are actually specific cases of more general results which can
be found in ~\citet{ehlers2002}.
\begin{figure}
    \centering
    \makebox{
    \includegraphics[width=0.8\textwidth]{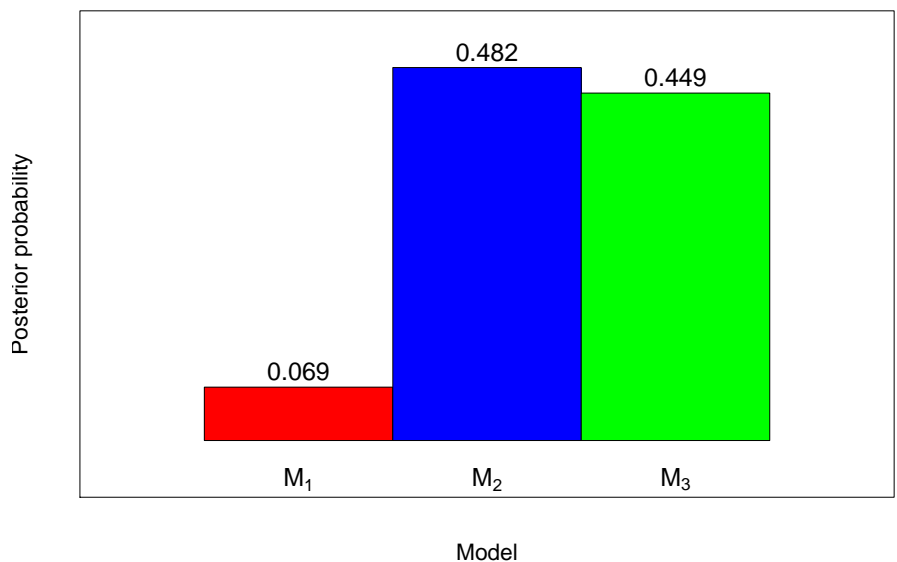}}
    \caption{\label{fig:ds3rj-eff}Posterior model probabilities for models 
      $M_1$, $M_2$, and $M_3$, second reversible jump implementation.}
\end{figure}
%%%
\begin{figure}
    \centering
    \makebox{
    \includegraphics[width=0.8\textwidth]{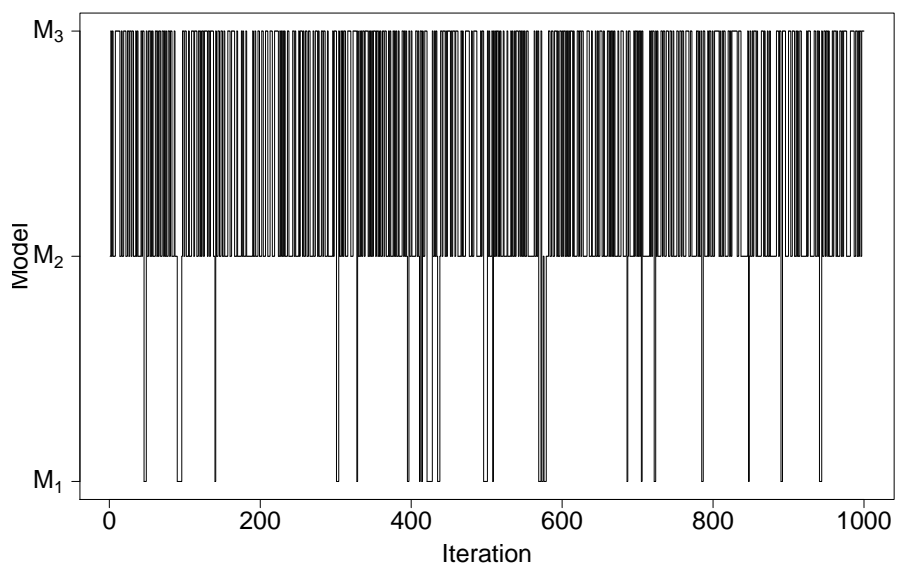}}
    \caption{\label{fig:ds3-model-eff}Trace plot of the model indicator, 
      second reversible jump implementation. The horizontal axis shows 
      the iteration number and the vertical axis shows the model indicator.}
\end{figure}

\subsection{Comparing the Model-move Schemes}
%%\note{Need to compute MC standard errors.}
%% \begin{figure}%
%%     \centering
%%     \includegraphics[width=0.8\textwidth]{ds3-stat-van.ps}
%%     \caption{Plot showing how the pilot tuned algorithm approaches
%%       stationarity.}
%%     \label{fig:ds3-stat-van}
%% \end{figure}
%% \begin{figure}%
%%     \centering
%%     \includegraphics[width=0.8\textwidth]{ds3-stat-eff.ps}
%%     \caption{Plot showing how the efficient proposals algorithm
%%       approaches stationarity.}
%%     \label{fig:ds3-stat-eff}
%% \end{figure}
\begin{figure}
    \centering
    \makebox{
    \includegraphics[width=0.8\textwidth]{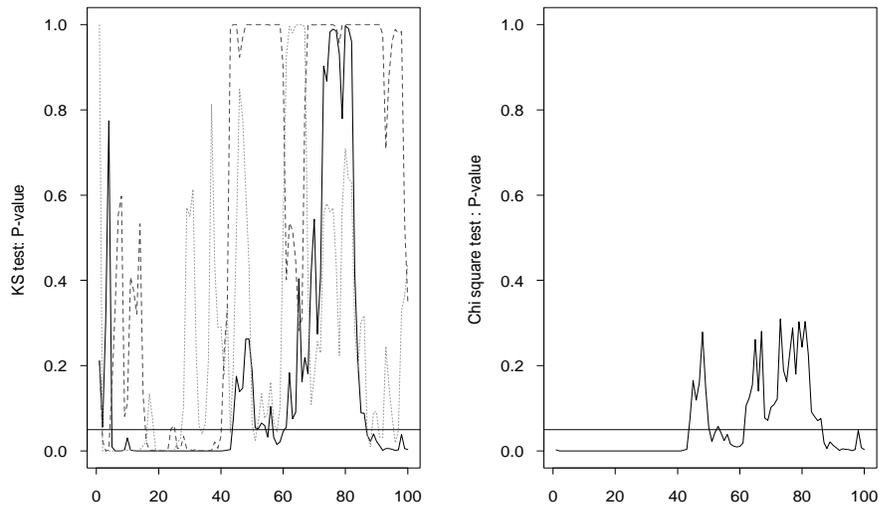}}
  \caption{\label{fig:ds3diag-van}Convergence diagnostics for the vanilla
    implementation. The horizontal axis times 1000 gives the iteration
    number.}
\end{figure}
\begin{figure}
  \centering 
  \makebox{
    \includegraphics[width=0.8\textwidth]{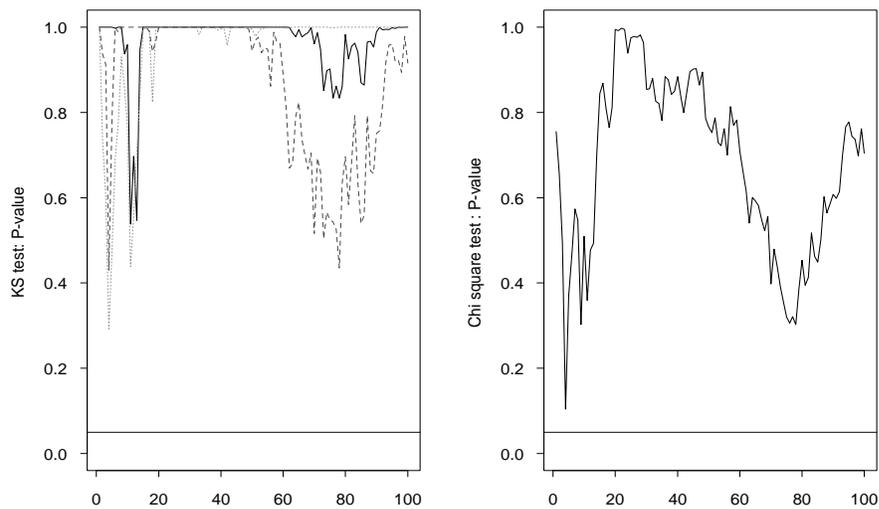}}
    \caption{\label{fig:ds3diag-eff}Convergence diagnostics for the 
      automatic proposals implementation. The horizontal axis times 
      1000 gives the iteration number.} 
\end{figure}
%%%

The empirical transition matrices for the vanilla reversible jump method,
$P^{\text{van}}$, and for the second order method, $P^{\text{eff}}$, are,
respectively
\begin{equation*}
  P^{\text{van}} =
  \bordermatrix{ & M_1 & M_2 & M_3 \cr
    M_1 & 0.703 &0.154 &0.142 \cr
    M_2 & 0.020 &0.758 &0.220 \cr
    M_3 & 0.021 &0.249 &0.729 \cr}
  \hbox{ and }P^{\text{eff}}=
  \bordermatrix{ & M_1 & M_2 & M_3 \cr
    M_1 & 0.597 &0.121 &0.281\cr
    M_2 & 0.018 &0.516 &0.465\cr
    M_3 & 0.043 &0.501 &0.456\cr}
\end{equation*}
The empirical transition matrices are computed by setting the $(i,j)$-element
equal to the proportion of times the model indicator $M_j$ follows the model
indicator $M_i$ for one long run of the reversible jump algorithm, in this
case for $1000000$ iterations.

They matrices clearly that between model (off-diagonal) transitions have
increased for $P^{\text{eff}}$ the transition matrix for the efficient
proposals method, except between models $M_1$ and $M_2$ where there were
small decreases. To assess convergence of the algorithm, we simulated 3
chains using different starting values and different random number seeds for
a total of $1000000$ iterations. In Section~\ref{sec:conv} we introduced two
methods of assessing convergence of reversible jump chains. Both the
$\chi$-square and Kolmogorov--Smirnov diagnostics are used to assess
convergence of our simulations. These diagnostics are plotted in
Figures~\ref{fig:ds3diag-van} and ~\ref{fig:ds3diag-eff} for the vanilla
reversible jump algorithm and efficient proposals implementations,
respectively. Clearly the efficient proposals implementation performs better
than the vanilla implementation

We summarise by giving the efficient proposals results applied to the models
discussed in Section~\ref{sec:ds3:revjump} and compare them with those obtained
using the vanilla reversible jump algorithm using the fine-tuned proposals
described in Section~\ref{sec:ds3:revjump}. We end this section by briefly
addressing convergence issues. The posterior model probabilities are shown in
Figure~\ref{fig:ds3rj-eff}, the posterior model probabilities are similar to
those obtained in Section~\ref{sec:ds3:revjump}. Model $M_1$ has posterior
probability 0.069, $M_2$ has posterior probability 0.482 and $M_3$ has
posterior probability 0.449. While the computing effort required to implement
this model is a bit greater than that required for the vanilla reversible
jump method, the improved mixing can also be seen by comparing
Figures~\ref{fig:ds3-model} and \ref{fig:ds3-model-eff}.
Figure~\ref{fig:ds3-model-eff} shows that the algorithm jumps between models
more frequently for the second implementation compared with the fine-tuned
proposals implementation shown in Figure~\ref{fig:ds3-model}. The within
model parameter estimates are almost identical to those obtained using the
implementation in Section~\ref{sec:ds3:revjump} and are not tabulated here.
The minor differences we attribute to Monte Carlo errors.

%% In the next chapter
%% we apply the methods used in this paper to more complex problems.
%% Specifically we analyse workers compensation insurance loss ratios using one
%% way and two way models. In a previous study, \citet{klugman1987} questioned
%% whether a two way model might offer improvements over a one way model. Using
%% the automatic proposals we compare models using the reversible jump
%%algorithm. For consistency also we continue or use of the $DIC$ criterion for
%% model comparison.
\section{Summary}
The reversible jump algorithm is presented as a method of computing posterior
model probabilities in a Bayesian setting. The vanilla reversible jump
algorithm although theoretically sound has some implementational problems.
One such problem is the choice of mapping function, another is the choice of
proposal density parameters. In this paper we have shown how recent
methodological advances in reversible jump computing can be applied to model
selection problems. This is particularly useful for actuarial practitioners
where the most appropriate choice of model is important.

\bibliographystyle{chicago}\bibliography{lratios}

\end{document}